\documentclass[12pt]{report}
\usepackage{amsfonts} 
\usepackage{a4}
\usepackage{epsfig}
\usepackage{latexsym}
\begin{document}

\title {Vacuum energy in the background of delta potentials}
\author{Marco Scandurra \thanks{e-mail: scandurr@itp.uni-leipzig.de} \\
  Universit\"at Leipzig, Fakult\"at f\"ur Physik und Geowissenschaften \\
  Institut f\"ur Theoretische Physik\\
  Augustusplatz 10/11, 04109 Leipzig, Germany} \maketitle

\begin{abstract}
The response of vacuum to the presence of external conditions is the subject of
this work. We consider a generalization of the Casimir effect in the presence of
curved boundaries on which a sharp potential is concentrated. The profile of the
potential is a delta function, which has some  features in common with a hard boundary and some with a smooth background field. The boundaries investigated  are:  i) a spherical shell, ii) a cylindrical shell, iii) a magnetic flux tube. The vacuum energy is calculated by means of the Jost function of the scattering problem related to the field equation. The energy is then renormalized by means of a zeta functional approach adopting the heat-kernel expansion. The heat kernel coefficients are calculated and a discussion of the UV-divergences of the model is presented. The renormalized vacuum energy $E^{ren}$ is then numerically studied and plotted. The sign of  $E^{ren}$ is found to be always negative in the case of the cylindrical shell, while in the case of the spherical shell and of the magnetic flux tube the sign depends on the value of the radius.  
\end{abstract}

\tableofcontents

\chapter{Introduction}
The investigation of vacuum energies in the context of quantum field theory under external conditions plays an important role in several areas of modern theoretical physics. The external conditions can be given by classical fields, like gravitational fields \cite{Trunov,Birrel}, electromagnetic fields, monopoles\cite{t'Hooft,Polyakov} and more exotic fields like sphalerons \cite{Klinkhamer} and electroweak skyrmions \cite{Skyrme}. External condition may also be given by the boundaries of a quantization volume, they can be surfaces with an arbitrary shape on which the zero point fluctuations are forced to obey some boundary conditions. In the latter case the calculation of the vacuum energy is  simple, at least for the geometries for which separation of the variables occurs. The most elementary boundary conditions are that of unpenetrable surfaces confining a quantum field in its vacuum state. In the case of two parallel plates in the electromagnetic vacuum the result is the well known Casimir effect, which has undergone a number of experimental verifications with increasing precision \cite{Sparnaay,Tabor,Lamoreaux,Mohideen}. For  other fields the unpenetrable surface is represented by Dirichlet boundary condition in the scalar field case and bag boundary conditions for the spinor field case, which find application in the bag model of quantum Chromodynamics \cite{bagmodel, repetita}.
A natural generalization of the Casimir effect are situations in which the boundaries become transparent at higher frequencies.  An approach to this generalization with ``softened'' boundaries was proposed in \cite{Actor,Joon-Il,Norberto}. This type of problems are interesting in order to study the response of vacuum to the presence of non-ideal boundaries, 
but also in view of more general investigations in the background of  a smooth external potential. The latter situation is the most difficult to approach. Usually one can perform analytical calculations only for some special examples like that of squared potential profiles. Some work in this direction has been done in \cite{Bordag smooth}, where the vacuum energy of a scalar field in the background of a square well potential and of a piecewise oscillatory potential were calculated in the one dimensional case. Some numerical evaluations for smooth potentials in three dimensions are also known \cite{BordagKirstenHellmund}. Given the difficulty to solve analytically the problem, it is desirable to investigate a simple model, which allows for  an explicit solution, while keeping the essential feature of the transparency of the boundary unchanged. 
In the present work we propose  a study of the vacuum energy in the presence of singular potentials given by delta functions. Delta functions are known to be a good idealization of a strong potential concentrated in a very small region \cite{Berezin,Albeverio}. We will use it to describe thin walls which show transmission properties. Such walls can be called ``semi-transparent''.  In the context of Casimir like problems a calculation with delta functions already exist \cite{Bordag Robaschik}, however only  for the case of parallel semi-transparent plates. In the present work we will consider more complicated geometries with spherical and cylindrical symmetry. 
In the third chapter a computation of the vacuum energy of a massive scalar field in the background of a semi-transparent spherical shell is performed. In the fourth chapter we compute the vacuum energy for a cylindrical shell. In the fifth and sixth chapter we extend the analysis to a scalar and to a spinor field in the presence of magnetic strings. 
Each of these situations has its own history which we describe briefly in the following.

After the first calculation of H.B.G. Casimir in 1948 \cite{Casimir1948} showed that two parallel mirrors in vacuum attract each oder with a force depending on the fourth power of their separation, the interest of the community turned to the problem of the conducting spherical shell in the electromagnetic vacuum. It was expected that the Casimir force would be attractive, since the spherical surface can ideally be divided into an infinite number of plane parallel surfaces. Following this idea, Casimir suggested in 1953 that an electron might be envisioned as a spherical shell of charge $e$ and that the  attractive force associated with the electromagnetic vacuum might exactly balance the outward repulsive force associated with the electrostatic self-energy. The purpose was to explain by means of the zero point fluctuations the stability of a charged particle. This fascinating model failed in 1968, when Boyer showed that the Casimir stress on a perfectly unpenetrable spherical surface is positive and that the vacuum tends to expand the shell. The result of Boyer, confirmed later by other authors \cite{B. Davies,Milton DeRaad Schwinger,Balian Duplantier}, put for the first time under the attention the strong dependence of the vacuum energy on the geometry of the system. Moreover, the work of Boyer showed that the simple subtraction of the empty Minkowski space is in general not sufficient to yield a finite result. New divergences, associated with the  curved surfaces, appear in the calculations. The treatment of this divergences needed new and  more efficient techniques which have been developed only in the last decades \cite{Dowker1,Wipf,Dowker,Leseduarte Romeo} and which are still susceptible of improvements. 

Unfortunately we have no general rule to predict the sign of the Casimir energy for a certain boundary shape. The research has gone forward in the last 50 years with investigations of special cases, from which we try to extract as more physical information as possible.
 A deep understanding of the phenomenology of quantum vacuum perturbed by external conditions is still far from being reached. 

Among the special cases, which allow for an analytical calculation, is the cylindrical configuration. Since the separation of the variable is possible in the wave equation with cylindrical coordinates and the solution for the field is known to be given by Bessel functions, the summation over the modes can be performed. The first complete calculation, realized in 1978 by DeRaad and Milton, \cite{Milton-cyl} showed that the energy of a conducting cylindrical shell is negative. 
While the investigation has turned recently to the problem of compact dielectric cylinders and spheres \cite{Brevik dispersive,Lambiase Nesterenko,Bordag dielectric,Klich}, a treatment of the semi-hard hollow cylinder and sphere  is still missing. Chapter 3 and 4 of our work, are devoted to the calculations for this kind of  shells. 
 
There is an evident analogy between a cylindrical material boundary and a magnetic flux tube. A study of the ground state energy in the background of a magnetic string is motivated not only by the necessity of completing the class of problems mentioned above (for instance with a multiple comparison of the sign of the energy), but also in order to gain  some insight in to the  ``interaction'' of vacuum with purely magnetic fields. Spinor fields in the background of homogeneous magnetic strings are a well know in the literature \cite{Heisenberg,Weisskopf}. Computations of the ground state energy  were performed by several authors \cite{Serebryany,Gornicki,Sitenko} which extensively considered  the case of a string with infinitesimal  radius. This model has the difficulty that the energy density per unit volume cannot be integrated to give the energy density per unit length of the string. Another matter of debate is the stability of the string in vacuum: if there exist a value of the radius for which the vacuum energy is minimized. Calculation for an homogeneous magnetic field inside a flux tube \cite{master} and for a planar field with a special profile in the $x$-direction and homogeneous in the $y$-direction \cite{Cangemi} showed the ground state energy to be without minimum. In order to shed some light into these topics, we propose here a calculation of the vacuum energy of a magnetic flux tube with finite radius and with an inhomogeneous potential given by a delta function. The calculation is performed for a massive scalar field (chapter 5) and for a massive spinor field (chapter 6). 

The technique we adopt for the computation of the vacuum energy makes use of the Jost function of the scattering problem related to the background. The ground state energy
\begin{equation}
E_B\ =\ \frac 12 \sum_{(n)} \omega_{(n)},
\end{equation}
where the $\omega_{(n)}$ are the discrete values of the Hamilton operator associated with the background $B$, can be transformed into an integral over the momentum $k$ by means of the Cauchy theorem and  with the deformation of the integration contour on the imaginary axis. Then we can write\footnote{This equation is valid in a spherically symmetric background. For a cylindrical geometry the equation changes slightly.} 
\begin{equation}
E_{B}\ =\ -\frac{\cos \pi s}{\pi} \mu^{2s} \sum_l \int_{m}^\infty dk (k^2-m^2)^{1/2-s}
\frac{\partial}{\partial k} \ln f_l(ik)\ ,
\end{equation}
here $\int dk$ and  $\sum_l$ are respectively  the integration over the momentum $k$  and the summation over all the remaining  quantum numbers $l$; $m$ is the mass of the quantum field, $s$ is a regularization parameter and $\mu$ a mass parameter which one introduces to maintain the correct dimensions of the energy. The Jost function $f_l(ik)$ is given for each problem by the coefficients of the regular solution of the wave equation. A nice feature of this integral representation is that the contribution of the possible bound states is ``automatically'' included. 

The renormalization of (1.2) is performed with the zeta functional method (for a review see \cite{zeta function}). In this approach the divergences are completely described by the lowest heat-kernel coefficients associated with the field equation. We expand the regularized vacuum energy  in powers of the mass of the field. This expansion contains the heat-kernel coefficients. The renormalization is then performed by subtraction of the pole terms and of the terms proportional to the non-negative powers of the mass. The reason for this, is that the renormalized vacuum energy must satisfy the condition
\begin{equation}
E_B^{ren}\ =\ 0\  \ \ \ \ {\rm for}\ \ \ \ \ m\rightarrow\infty
\end{equation}
which is natural under the physical point of view. This condition defines the renormalized vacuum energy removing the ambiguity introduced with the mass parameter $\mu$. 

For each of the configurations considered in this work we will calculate the relevant heat-kernel coefficients. Their importance resides not only in the calculation of the UV-divergences, but also in the study of the asymptotic behaviour of the energy, which is provided by the lowest half-integer heat-kernel coefficients.

Other mathematical tools are employed in the course of the calculations, like the introduction of the uniform asymptotic expansion of the Jost function  to perform the analytical continuation $s\rightarrow 0$; and the application of the Abel-Plana formula for half-integer variables \cite{Mostepanenko}. The latter technique allows for an analytical elaboration of the vacuum energy toward a more compact form.  After the renormalization the vacuum energy is evaluated numerically as a function of the relevant parameters of the background, namely the radius of the shell and the coupling.
At the end of each chapter we display plots of the renormalized vacuum energy. The plots are briefly discussed in a special section for each of the 4 problems. The global results of the work are then summarized and commented in the conclusion. We would like to begin with a short introductory description of vacuum in  QFT  and of the zeta functional renormalization.

\chapter{Vacuum energy and renormalization}

\section{Vacuum energy in quantum field theory}
Quantum field theory is a formulation of the laws of nature which allows to predict and describe the phenomena of particle physics. The canonical quantization of a field (or second quantization) follows a standard procedure, which we summarize briefly in the following for the case of a massive scalar field.
The Lagrangian density of a real field $\phi(x)$ with spin 0
 and mass $m$, is
\begin{equation}\label{lagrangean}
{\cal L}(x)\ =\ \frac 12 \frac{\partial\phi}{\partial x_\mu}\frac{\partial\phi}{\partial x^\mu}-\frac 12 m^2 \phi^2(x)\ ,
\end{equation}
where natural units $\hbar=c=1$ are used.
From the Euler-Lagrange equation $\frac{\partial{\cal L}}{\partial \phi}=\frac{\partial}{\partial x_\mu}\frac{\partial}{\partial(\partial^\mu \phi)}$ one arrives at the Klein-Gordon equation
\begin{equation}\label{Klein-Gordon}
(\Box + m^2)\phi(x)\ =\ 0.
\end{equation}
 With the definition of  the canonical conjugate field 
\begin{equation}\label{canonical-field}
\pi(x)\ =\ \frac{\partial{\cal L}}{\partial\dot{\phi}(x)}\ =\ \dot{\phi}(x)
\end{equation}
 the  Hamiltonian density  in terms of canonical variables reads
\begin{equation}\label{Hamiltonian-density}
{\cal H}\ =\ \pi(x)\dot{\phi}-{\cal L}\ =\ \frac 12\left(\pi^2(x)\ +\ (\vec{\nabla}\phi(x))^2\ +\ m^2\phi^2(x)\right)\ .\end{equation}
In order to quantize the system the canonical variables are transformed into the operators $\hat{\phi}$ and $\hat{\pi}$ obeying the commutation relations at equal time
\begin{equation}\label{commutation-phi-pi}
[\hat{\phi}(\vec{x},t),\hat{\pi}(\vec{x} ' ,t)] \ =\  i\delta^{(3)}(\vec{x}-\vec{x} ' )
\end{equation} 
\begin{equation}\label{commutation-phi-phi}   
[ \hat{\phi}(\vec{x},t),\hat{\phi}(\vec{x} ' ,t)] \ =\  [\hat{\pi}(\vec{x},t),\hat{\pi}(\vec{x} ' ,t)] \ =\ 0\ .
\end{equation}
Then, the quantized Hamiltonian reads
\begin{equation}\label{Hamilton-operator}
\hat{H}\ =\ \int d^3x \frac 12 \left( \hat{\pi}^2(\vec{x},t)\ +\ (\vec{\nabla}\hat{\phi}(\vec{x},t))^2\ +\ m^2\hat{\phi}^2(\vec{x},t)\right)\ 
\end{equation}
Now, we decompose the field and its canonical conjugate using a basis of eigenfunctions $e^{ipx}$ and introducing the operators $\hat{a}_{\vec{p}}$ and $\hat{a}^{\dag}_{\vec{p}}$ 
\begin{equation}\label{basis-phi}
\hat{\phi}(\vec{x},t)\ =\ \int  \frac{d^3p}{\sqrt{2\omega_p(2\pi)^3}}\left(\hat{a}_{p} e^{ipx}+\hat{a}^{\dag}_{p} e^{-ipx}\right)
\end{equation}
\begin{equation}\label{basis-pi}
\hat{\pi}(\vec{x},t)\ =\ \int  \frac{d^3p}{\sqrt{2\omega_p(2\pi)^3}}(-i\omega_p)\left(\hat{a}_{p} e^{ipx}-\hat{a}^{\dag}_{p} e^{-ipx}\right),
\end{equation}
where $\omega_p=\sqrt{\vec{p}^2+m^2}$ is the energy of a particle with momentum $\vec{p}$ and the operators $\hat{a}_{\vec{p}}$ and $\hat{a}^{\dag}_{\vec{p}}$ obey the commutation relations  
\begin{equation}\label{commutation-a-a+}
[ \hat{a}_{\vec{p}},\hat{a}^{\dag}_{\vec{p} ' }]  \ = \ \delta^{(3)}(\vec{p}-\vec{p} ' )
\end{equation}
\begin{equation}\label{commutation-a-a}
[ \hat{a}_{\vec{p}},\hat{a}_{\vec{p} ' }]  \ =\ [ \hat{a}^{\dag}_{\vec{p}},\hat{a}^{\dag}_{\vec{p} ' }]\ =\ 0\ ,        \end{equation}
which can be proven by means of (\ref{commutation-phi-phi}) and (\ref{commutation-phi-pi}). The relations (\ref{commutation-a-a+}) and (\ref{commutation-a-a}) represent a Fock space in which the  $\hat{a}_{\vec{p}}$ and $\hat{a}^{\dag}_{\vec{p}}$ play the role of creation and annihilation operators respectively. Inserting (\ref{basis-phi}) and (\ref{basis-pi}) in (\ref{Hamilton-operator}) we can express the Hamiltonian  in terms of these  operators  
\begin{equation}\label{Hamiltonian-continuous}
\hat{H}\ =\ \frac 12 \int d^3p\ \omega_p (\hat{a}^{\dag}_{\vec{p}}\hat{a}_{\vec{p}}\ +\ \hat{a}_{\vec{p}}\hat{a}^{\dag}_{\vec{p}})\ .
\end{equation}
To give a physical meaning to this expression we enclose the system in a large quantization box of dimensions $L$x$L$x$L$ on whose surface the field obeys periodic boundary conditions. Then the integration over the momentum in the continuous is replaced by a sum over the spectrum of the discrete eigenvalues or ``modes'' of the field and the Hamiltonian becomes
\begin{equation}\label{Hamiltonian-discrete}
\hat{H}\ =\ \frac 12 \sum_{j}\omega_{j}(\hat{a}^{\dag}_j\hat{a}_j\ +\ \hat{a}_j\hat{a}^{\dag}_j)\ .
\end{equation}
The commutation relation (\ref{commutation-a-a+}) becomes
\begin{equation}\label{discrete-commutation}
[ \hat{a}_j,\hat{a}^{\dag}_{j ' }]  \ = \ \delta_{jj ' }.
\end{equation}
With the help of (\ref{discrete-commutation}) the Hamiltonian is rewritten as
\begin{equation}\label{Hamiltonian-final}
\vec{H}\ =\ \sum_{j}\omega_{j}(\hat{a}^{\dag}_j\hat{a}_j\ +\ \frac 12)\ .
\end{equation}
In this form the Hamilton operator can be easily interpreted: in each sate $j$ there are some particles whose number is given by $\hat{n}_j\ =\ \hat{a}^{\dag}_j\hat{a}_j$ and the energy quantum is $\omega_j$\ . Moreover it appears in (\ref{Hamiltonian-final}) a contribution $\frac 12 \omega_j$ which is independent of the occupation number. It is the so called  zero point energy or vacuum energy or, if one wishes,  ground state energy, in analogy with the lowest energy level of an harmonic oscillator in quantum mechanics. This contribution is not a peculiarity of the Klein-Gordon field. It appears in the Hamilton operator of each quantum field (complex scalar fields, Dirac fields,  Maxwell fields, gauge fields in Chromodynamics and so on). One must only be careful with the commutation relation (\ref{discrete-commutation}), which changes in each case. For instance in the quantization of a spinor field the addend $\frac 12$ in (\ref{Hamiltonian-final}) appears with a sign minus, because of the use of anticommutators in the Fermi-Dirac statistics. Furthermore the index $j$ must take into account all oder possible degree of freedom of the field. So it is usually replaced by a subscript $(n)$ which includes all the quantum numbers in a compact form. The vacuum energy 
\begin{equation}\label{zero-point-energy}
E_0\ =\ \frac 12 \sum_{(n)}\omega_{(n)}
\end{equation}
is a divergent quantity. Since physically we can only observe energy differences, the quantity (\ref{zero-point-energy}) as a constant of the energy, cannot be physically observed. However a variation of $E_0$ caused by some external conditions can be  measured in laboratory. We must not forget that result (\ref{Hamiltonian-final}) is obtained by imposing periodic boundary conditions on the surface of a quantization box of size $L$, which is a fictitious boundary.  If we introduce a physical boundary on which the field is forced to obey some conditions, the modes of the field will change and with them the vacuum energy. The most eminent example of this situation is the Casimir effect. This is the attraction of two conducting parallel plates in the vacuum of the electromagnetic field. The force is caused by a  difference in the spectrum of the electromagnetic wave vector $\vec{k}=(k_x,k_y,k_z)$ inside and outside the plates. Because of the boundary conditions:  $E_{\|}=0,\ B_{\bot}=0$ for an electromagnetic wave near a perfect conductor, the component of $\vec{k}$ perpendicular to the plates can only take the discrete values: $k_z=\pi n/d$ inside the cavity, where $d$ is the distance of the plates and $n=1,2,...$ All other components inside and outside the cavity have a number of allowed values which is so large that can be well approximated by a continuous spectrum. Then we have
\begin{eqnarray}\label{Casimir-difference}
\Delta E_0 & = & E_0^{in}\ -\ E_0^{out}\nonumber \\
           & = & \sum_{n} \int dk_x dk_y \sqrt{k_x^2\ +\ k_y^2\ +\ \left(\frac{\pi n}{d}\right)^2}\nonumber\\
           &   &    -\ \int d^3k\sqrt{k_x^2\ +\ k_y^2\ +\ k_z^2}\ ,
\end{eqnarray}  
where the two independent polarisation states have been taken into account. This difference was computed by Casimir \cite{Casimir1948} with the help of the Euler-McLaurin formula (see \cite{Abramowitz}, equation 3.6.28). The result is a negative energy:  $\Delta E_0\ =\ -\frac{\pi^3}{720 d^3}$ which gives rise to an attractive force between the plates.

The conducting plates are just an example of a wide class of  boundaries which influence the vacuum.
For a scalar field the boundary conditions can be Dirichlet boundary conditions 
\begin{equation}\label{Dirichlet}
\phi(x)|_{x=x_B}\ =\ 0\ ,
\end{equation}
where $x_B$ is the value of the spatial coordinates at the surface of the boundary, or Robin boundary conditions
\begin{equation}\label{Robin}
\partial_n(u(x)\phi(x))|_{x=x_B}\ =\ 0\ ,
\end{equation}
where $\partial_n$ is the derivative in the direction normal to the surface of the boundary and $u(x)$ is an arbitrary function of the coordinates. In the case $u(x)=1$ we have the Neuman boundary conditions. For a spinor field  local boundary conditions are usually studied
\begin{equation}\label{local}
(1+\vec{n}\vec{\gamma})\psi(x)|_{x=x_B}\ =\ 0\ ,
\end{equation}
 like in the case of the bag-model for the hadrons. The variation of the spectrum (\ref{zero-point-energy}) can also be generated by the presence of a classical background field. This is  usually introduced with a potential $V(x)$ which can couple to the mass in the case of a Newtonian background field or can enter  the covariant derivative, as in the case of a gauge potential $A_\mu(x)$ which yields the operator $\nabla_\mu\rightarrow \partial^\mu+iA_\mu(x)$ in the field equation. In problems with scalar fields the potential can be given in the simple form $\alpha/R$ where $\alpha$ is a generic coupling constant and $R$ is a geometric parameter. In the context of general relativity the metric $g_{\mu\nu}$ can also play the role of the background potential.

In each of the above mentioned situations the calculation of the zero point energy requires the knowledge of the spectrum $\omega_{B(n)}$ where the label $B$ represents here a boundary condition or a background potential. The calculation of the $\omega_{B(n)}$  can be a hard task, when the geometry is non-planar and when the potential is a smooth function. The other difficulty related with the calculation of the vacuum energy, is the renormalization. In the Casimir effect, the subtraction performed in (\ref{Casimir-difference}) is just an elementary example of renormalization. Namely it is subtracted the energy of the empty unbounded space (Minkowski space contribution). This simple procedure does not give a finite result  in spherical or cylindrical configurations. For these situations the isolation of the UV-divergences is performed by asymptotic expansions of the vacuum energy in terms of a cut-off function. In recent years the zeta functional technique is used for the renormalization. We briefly describe the technique in the next section. For a more exhaustive treatment the reader is referred to \cite{zeta function}.

\section{The zeta functional method}

Zeta functions provide an elegant re-formulation of the sum (\ref{zero-point-energy}). The sum is regularized by a complex parameter $s$ which must vanish after the renormalization. The regularized energy has the form of a zeta function, whose poles are well known and can be subtracted to yield a finite energy. Moreover, a zeta function has a variety of interesting properties, like the integral representation and the heat-kernel expansion, which allow for easy mode summation and for a deeper insight into the structure of the UV-divergences of the investigated model.

\subsection{Integral representation}
Let us start with a general review of the analytical properties of zeta functions. The most simple zeta function is 
\begin{equation}\label{Euler}
\zeta(s)\ =\ \sum_n^\infty\frac{1}{n^s}\ .
\end{equation}
This is commonly known as the Riemann zeta function $\zeta_R(s)$, since Riemann studied the general case of a complex variable $s$ and found the analytical continuation in the region $\Re s<1$, where representation (\ref{Euler}) does not converge. The analytical continuation is of interest also for physical  applications. In quantum field theory one generally adopts the zeta function related to a differential operator. If one has to consider a field $\phi$ and the wave equation
\begin{equation}\label{eigenvalues}
D\phi_n\ =\ \epsilon_n\phi_n\ , 
\end{equation}
where $D$ is a generic differential operator (like a Laplace operator) and $\epsilon_n$ are its eigenvalues, then the zeta function of the operator $D$ with spectrum ${\cal D}$ is defined as \cite{Carlemann}
\begin{equation}\label{zeta-operator}
\zeta_D(s)\ =\ \sum_{n\in{\cal D}}\left(\epsilon_{n}\right)^{-s}\ .  
\end{equation}
The Hamilton operator of a system is then given by the zeta function of the corresponding differential operator with a simple function of $s$ in the argument. For instance, in the problem of a spherical shell which we will analyze in the next chapter, the regularized vacuum energy is given by $\zeta (s-1/2)$, while in the case of the cylindrical shell the relevant zeta function will be $\zeta (s-1)$. They are elementary  generalizations of the Riemann zeta function. The latter can undergo at least two types of transformation into an integral. The first makes use of the integral representation of the Gamma function
\begin{equation}\label{Gamma-function-integral}
\Gamma(s)\ =\ \int_0^\infty dt\ t^{s-1} e^{-t}\ ,
\end{equation}
from which it follows
\begin{equation}\label{formula-kirsten}
n^{-s}\ =\ \frac{1}{\Gamma(s)}\int_0^\infty dt\ t^{s-1} e^{-nt}\ ,
\end{equation}
and
\begin{equation}\label{Gamma-representation-Riemann}
\zeta_R(s)\ =\ \sum_{n=1}^\infty \int_0^\infty dt\ \frac{t^{s-1}}{\Gamma(s)} e^{-nt}\ =\ \int_0^\infty dt\ \frac{t^{s-1}}{\Gamma(s)}\ \frac{1}{e^t-1}\ ,
\end{equation}
which in the case of the function $\zeta_D(s)$ becomes
\begin{equation}\label{Gamma-representation-operator}
\zeta_D(s)\ =\ \int_0^\infty dt\ \frac{t^{s-1}}{\Gamma(s)}\sum_n e^{-\epsilon_n t}\ .
\end{equation} 
The second representation makes use of the Cauchy theorem, which transforms the sum (\ref{Euler}) into a contour integral
\begin{equation}\label{contour-Riemann}
\zeta_R(s)\ =\ \int_\gamma  dt \frac{t^{-s}}{e^{2\pi it}-1}\ .
\end{equation}
The contour $\gamma$ encloses, in the complex plane, all the positive integers lying on the axis $\Re t$, as it is shown in the figure below. Each integer is a pole of the integrand in (\ref{contour-Riemann}) and the residue contributes to the initial sum (\ref{Euler}).     

\begin{figure}[ht]\unitlength1cm
\begin{picture}(6,4)
\put(2.5,0){\epsfig{file=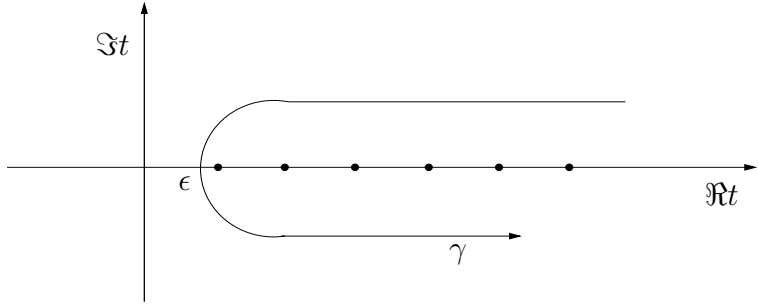,width=10cm,height=4cm}}
\put(8.4,0.6){$\gamma $}
\put(11.8,1.3){$\Re t $}
\put(4.8,1.5){$ \epsilon $}
\put(3.7,3.3){$\Im t$}
\end{picture}
\caption{\small The integration contour of (\ref{contour-Riemann}).} 
\end{figure}

The integral (\ref{contour-Riemann}) is convergent only for $\Re s>1$. We are interested in a representation in which the continuation for small $s$ is possible. To this aim we split the integration contour in two pieces: an upper contour $\gamma_+$ for $\Im t\geq 0$ and a lower contour $\gamma_-$ for $\Im t<0$. With the help of: $1/(e^{2i\pi t}-1)=-1+1/(e^{-2i\pi t}-1)$,  we get
\begin{equation}\label{contour-intermediate-step}
\zeta_R(s)\ =\ \int_{\gamma_+}  dt\ t^{-s} \left(-1+\frac{1}{1-e^{-2\pi it}}\right)\ +\ \int_{\gamma_-} dt\ t^{-s}\frac{1}{e^{2\pi it}-1}\ .
\end{equation}
The integration over $\gamma_+$ is split into two parts, the first of which can be explicitly calculated and it gives $\-\epsilon^{1-s}/(1-s)$, where $\epsilon\in (0,1)$ is the point at which the initial contour $\gamma$ crosses the real axis. In the second part we deform the integration path until it runs along to the positive imaginary axis from $\epsilon$ to $i\infty$. On this path the integrand function falls exponentially. In the integral  over $\gamma_-$ we deform the  contour downwards until it runs from  $\epsilon$ to $-i\infty$. By this means we find a result in which the analytical continuation to $s<1$ is possible. After setting $\epsilon=0$ we find 
\begin{equation}\label{sin-integral}
\zeta_R(s)\ =\ 2 \sin (s\pi/2)\int_0^\infty  dt \frac{t^{-s}}{e^{2\pi t}-1}\ .
\end{equation}
This expression can be generalized for a function $\zeta_D(s)$, as we will see in each of the following chapters, providing a useful tool for the calculation of the ground state energy.

\subsection{Renormalization via heat-kernel expansion}
The regularized vacuum energy must be at the end a finite quantity for $s=0$. To achieve this result,  the divergent contribution to the zeta function must be isolated. This is done by means of the so called global heat-kernel $K(t)$, given by
\begin{equation}\label{global}
K(t)\ \equiv\ \sum_ {n=1}^\infty e^{-nt}\  .
\end{equation}
This expression appears already in (\ref{Gamma-representation-Riemann}). It its expansion for small $t$ which is of interest:
\begin{equation}\label{global-expansion}
K(t)\ \stackrel{t\rightarrow 0}{\sim}\ \frac{1}{(4 \pi t)^{d/2}} \sum^{\infty}_{j=0} a_j t^j. 
\end{equation}
Here $d$ is the dimension of the manifold and the index $j$ assumes also fractionary values in the case of a manifold with boundary. The coefficients $a_j$ are called the {\itshape heat-kernel coefficients}. If we insert (\ref{global-expansion}) in (\ref{Gamma-representation-Riemann}), we get a formula in which  the divergences appear associated with the lowest heat-kernel coefficients. To better enlighten the problem and also to restrict us to the renormalization procedure which we will adopt in this work, let us consider a physical system given by a massive\footnote{We will work throughout this thesis with massive fields.} scalar field and by a classical background described by a potential $V_b(x)$ concentrated on a surface with arbitrary shape. The ground state energy of the system is given by
\begin{equation}\label{esempio}
E_b\ =\ \frac 12 \sum_{(n)}\sqrt{p_{(n)}^2+m^2}\ , 
\end{equation}
where $\sqrt{p_{(n)}^2+m^2}$ are the energy eigenvalues of a particle of the field and $m$ is the mass of the field. We regularize this expression with the introduction of a parameter $s$, to be put to zero after the renormalization, and we rewrite (\ref{esempio}) in terms of a zeta function
\begin{equation}\label{esempio-regularized}
E_b\ =\ \frac 12 \mu^{2s}\sum_{(n)} (p_{(n)}^2+m^2)^{1/2-s}\ =\ \frac 12 \mu^{2s}\zeta_b(s-1/2)\ ,
\end{equation} 
where $\mu$ is an arbitrary parameter with the dimension of a mass. Then we have
\begin{equation}
\frac 12 \mu^{2s}\zeta_b(s-1/2)\ =\ \frac 12 \mu^{2s} \frac{1}{\Gamma(s-1/2)}\int dt\ t^{s-3/2} K_b(t)\ .
\end{equation}
With the help of
\begin{equation}
 K_b(t)\ =\ \sum_{(n)}e^{-t(p_{(n)}^2+m^2)}\ \stackrel{t\rightarrow 0}{\sim}\ \frac{e^{-tm}}{4\pi t^{3/2}}\sum_j a_jt^j\ , 
\end{equation}
we can expand the zeta function in a series in which the poles $1/s$, contributing to the divergences, are well isolated 
\begin{eqnarray}\label{esempio-divergent}
E_b=\frac {\mu^{2s}}{2}\zeta_b(s-1/2) & \sim & -\frac{m^4}{64 \pi ^4}\left( \frac 1s + \ln \frac{4\mu ^2 }{m^2} - \frac 12\right) a_0\  -\frac{m^3}{24\pi^{3/2}}a_{1/2} \nonumber \\
                        &     &  +\frac{m^2}{32 \pi ^4}\left( \frac 1s + \ln \frac{4\mu ^2 }{m^2} - \ 1\right) a_1 \ + \frac{m}{16 \pi^{3/2}}a_{3/2}  \nonumber \\
                        &     &  -\frac{1}{32 \pi ^2}\left( \frac 1s + \ln \frac{4\mu ^2 }{m^2} - \ 2\right) a_2\ +{\cal O}(m^{-1})\ . 
\end{eqnarray}
Here the terms of order ${\cal O}(s)$ have been dropped. The renormalization of the ground state energy is then performed by simple subtraction of the pole terms $\sim 1/s$, corresponding to the lowest heat-kernel coefficients. This subtraction is physically interpreted as a re-definition of the classical parameters of the system. In fact the heat-kernel coefficients depend on the geometric  features of background.  For instance, if we calculate the vacuum energy inside and outside a spherical boundary, the coefficients $a_j$ will depend on powers of the radius $R$ of the sphere. Now, it is possible to express also the classical energy of the sphere in powers of the radius \footnote{Such a definition  appeared for the first time in \cite {Wipf}, see also section 3.1 for other details.} 
\begin{equation}\label{classical-energy}
E_b^{class}\ =\ \left(pV + \sigma S + FR +k +\frac hR\right)\ ,
\end{equation}
where $V$ is the volume of the sphere, $p$ is the pressure, $\sigma$ is the surface tension and $F$, $k$ and $h$ are other parameters with no special names. Then, the renormalization of the vacuum energy  requires a renormalization of the parameters defined in (\ref{classical-energy})
\begin{eqnarray}\label{re-definition}
pV & \rightarrow &  pV -\frac{m^4}{64 \pi ^4}\left( \frac 1s + \ln \frac{4\mu ^2 }{m^2} -\frac 12\right)\ a_0\ ;\nonumber \\
\sigma S & \rightarrow & \sigma S -\frac{m^3}{24\pi^{3/2}}\ a_{1/2}\ ; \nonumber \\
F R & \rightarrow &  F R +\frac{m^2}{32 \pi ^4}\left( \frac 1s + \ln \frac{4\mu ^2 }{m^2} - \ 1\right)\ a_1\ ; \nonumber \\
k & \rightarrow &  k + \frac{m}{16 \pi^{3/2}}\ a_{3/2}\ ; \nonumber \\
h/R & \rightarrow &  h/R -\frac{1}{32 \pi ^2}\left( \frac 1s + \ln \frac{4\mu ^2 }{m^2} - \ 2\right)\ a_2. 
\end{eqnarray}
By means of this procedure, the total energy of the system remains unchanged. The transition from (\ref{classical-energy}) to (\ref{re-definition}) does not correspond to a physical phenomenon. The parameters (\ref{classical-energy}) are ``naked'' quantities which do not take in to account quantum corrections. In an experimental measurement only the parameters given in (\ref{re-definition}) are observed. An important question is if the renormalization procedure described here is unique. In the case of a massive field, it is not clear which value the mass parameter $\mu$ (introduced for dimensional reasons) should take. Generally this ambiguity is removed by the requirement that the renormalized vacuum energy must only depend on negative powers of the mass, which is sufficient to fix univocally the meaning of the vacuum energy. In this case the terms proportional to $m^3$ and  $m^1$ in (\ref{esempio-divergent}) must also be subtracted. They correspond to the heat-kernel coefficients with fractionary index.

The heat kernel coefficients play a crucial role in this technique. An explicit knowledge of the lowest coefficients fully solves the renormalization problem for a given configuration. The coefficients $a_1$, $a_2$ and $a_3$ are known, at least in principle, \cite{firstball} for a generic background potential $V_b$:
\begin{eqnarray}\label{general-heat-kernels}
a_0  & = & \int d^3x\ , \nonumber \\
a_1  & = & -\int d^3x V_b(x)\ ,\nonumber \\
a_2  & = & \frac12 \int d^3x V_b^2(x)\,.
\end{eqnarray}
The first coefficient represents the contribution of the flat empty space, which, as we saw, is subtracted for the derivation of the Casimir effect. In the simplest example of  two parallel plates all other coefficients vanish. In general the coefficients $a_1$ and $a_2$ are not zero and must be calculated. Unfortunately  formula (\ref{general-heat-kernels}) cannot always be used. For instance, in the problems that we analyze in this thesis the background potential contains a delta function and the equation for $a_2$ would be ill-defined.  One last comment must be made about the limit of a massless quantum field. In this case equation (\ref{esempio-divergent}) does not apply and another type of expansion is needed. Furthermore the requirement  that the vacuum energy vanishes for $m\rightarrow \infty$ would fail and the renormalized energy would not have a unique meaning. However the present work is restricted to fields with finite mass. In the following chapters the mathematical tools  summarized here will be applied to four different exercises. The vacuum energy will be calculated by mode summation with integral representations of the type (\ref{sin-integral}). The renormalization will follow formula (\ref{esempio-divergent}). Modifications in the procedure, due to the specific features of each exercise, will be displayed and explained in each chapter.

\chapter{Vacuum energy for a semi-transparent spherical shell}

The interest in the zero point energy for spherically symmetric configurations has been renewed in the last years by the phenomenon of sonoluminessence\footnote{The emission of short intense pulses of light by collapsing bubbles of air in water \cite{sonoluminessenz}.}, which, in the suggestion of some authors \cite{Schwinger,Eberlein,Milton,Molina-Paris}, could be a macroscopic manifestation of vacuum.  This hypothesis, which has been lively discussed by the community, could not be completely accepted or ruled out, since an exact model for the dielectric ball in vacuum is still missing\footnote{The compact dielectric ball was also in the center of interest at the workshop ``Quantum field theory under external conditions'' held in Leipzig on 14-18 September 1998.}. The model of a hollow spherical surface has recently attracted the interest of physicists in the context of the MIT bag-model for the hadronic structure. In this model, confinement is provided by special boundary conditions requiring the vanishing of the quark and gluon currents through the boundary (see eq. (\ref{local}), and  \cite{bagmodel}). Massive scalar fields bounded by a Dirichlet spherical shell  were considered recently in \cite{Bordag Elizalde}, with the result that the sign of the ground state energy is positive. 

In the present chapter we calculate the ground state energy of a massive scalar field in the background of spherical shell with delta function. The technique developed in \cite{firstball} for generic background fields is used and explained with details. The calculation of the ground state energy is based on the knowledge of the Jost function of the associated scattering problem, which is given by a simple formula in terms of modified Bessel functions. The renormalization is performed via zeta function. The heat-kernel coefficients whose knowledge is required for the renormalization, were first derived in \cite{Bordag dielectric}. Here they are rederived in the course of calculations.  

\section{The model}
We want to study the ground state energy (GSE) of the scalar field
$\varphi(t,\vec{x})$ to be quantized in the background of a potential $V(r)$ concentrated on the surface of a sphere.  We start with the following field equation\footnote{Natural units are used}
\begin{equation}\label{S-1}
(\Box +m^2+V(r))\ \varphi(t,\vec{x})\ =\ 0,
\end{equation}
where $m$ is the mass of the field and $r=\sqrt{x^2+y^2+z^2}$. This equation will be rewritten later in polar coordinates.  The sphere  is a geometrical object with radius $R$ and a surface $S$, to whom it can be associated a
classical energy in terms of classical parameters as we mentioned in the previous chapter.  The total energy of the system reads
\begin{eqnarray}\label{S-2}
E_{TOT} & = & E_{class}+E_{quant}\nonumber \\
        & = & \left(pV + \sigma S + FR +k +\frac hR\right) +
\left({1\over 2}\sum _{n} \omega _{n}\right),
\end{eqnarray} 
 The classical part of the energy is expressed in a general form in which
 the dependence on powers of $R$ is explicit. This definition is suitable for its renormalization, it has been introduced in
\cite{Wipf} and used in many works concerning the bag model and the Casimir
energy for fermionic and scalar fields with spherical boundaries \cite{7}. The
quantum contribution in (\ref{S-2}) is the expression for the vacuum
energy of a scalar field whose energy eigenvalues are $\omega_n$. To render
the eigenvalues of the energy discrete we take temporarily a finite
quantization volume with radius $L \gg 1$.

The classical shell is static and spherically symmetric. It is described by a
potential
\begin{equation}\label{S-3}
V(r)= \frac{\alpha}{R} \delta (r-R),
\end{equation}                 
where $\alpha $ is the strength of the potential. The continuity of the field on the boundary will be discussed later. The potential could be expressed also in
other forms involving the mass, for instance as
\[
V(r)=\alpha \ m \ \delta (r-R) 
\] 
since both § $m$ and $R$ are dimensional parameters. However the choice of $R$
is the most natural since the mass concerns the quantum field while the radius
concerns the background potential which is independent from the field.  

The quantum contribution to the total energy is divergent, for the
regularization we adopt the zeta function technique . We define the regularized
ground state energy
\begin{equation}\label{S-4}
E_{\varphi}={1\over 2}\sum _{(n)} (\lambda _{(n)}^2 + m^2)^{1/2-s} 
\mu ^{2 s} \,,
\end{equation}
where $\mu$ is an arbitrary mass parameter, $s$ is the regularization
parameter which we will put to zero after renormalization and $\lambda_{(n)}$
are the eigenvalues of the wave equation
\begin{equation}\label{S-5}
[-\Delta + V(r)] \varphi _{(n)}(x)=\lambda _{(n)}^2  \varphi _{(n)}(x) .
\end{equation}
Now we introduce a zeta function. The zeta function of the wave operator with
potential $V(r)$ as defined in (\ref{S-5})  is
\begin{equation}\label{S-6}
\zeta_V(s)=\sum_{(n)}(\lambda _{(n)}^2 + m^2)^{-s}\,.
\end{equation}
We can express the ground state energy in terms of this zeta function
\begin{equation}\label{S-7}
E_{\varphi}=\frac12\zeta_V(s-\frac12)\mu^{2s}.
\end{equation}
Using
\begin{equation}\label{S-8}
x^{-s}=\frac{1}{\Gamma (s)}\int^{\infty}_0 dt\ t^{s-1} e^{-xt},
\end{equation}
we can write eq.(2.6) in the following form
\begin{equation}\label{S-9}
\zeta_V(s)=\frac{1}{\Gamma(s)}\int^\infty _0 dt\ t^{s-1}e^{-m^2t}\sum_{(n)} 
e^{-\lambda ^2 _{(n)}t},
\end{equation}
that is 
\begin{equation}\label{S-10}
\zeta_V(s)=\frac{1}{\Gamma (s)}\int^{\infty}_0 dt\ t^{s-1}e^{-m^2t} K(t),
\end{equation}
where function $K(t)$ is the heat kernel.  Taking its the asymptotic expansion
for $t\rightarrow 0$,
\begin{equation}\label{S-11}
K(t)= \sum _{(n)} \exp (- \lambda _{(n)}^2 t)\stackrel{t\rightarrow 0}{\sim}
\left(\frac{1}{4 \pi t}\right) ^{3/2}  \sum^{\infty}_{j=0} A_j t^j; \ \ \ \
j=0,\frac12,1,\dots ,
\end{equation}
and making the substitution $s\rightarrow s-1/2$ in eq.(\ref{S-10}), we get an
expansion of the type (\ref{esempio-divergent}), in which it is easy to recognize the pole terms. This makes it possible to define the total divergent
contribution to the ground state energy by
\begin{eqnarray}\label{S-12} 
E_{\varphi}^{div} &  =  & -\frac{m^4}{64 \pi ^4}\left( \frac1         
                       s + \ln \frac{4\mu ^2 }{m^2} -                      
           \frac 12\right) A_0\  -\frac{m^3}{24\pi^{3/2}}A_{1/2} \nonumber \\
                        &     &  +\frac{m^2}{32 \pi ^4}\left( \frac         
                       1s + \ln \frac{4\mu ^2 }{m^2} - \                    
             1\right) A_1 \ + \frac{m}{16 \pi^{3/2}}A_{3/2}  \nonumber \\
                        &     &  -\frac{1}{32 \pi ^2}\left( \frac 1          
                      s + \ln \frac{4\mu ^2 }{m^2} - \                       
          2\right) A_2. 
\end{eqnarray}
The quantities $A_j$ are the heat kernel coefficients of the spherical $\delta$-shell. In the definition (\ref{S-12})
we have included the terms $A_{1/2}$ and $A_{3/2}$ with half integer index,
which do not contain poles, to satisfy the  normalization condition  which we formulate explicitly below.
 We define the renormalized zero point energy by
\begin{equation}\label{S-13}
E_{\varphi}^{ren}=E_{\varphi}-E_{\varphi}^{div}.
\end{equation}
To keep the total energy of the system unchanged one must add the subtracted object
$E_{\varphi}^{div}$ to the classical energy.
Then one has also a definition of a new classical energy
\begin{equation}\label{S-14}
\epsilon_{class}=E_{class}+E_{\varphi}^{div}.
\end{equation}
The transition from $E_{class}$ to $\epsilon_{class}$ consists in the 
renormalization of the classical parameters  in a way described in (\ref{re-definition}). However in our particular case both the $A_0$ and $A_{1/2}$ coefficients,
corresponding respectively to $p$ and $\sigma$, will turn out to be
zero\footnote{More exactly, the contribution of the Minkowski space $A_{0}$ does not depend on the
  background and can be simply ignored.}, then only the last three terms in
(\ref{re-definition}) will undergo renormalization. Now we have
\[
E_{TOT}\ =\ \epsilon_{class}\ +\  E_{\varphi}^{ren}\ .
\]
The old classical energy $E_{class}$ as defined in(\ref{S-2})  is an unphysical  quantity, since experimentally we can observe only an energy which
includes the vacuum fluctuations.  The term $h/R$ in (\ref{re-definition})  deserves a
particular attention. In fact, in the case of a massless quantum field the
vacuum energy takes the form $\sim 1/R$. Therefore the classical and the
quantum contributions would not be distinguishable and the calculation of
$E^{ren}_{\varphi}$ would lose its predictive power. This difficulty makes it
impossible  to apply our procedure in 
the limit $m\rightarrow 0$.  Furthermore, we must note that the ground state energy
proposed in (\ref{S-13}) has not yet a unique meaning. For the uniqueness of $
E_{\varphi}^{ren}$ we impose the normalization condition
\begin{equation}\label{S-15}
\lim_{m\rightarrow \infty} E_{\varphi}^{ren} \ =\ 0,
\end{equation}
which physically means that for a field of infinite mass we have no quantum
fluctuations. We fulfil this requirement by subtracting all the contributions
in $E_{\varphi}^{div} $ proportional to non negative powers of the mass. That
is we subtract also terms with fractionary indexes up to and including the
term resulting from the heat kernel coefficient $A_2$. The remaining part,
containing only negative powers of $m$, will go to zero for $m\rightarrow
\infty$.  Note that condition (\ref{S-15}) does not apply to a massless field.

\section{Representation of the ground state energy in terms of the Jost 
  function} 

We adopt the approach appeared for the first time in \cite{firstball} for the
calculation of the GSE in the background of a smooth potential. The method for
the calculation of the heat-kernel coefficients for different boundary
conditions was developed in an earlier work \cite{11}.

With the ansatz of the separation of the variables in spherical coordinates, one finds the following  radial Schr\"odinger
equation
\begin{equation}\label{S-16}
\left( \frac{d^2}{dr^2} -\frac{l(l+1)}{r^2}-V(r)+
\lambda^2_{n,l}\right)\varphi_{n,l}(r)=0,
\end{equation}
where $l$ is the angular momentum.  In the general scattering theory with a
continuous spectrum $p$ we have the ``regular solution'' \cite{Taylor} defined as
\begin{equation}\label{S-17}
\varphi_{p,l}(r)\stackrel{r\rightarrow 0}{\sim} j_l(pr) ,
\end{equation}
where $j_l(pr) $ is the Riccati Bessel function. The asymptotics of the
regular solution is expressed in terms of the Jost function $f_l(p)$
\begin{equation}\label{S-18}
\varphi_{p,l}(r)\stackrel{r\rightarrow \infty}{\sim}  
\frac i2\left( f_l(p)\hat{h}^-_l(pr)-  f^{\star}_l(p)\hat{h}^+_l(pr)\right),
\end{equation}
where $\hat{h}^{\pm}_l(pr)$ are the Riccati-Hankel functions. Now we examine
the field at the boundary of our quantization volume. As the potential has a
compact support, at the boundary, expression (\ref{S-18}) becomes an exact
equation. It can be considered as an equation for the eigenvalues
$p=\lambda_{n,l}$. Now, taking for instance Dirichlet boundary conditions at $L$:
$\varphi_{p,l}(L)=0$, we get
\begin{equation}\label{S-19}
\left( f_l(p)\hat{h}^-_l(pL)-  f^{\star}_l(p)\hat{h}^+_l(pL)\right)=0. 
\end{equation}
Since eq.(\ref{S-19}) is satisfied for $p=\lambda_{l,n}$, we can rewrite
the sum in (\ref{S-4}) as a contour integral using the Cauchy theorem
\begin{eqnarray}\label{S-20}
E_{\varphi} & = & \mu^{2 s}\sum ^\infty _{l=0}(l+ 1/2)\int_{\gamma} 
\frac{dp}{2\pi i} (p^2+m^2)^{1/2-s} \nonumber \\
            &   & \frac{\partial}{\partial p} \ln \left(f_l(p)\hat{h}^-_l(pL)-  
f^{\star}_l(p)\hat{h}^+_l(pL)\right),
\end{eqnarray}
where the contour $\gamma$ encloses all the solutions of eq.(\ref{S-19}) on the
positive real $p$ axis and also the bound state solutions in the limit
$L\rightarrow \infty$, which lie on the imaginary axis. We further simplify eq.(\ref{S-20}) by separating the contour into two pieces $\gamma_1$ and $\gamma_ 2$ and
expanding the Hankel functions for large $L$. Then it is possible to recognize
in the integrand a term $ipL$ which corresponds to the Minkowski space
contribution. This term can be dropped. Now we shift the two contours
$\gamma_1$ and $\gamma_ 2$ to the imaginary axis and substitute $p\rightarrow
ik$. We then obtain
\begin{equation}\label{S-21}
E_{\varphi}=-{\cos{\pi s}\over \pi} \mu^{2 s}
\sum ^\infty _{l=0}(l+ 1/2)\int^\infty_m dk [k^2 -m^2]^{1/2 -s} 
\frac{\partial}{\partial k} \ln f_l(ik).
\end{equation}
Since at the end our quantization volume will go to infinity (
$L\rightarrow\infty$ ) this equation will be independent from the boundary
condition chosen for the quantization volume. Eq.(\ref{S-21}) is a very general and
useful representation of the ground state energy, where all the information
about the background potential is contained in $ f_l(ik)$, possible bound
states as well.

In order to perform the analytical continuation to $s=0$ and the subtraction
proposed in (\ref{S-13})  we split $ E^{ren}_\varphi$ into two suitable parts, 
which are convergent in the limit $s\rightarrow 0$. We obtain this by adding and subtracting the uniform
asymptotic expansion of the Jost unction (for more details on this procedure
see \cite{Bordag-Geyer}). We define
\begin{equation}\label{S-22}
E^{ren}_\varphi=E_f + E_{as},
\end{equation}
\begin{equation}\label{S-23}
E_f=-{\cos{\pi s}\over \pi} \mu^{2 s}\sum_{l}(l+\frac12)
\int^\infty_m dk [k^2 -m^2]^{1/2 -s} 
\frac{\partial}{\partial k} [\ln f_l(ik)- \ln f^{as}_l(ik)] 
\end{equation}
and
\begin{equation}\label{S-24}
E_{as}=-{\cos{\pi s}\over \pi} \mu^{2 s}\sum_{l}(l+\frac12)
\int^\infty_m dk [k^2 -m^2]^{1/2 -s} 
\frac{\partial}{\partial k}  \ln f^{as}_l(ik)-E^{div}_{\varphi} ,
\end{equation}
where $ f^{as}_l(ik)$ is the uniform asymptotic expansion of the Jost function (taken for $\nu$ and $k$ equally large)
which we will take up to the third order in $\nu\equiv l+1/2$. In fact,
three orders are sufficient to make (\ref{S-23}) converge and they allow to put $s=0$ under
the sign of the sum and the integral. Higher orders could be included to speed
up the convergence, however, the final result $E^{ren}_{\varphi}$ remains unchanged in
whatever order ($>3$) $\ln f^{as}_l(ik)$ is taken. This is obvious, since the quantity subtracted in the integrand of (\ref{S-23}) is added again in (\ref{S-22}).

Now we need the Jost function corresponding to our scattering problem, so we turn to study the field at the surface of the shell.

\section{Jost function of the spherical $\delta$-shell}
The initial field equation 
\begin{equation}\label{S-25}
\left( \frac{d^2}{dr^2} -\frac{l(l+1)}{r^2}-V(r)+
\lambda^2_{n,l}\right)\varphi_{n,l}(r)=0,
\end{equation}
valid for $-\infty<r<\infty $
can be divided into two parts: an equation for the free field
\begin{equation}\label{S-26}
{\rm at}\ \  r\neq R\  \longrightarrow\ (\Box + m^2)\ \varphi(t,\vec{x})\ =\ 0,
\end{equation} 
and an equation for the field on the shell, which includes the matching conditions based on the continuity of the field at the boundary
\begin{equation}\label{S-27}
{\rm at}\ r=R\left\{ \begin{array}{ll}\varphi\   {\rm continuous.}\\
\varphi '(R+0) - \varphi '(R-0) =\ \frac{\alpha}{R}\varphi (R) \end{array}\right.
\end{equation}
where the prime indicates derivative with respect to $r$.
We take  the regular solution for the delta potential
\begin{equation}\label{S-28}
\varphi_{k,l}(r)= j_l(kR)\Theta (R-r)+\frac i2\left(f_l(k)\hat h^-_l(kR)-
f^{\star}_l(k)\hat h^+_l(kR)\right)\Theta (r-R)
\end{equation}
consisting of two pieces inside and outside the radius $R$, respectively. As above, $ j_l(kR)$ is the Riccati-Bessel function and $ h^{\pm}_l(kR)$ are the Riccati-Hankel functions.  Combining eq.(\ref{S-27}) with eq.(\ref{S-28}) we get
\begin{equation}\label{S-29}
\left\{ \begin{array}{ll}
j_{l}(kR)\ =\   \frac i2 \left(f_l(k)\hat h^-_l(kR)-
f^{\star}_l(k)\hat h^+_l(kR)\right), \nonumber \\  
\frac{\alpha}{R}\ j_{l}(kR)\ =\   k\left(\frac i2 
\left(f_l(k)\hat h'^-_l(kR)-f^{\star}_l(k)\hat h'^+_l(kR)\right)-j'_l(kR) \right).\end{array}\right.
\end{equation}
We solve for $f_l(k)$, keeping in mind that the Wronskian determinant of $\hat
h^{\pm}_l$ is $2i$.  We find
\begin{equation}\label{S-30}
f_l(k)=\frac{1}{2i}\left(-2i (-1) + 2i 
\frac{\alpha}{kR}j_l(kR)\hat h^+_l(kR)\right)
\end{equation}
or
\begin{equation}\label{S-31}
f_l(k)= 1+ \frac{\alpha}{kR}j_l(kR)\hat h^+_l(kR)\,.
\end{equation}
For the Jost function on the imaginary axis we get 
\begin{equation}\label{S-32}
f_\nu (ik)= 1+ \alpha I_\nu(kR) K_\nu(kR),
\end{equation}
which is in terms of the modified Bessel functions $I_\nu$ and $K_\nu$ where
$\nu=l+ 1/2 $ . We need also the asymptotics of the Jost function: $f_\nu
^{as}(ik)$, or more exactly the logarithm of $f_\nu ^{as}(ik)$. The
expansion of the product of the two Bessel functions in (\ref{S-32}), for $k$ and
$\nu$ equally large, is easily obtained with the help of \cite{Abramowitz}. Then we
find the needed asymptotics as a sum of negative powers of $\nu$ with
coefficients $X_{j,n}$ depending on $\alpha$. We define
\begin{eqnarray}\label{S-33}
\ln f_\nu ^{as}(ik) & \equiv & \sum_{n=1}^3
\sum_j X_{j,n}\frac{t^j}{\nu^n}\nonumber \\
&=& \frac{\alpha}{2}\frac{t}{\nu} - \frac{\alpha^2}{8}
\frac{t^2}{\nu^2} +\frac{\alpha}{16}\frac{t^3}{\nu^3}+
\frac{ \alpha^3}{24}\frac{t^3}{\nu^3} - \frac{3 \alpha}{8}
\frac{t^5}{\nu^3} +  \frac{5 \alpha}{16}\frac{t^7}{\nu^3}
\end{eqnarray}
with $t=1/\sqrt{1+\frac{k^2R^2}{\nu^2}}$. Now, inserting (\ref{S-32}) and (\ref{S-33}) in (\ref{S-23})
and (\ref{S-24}) the renormalized ground state energy $ E^{ren}_\varphi =E_f+E_{as}$ can be calculated.

\section{Analytical simplification and heat-kernel coefficients}

 Let us work on $E_{as}$. We transform the sum over $l$ into an integral with the help of the known Abel-Plana formula \cite{sacrarium}
\begin{equation}\label{S-34}
\sum^{\infty}_{l=0}F(l+\frac12)=\int^\infty _0 d\nu F(\nu) +
\int^\infty _0\frac{d\nu}{1+e^{2\pi\nu}}\frac{F(i\nu)-F(-i\nu)}{i}.
\end{equation}
In our case we have
\begin{equation}\label{S-35}
F(\nu)=\int^\infty_m dk\ \nu\  [k^2 -m^2]^{1/2 -s} 
\frac{\partial}{\partial k}  \ln f^{as}_\nu (ik),
\end{equation}
which analytically satisfies the validity conditions for eq.(\ref{S-34}). $E_{as}$ is split into
two addenda:
\begin{equation}\label{S-36}
E^{(1)}_{as}=-{\cos{\pi s}\over \pi} \mu^{2 s}\int^\infty_0 d\nu
\ F(\nu)
\end{equation}
and
\begin{equation}\label{S-37}
E^{(2)}_{as}=-{\cos{\pi s}\over \pi} \mu^{2 s}\int^\infty_0  
\frac{d\nu}{1+e^{2\pi\nu}} \frac{(F(i\nu)-F(-i\nu))}{i}\ .
\end{equation}
First we calculate $E^{(1)}_{as}$; for the $k$ and $\nu$-integrations we use
the formula
\begin{equation}\label{S-38}
\int^\infty _0 d\nu \ \nu\int^\infty_m dk\ [k^2 -m]^{1/2 -s} 
\frac{\partial}{\partial k} \frac{t^j}{\nu^n}=-m^{1-2s}
\frac{\Gamma (\frac32-s)\Gamma (1+\frac{j-n}2)
\Gamma (s+\frac{n-3}2)}{2(mR)^{n-2}\Gamma (\frac j2)}, 
\end{equation}
then
\begin{eqnarray}\label{S-39}
E^{(1)}_{as} & = & -{\cos{\pi s}\over \pi} \mu^{2 s} \int^\infty _0 d\nu\ 
\nu \int^\infty _m dk\ [k^2 -m]^{1/2 -s} 
\frac{\partial}{\partial k} \ln f^{as}_\nu(ik)) \\
             & = & \left(\frac{m^{1-2s}\mu^{2 s}}{\pi}\right) \sum_{j,n}
X_{j,n}\frac{(mR)^{2-n}}2\frac{\Gamma (\frac32-s)\Gamma (1+\frac{j-n}2)
\Gamma (s+\frac{n-3}2)}{\Gamma (\frac j2)}. \nonumber 
\end{eqnarray}
Here, inserting the coefficients of (\ref{S-33}) and expanding up to the first order
in $s$ all the terms which depend on the renormalization parameter we get
\[
E^{(1)}_{as}   =   \frac{2\alpha^3-\alpha}{96\pi R}\left(\frac1s +
\ln \frac{4\mu^2}{m^2} -2\right) -\frac{Rm^2\alpha}{8\pi}\left(\frac1s 
+\ln \frac{4\mu^2}{m^2} -1\right) 
               +\frac{m\alpha^2}{16}. 
\]
The terms poles contribute to the divergence of the
energy. They are used to calculate the heat-kernel coefficients in (\ref{S-12}) and
will disappear after the subtraction of $E_\varphi^{div}$. The term
proportional to $m$ corresponds to the $A_{3/2}$ term of the heat kernel
expansion; although this term generates no divergence it will be as
well subtracted because of the  normalization condition. 

Now we calculate $E^{(2)}_{as}$. The integration over $k$ is carried out with
the formula
\begin{equation}\label{S-40}
\int^\infty_m dk\ [k^2 -m]^{1/2 -s} 
\frac{\partial}{\partial k} t^j=-m^{1-2s}\frac{\Gamma (\frac 32-s)
\Gamma (s+\frac{j-1}{2})}{\Gamma (\frac j2)}\frac{\left(\frac{\nu}{mR}
\right)^j}{\left(1+\left(\frac{\nu}{mR}\right)^2\right)^{s+\frac{j-1}{2}}}.
\end{equation}
We obtain
\begin{eqnarray}\label{S-41}
E^{(2)}_{as}={\cos{\pi s}\over \pi} \mu^{2 s}m^{1-2s}\sum_{j,n}X_{j,n} 
\frac{\Gamma (\frac 32-s)\Gamma (s+\frac{j-1}{2})}{\Gamma (\frac j2)}
\frac{1}{(Rm)^j}\int^\infty _0\frac{d\nu\ \nu}{1+e^{2\pi\nu}}\nonumber \\
\cdot\left(\frac{(i\nu)^{j-n}}{\left(1+\left(\frac{i\nu}{mR}\right)^2
\right)^{s+\frac{j-1}{2}}}+\frac{(-i\nu)^{j-n}}{\left(1+\left(
\frac{-i\nu}{mR}\right)^2\right)^{s+\frac{j-1}{2}}}\right)\,.
\end{eqnarray}
This expression can be transformed into
\begin{eqnarray}\label{S-42}
E^{(2)}_{as} &  =  & -\frac{1}{32\pi^2}\left(\frac1s +
\ln \frac{4\mu ^2 }{m^2}-2\right) -\frac{\alpha}{2\pi R}\ b_1(Rm)-
\frac{\alpha^2}{8R^2m}\  b_2(Rm)\nonumber \\
             &     & + \frac{2\alpha^3+3\alpha}{48\pi R}\ 
 b_3(Rm)-\frac{\alpha}{8\pi R}\  b_4(Rm) +\frac{\alpha}{48\pi R}\  b_5(Rm), 
\end{eqnarray}
where integration by parts was used and the following functions containing the
$\nu$-integrations of eq.(\ref{S-41}) are introduced:
\begin{eqnarray}\label{S-43}
 b_1(x)         &  =  & \int^\infty _0 d\nu 
\frac{\nu}{1+e^{2\pi\nu}} \ln |1-\frac{\nu ^2}{x^2}|\ ; \nonumber \\ 
 b_2(x)         &  =  & \int^{x}_0 d\nu \frac{\nu}{1+e^{2\pi\nu}} 
\frac{1}{\sqrt{1-\frac{\nu ^2}{x^2}}}\ ; \nonumber \\ 
 b_3(x)         &  =  & \int^\infty_0 d\nu \ln|1-\frac{\nu^2}{x^2}|
\left(\frac{\nu^2}{1+e^{2\pi\nu}}\right)'\ ;\nonumber \\
b_4(x)         &  =  & \int^\infty_0 d\nu \ln|1-\frac{\nu^2}{x^2}|
\left(\frac1\nu\left(\frac{\nu^2}{1+e^{2\pi\nu}}\right)'\right)' \ ;\nonumber \\
b_5(x)         &  =  & \int^\infty_0 d\nu \ln|1-\frac{\nu^2}{x^2}|
\left(\frac1\nu\left(\frac1\nu\left(\frac{\nu^2}{1+e^{2\pi\nu}}
\right)'\right)'\right)'\ .
\end{eqnarray}
We see that also $E^{(2)}_{as} $ has a pole of the form $\frac1s$ for
$j=n=1$. This pole contributes to the heat kernel coefficient $A_2$.  

Now, since $E_f$ contains no poles, we are able to write down the complete
heat kernel coefficients $A_j$, up to the order $j\leq 2$:  
\begin{eqnarray}
 & A_0= 0;              & \  A_{1/2}=0;\nonumber \\  
 & A_1=- 4 \pi R\alpha; & \  A_{3/2}=\pi^{3/2}\alpha^2;\ \  
A_2=- \frac23\pi\frac{\alpha^3}{R}\ .\nonumber
\end{eqnarray}
This coefficients are the same as in paper \cite{Bordag dielectric}.  After performing the subtraction
\[
(E^{(1)}_{as}+E^{(2)}_{as})\  -\ E^{div}
\]
$E^{(1)}_{as}$ cancels completely  and only  $E^{(2)}_{as}$ (without its
divergent portion) contributes to the total energy. So we have finally, 
\begin{eqnarray}\label{S-44}
E_{as} \mid_{s=0} &  = & -\frac{\alpha}{2\pi R}\ b_1(Rm)-
\frac{\alpha^2}{8R^2m}\  b_2(Rm)+ \frac{2\alpha^3+3\alpha}{48\pi R}\ 
 b_3(Rm)\nonumber \\
                                 &    & -\frac{\alpha}{8\pi R}\ 
 b_4(Rm)+\frac{\alpha}{48\pi R}\  b_5(Rm)\ . 
\end{eqnarray}

\section{Asymptotics of $E_{as}$}
It is interesting to check (analytically as far as possible) the
behaviour of  $E_{as}$ for small and for large values of $R$. To this aim, we first calculate the corresponding asymptotics of the functions $b_n(x)$. 

In the limit  $R\rightarrow 0$ we find:
\begin{eqnarray}\label{S-45}
\lim_{R\rightarrow 0} b_1(Rm) & \sim & \frac{1}{48}\ln (Rm)\ +\ C_1\  ;\nonumber\\
\lim_{R\rightarrow 0} b_2(Rm) & \sim &  R^2m^2+ C_2\ ;\nonumber\\
\lim_{R\rightarrow 0} b_3(Rm) & \sim & \ln (Rm)\ +\  C_3\ ;\nonumber\\
\lim_{R\rightarrow 0} b_4(Rm) & \sim & 2\ln (Rm)\ +\  C_4\   ;\nonumber\\
\lim_{R\rightarrow 0} b_5(Rm) & \sim  & 8\ln (Rm)\ +\  C_5\  ,
\end{eqnarray}
where the $C_n$'s are numbers resulting from the $\nu$-integrations.
Then we have 
\begin{equation}\label{S-46}
\lim_{R\rightarrow 0} E_{as}\sim -\frac{\alpha C_0}{16\pi R}-
\frac{\alpha^3}{24\pi R}\left(\ln\frac{1}{Rm}-C_3\right),
\end{equation}
where $C_0=(-8C_1+C_3-C_4+ C_5/3)\sim 0.224$ and $C_3\sim 1.96$. Therefore, the asymptotic part of the energy behaves logarithmically for small radii of the spherical shell. This behaviour will be observed numerically in the complete renormalized ground state energy.

For $R\rightarrow \infty$ we have
\begin{eqnarray}\label{S-47}
\lim_{R\rightarrow \infty} b_1(Rm) & \sim & -\frac{7}{1920}\frac{1}{m^2R^2}\ ;\nonumber\\
\lim_{R\rightarrow \infty} b_2(Rm) & \sim &  \frac{1}{48}\ ;\nonumber\\
\lim_{R\rightarrow \infty} b_3(Rm) & \sim &  \frac{1}{24}\frac{1}{m^2R^2}\ ;\nonumber\\
\lim_{R\rightarrow \infty} b_4(Rm) & \sim &  -\frac{7}{480}\frac{1}{m^4R^4}\ ;\nonumber\\
\lim_{R\rightarrow \infty} b_5(Rm) & \sim &  -\frac{31}{16128}\frac{16}{m^6R^6}\ ,
\end{eqnarray}
and we find
\begin{equation}\label{S-48}
\lim_{R\rightarrow \infty} E_{as}\sim -\frac{1}{384}
\frac{\alpha^2}{m^2R^2}.
\end{equation}

\section{Numerical results} 
To numerically study  $E_{as}$ we rewrite it in the form
\begin{eqnarray}\label{S-49}
E_{as} & =  & \frac{1}{16\pi R} \left(\alpha v_1(Rm)\ +\ \alpha^2 v_2(Rm)\ +\ \alpha^3 v_3(Rm)\right),
\end{eqnarray}
where the three functions $v_n(x)$ are given by
\begin{eqnarray}\label{S-50}
v_1(x) & = & -8b_1(x) +b_3(x) -2b_4(x) + \frac13 b_5(x)\ ;\nonumber \\
v_2(x) & = & -2\frac{b_2(x)}{x}\ ; \nonumber \\
v_3(x) & = & \frac23 b_3(x).
\end{eqnarray}
As it is clear from (\ref{S-49}) , for small values of $\alpha$ \ $E_{as}$  behaves like function $v_1(x)$. For large values of $\alpha$,
$E_{as}$ behaves like  function $v_3(x)$.
The plots of the three $v_n(x)$ functions are shown below in Fig. 3.1. In this
as in all the following plots $m$ is set to be equal §$1$.
\begin{figure}[h]\unitlength1cm
\begin{picture}(6,6)
\put(-0.5,-5){\epsfig{file=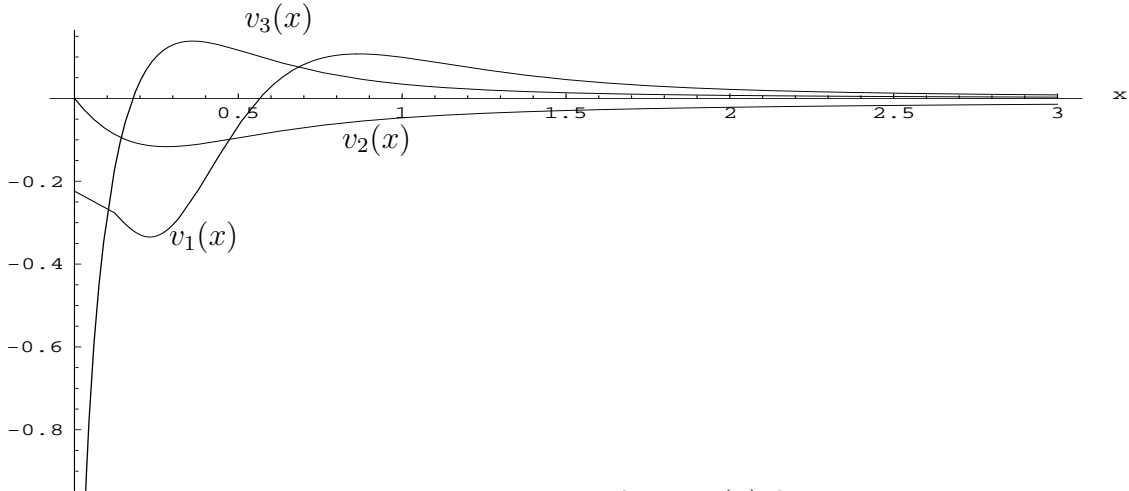,width=15cm,height=15cm}}
\put(2.7,5.6){$v_3(x)$}
\put(1.7,2.7){$v_1(x)$}
\put(4,4){$v_2(x)$}
\end{picture}
\caption{ The curves of the  $v_n(x)$ functions.} 
\end{figure}
For the complete quantum energy we still need the  contribution $E_f$. In the
expression (\ref{S-23}), after putting $s=0$, we integrate by parts and we
obtain
\begin{equation}\label{S-51}
E_f= \frac1\pi\sum^\infty_{l=0}(l+\frac12)\int^\infty_m
\frac{k}{\sqrt{k^2-m^2}}\left(\ln f_\nu (ik) - \ln f^{as}_\nu (ik)\right) 
dk \,.
\end{equation}
This quantity cannot be further analytically simplified. Below we show (Fig. 3.2)
a plot of $ R\cdot E_f$ as function of $R$ \  for $\alpha=1$.
\begin{figure}[h]\unitlength1cm
\begin{picture}(7,7)
\put(0,-4){\epsfig{file=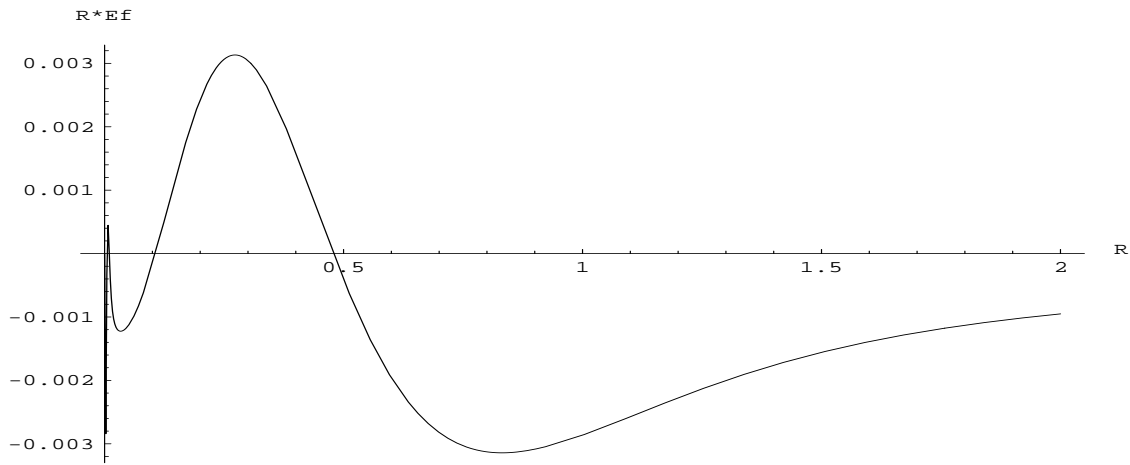,width=15cm,height=14cm}}
\end{picture}
\caption{ The curve of $R\cdot E_f(R)$ for a strength of the potential 
equal to 1. For $R=0$ the curve converges to a finite value.} 
\end{figure}

For the total ground state energy as a function of the radius of the shell we 
get the curves shown below (Fig. 3.3-3.5) for different values of the strength of 
the potential $\alpha$.

\vspace{2cm}

\begin{figure}[h]\unitlength1cm
\begin{picture}(4,4)
\put(0,-4){\epsfig{file=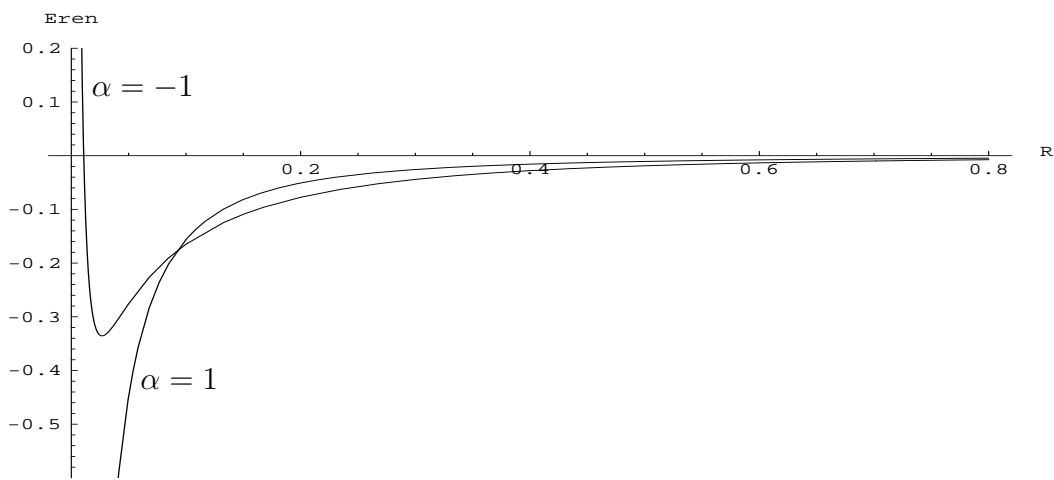,width=14cm,height=14cm}}
\put(1.8,1.1){$\alpha=1$}
\put(1.15,5){$\alpha=-1$}
\end{picture}
\caption{The renormalized vacuum energy $E^{ren}_\varphi (R)$ for positive and 
negative values of the potential.} 
\end{figure}

\begin{figure}[h]\unitlength1cm
\begin{picture}(6,7)
\put(0,-4.1){\epsfig{file=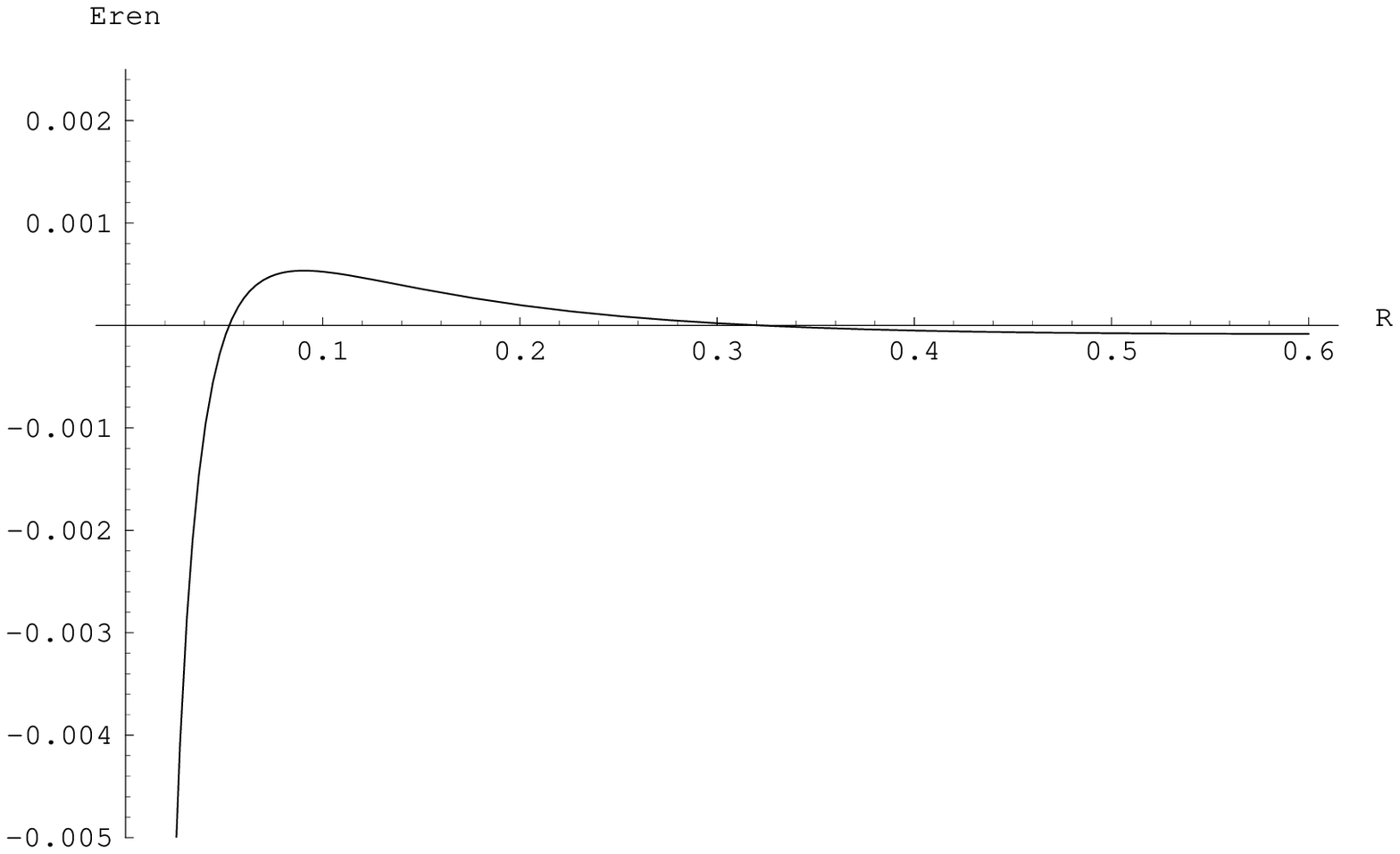,width=14cm,height=14cm}}
\end{picture}
\caption{The renormalized vacuum energy $E^{ren}_\varphi (R)$ for $\alpha$=0.3} 
\end{figure}

\begin{figure}[h]\unitlength1cm
\begin{picture}(6,6)
\put(0,-4){\epsfig{file=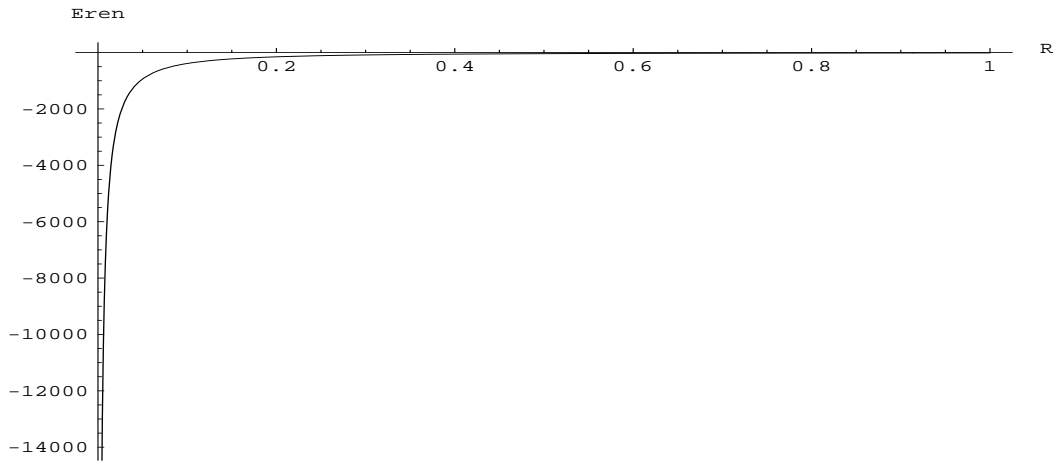,width=14cm,height=14cm}}
\end{picture}
\caption{The renormalized vacuum energy $E^{ren}_\varphi (R)$ for $\alpha$=10} 
\end{figure}

\section{Discussion}
We have obtained a representation of the renormalized ground state energy of a
scalar massive field in the background of a semi-transparent shell containing
convergent integrals of simple functions. This expression is given by the sum
of (\ref{S-49}) and (\ref{S-51}) and it depends only on the two parameters of the classical
system, namely the radius and the strength of the potential of the spherical
shell.  The plots of $E^{ren}_\varphi $ as a function of the radius show that
for repulsive potentials the renormalized ground state energy is positive only
in some limited intervals of the radius axis and only when $\alpha$ is smaller
than 1. For a strength of the potential larger than 1 the energy is always
negative. This is the most striking conclusion of this chapter. For very large
values of $\alpha$ the shell should be no more transparent and the problem
should formally become a Dirichlet boundary condition problem. One could check
this in the equation (\ref{S-32}) for the Jost function: here inserting a large
$\alpha$ the addend $1$ becomes negligible, then one would just have the
product of the two modified Bessel $I$ and $K$ functions; such a Jost function
is exactly the one for a perfectly reflecting spherical shell (Dirichlet
boundary conditions). In that case the ground state energy is simply the sum
of the energies inside and outside the shell. Then formally we should have:
\begin{equation}\label{S-52}
\lim_{\alpha \rightarrow +\infty}\ \  \ GSE^{semi-trans.} = \ \ \ GSE^{mirror}.
\end{equation}
Now it is shown in \cite{Bordag Elizalde} that the ''mirror'' configuration, in the massive
field case, has always positive ground state energy for repulsive potentials.
This is in contradiction with our plots which show an opposite sign.
Furthermore, the $A_2$ coefficient in paper \cite{Bordag Elizalde} is zero. In our work
$A_2$ remains a non zero coefficient also in the limit $\alpha \rightarrow
+\infty$ demonstrating that the transition hypothesized in (\ref{S-52}) is singular. For
flat parallel semi transparent boundaries with delta function potential in the
vacuum of a scalar massive field, the transition is actually fulfilled as
shown in paper \cite{Bordag Robaschik}. In the configuration analyzed in this paper the
limit (\ref{S-52})  works only for the regularized GSE, as mentioned above, but after
the renormalization the limit is no more valid. This means that the
subtraction of the divergent part of the energy and the limit $\alpha
\rightarrow +\infty$ are two non commutative operations.  We remark again that
the results of this work cannot be directly applied to the case of a massless
field, since the initial normalization condition would fail and the vacuum
energy would not be univocally defined.

\chapter{Vacuum energy for a semi-transparent cylindrical shell}

 Calculations with cylindrical geometries  began historically with \cite{Milton-cyl}, after the suggestion that a cylinder, as a kind of intermediate shape between the parallel plates and the  sphere, could possess a zero Casimir stress \cite{pazzo}. However paper \cite{Milton-cyl} showed that a perfectly conducting cylinder in the electromagnetic vacuum has a negative Casimir energy. A number of papers on the dielectric cylinder  \cite{Nesterenko},  \cite{Klich}, have recently showed the interesting result of a vanishing vacuum energy in the dilute case. 
  We find of interest to carry on this research with the analysis of massive fields in the background of non-ideal cylindrical boundaries. In this chapter we consider a hollow cylinder with radius $R$, having a delta function $\delta(r-R)$ as a potential profile.  This model can be considered as a ``scalar''  version of the dielectric background. A non singular potential would be, of course, more realistic under the physical point of view, but the calculations would be considerably more complicated. We hope this work can contribute in the understanding of the structure and meaning of vacuum. The cylindrical shell could, besides, find physical applications in the calculation of the quantum corrections to vortices in QCD or in the electroweak theory. Another interesting perspective is in the recent discovery of the so called nanotubes \cite{nanotubes1}, \cite{nanotubes2}, which are large carbon molecules generated in laboratory offering the intriguing possibility of measuring quantum effects on small cylindrical objects.

The setup of the problem is  analogous to that of the spherical shell. We will first approach the quantum field theoretical problem and discuss the renormalization. Then, the Jost function will be calculated. In the remaining part of the chapter, the numerical computation will be performed and graphically represented. 

\section{Vacuum energy in terms of the Jost function and renormalization }

Let us work  with a real massive scalar field $\phi(x)$ with mass $m$ and let us quantize it in the background of a cylindrical potential. The field equation in cylindrical coordinates $r,\phi, z$, after separation of the variables, reads
\begin{equation}
\left(p_0^2-m^2-p_z^2-\frac {l^2}{r^2}-V(r)+
\frac 1r\partial_r +\partial_r^2\right)\phi_l(p_0,p_z,r)\ \ =\ 0 \ ,
\end{equation}
where $p_\mu $ is the momentum four vector being $p_z$ its component along the longitudinal axis of the cylinder and $l$ is the angular momentum quantum number. $V(r)$ is the background potential given by 
\begin{equation}
V(r)\ =\ \frac{\alpha}{R}\delta(r-R)\ ,
\end{equation}    
it represents an infinitely thin cylindrical shell whose profile is a delta function. The shell has a circular section of radius $R$ and it extends from $-\infty$ to $+\infty$ in the $z$ direction. $\alpha$ is the dimensionless parameter giving the strength of the  potential.

Considering eq.(4.1) as a scattering problem, we choose the ``regular solution''  which is given by
\begin{equation}
\phi(r)\ =\ J_l (kr)\Theta(R-r)+ \frac 12 (f_l(k)H^{(2)}_{l}(kr)+f^*_l(k)H^{(1)}_{l}(kr))\Theta(r-R) \ ,
\end{equation}
where $k=\sqrt{p_0^2-m^2-p_z^2}$, $J_m (kr)$ is a Bessel function of the first kind, $H^{(1)}_l(kr)$ and $H^{(2)}_l(kr)$ are the Hankel functions of the first and of the second kind and the coefficients $f_l(k)$ and $f^*_l(k)$ are the Jost function and its complex conjugate respectively. $\Theta(R-r)$ is a theta function. The field is therefore free in the regions $0<r<R$ and $R<r<\infty$. At $r=R$ the field is continuous and we have the following matching conditions
\begin{equation}
\left\{ \begin{array}{ll}\phi '(R+0) - \phi '(R-0) =\ 
\frac{\alpha}{R}\phi (R)\\ 
\phi(R+0)\ =\  \phi(R-0)\ , \end{array} 
\right.
\end{equation}
where the prime indicates derivative with respect to $r$.\\
The quantum field is in its vacuum state, its energy is given by half of the sum over all possible eigenvalues $\omega_{(n)}$ of the Hamilton operator related with the wave equation (4.1). We define a regularized vacuum energy
\begin{equation}
E\ =\ \frac {\mu^2 s}{2}\sum_{(n)} \omega_{(n)}^{1-2s}\ ,
\end{equation}
where $s$ is the regularization parameter which we will put to zero after the renormalization, $\mu$  is the mass parameter introduced for dimensional reasons and $(n)$ includes all possible quantum numbers. We will calculate the energy density per unit length of the cylinder given by
\begin{equation}
{\cal E}\ =\ \frac 12 \mu^{2s}\int^{\infty}_{-\infty}\frac{dp_z}{2\pi}
\sum_{(n)} (p_z^2+ \epsilon_{(n)}^2)^{1/2-s}\ ,
\end{equation}
where $\epsilon_{(n)}$ are the eigenvalues of the operator contained in (4.1) without $p_z$ and with $k=\sqrt{p_0^2-m^2}$. We carry out the integration over $p_z$ and we arrive at
\begin{equation}
{\cal E}\ =\ \frac 14 \mu^{2s}\frac{\Gamma(s-1)}{\sqrt{\pi}\Gamma(s-1/2)}\sum_{(n)}(\epsilon_{(n)}^2)^{1-s}\ .
\end{equation}
Following a  procedure described in the last chapter\footnote{ The procedure is explained with details in \cite{firstball}.}  we transform the sum in (4.7) into a contour integral and, dropping the Minkowski space contribution, we arrive at
\begin{equation}
{\cal E} \ =\ -\frac 14 C_s\ \sum^\infty_{l=-\infty}\int^\infty_m dk\ (k^2+m^2)^{1-s} \partial_k \ln f_l(ik)\ ,
\end{equation}
 where $C_s=(1+s(-1+2\ln(2\mu)))/(2\pi)$ and $f_l(ik)$ is the Jost function defined in (4.3) on the imaginary axis. It contains all the information about the background potential under examination. We will find $ f_l(ik) $ explicitly in the following section. The energy defined in (4.8) is renormalized by direct subtraction of its divergent part
\begin{equation}
{\cal E}_{ren}\ =\ {\cal E}\ -\ {\cal E}_{div}\ 
\end{equation} 
with the normalization condition demanding that the vacuum fluctuations vanish for a field of infinite mass
\begin{equation} 
\lim_{m \rightarrow \infty }{\cal E}_{ren} = 0\ .                       
\end{equation} 
This normalization condition eliminates the arbitrariness of the mass parameter $\mu$ (the interested reader can find more details in  \cite{Boston}). 
The isolation of ${\cal E}_{div}$ is done with the use the heat-kernel expansion of the ground state energy 
\begin{equation}
{\cal E}\ =\ \sum_j \frac{\mu^{2s}}{32\pi^2}\frac{\Gamma(s+j-2)}{\Gamma(s+1)}m^{4-2(s+j)}A_j\ \ ,\ \ \ \  j=0, \frac 12, 1,...\ ,     
\end{equation}
where the $A_j$ are the heat kernel coefficients related to the background.
Then it is possible to define
\begin{eqnarray} 
{\cal E}_{div} &  =  & -\frac{m^4}{64 \pi ^2}\left( \frac 1s + \ln
\frac{4\mu ^2 }{m^2} - \frac 12\right) A_0\ -\frac{m^3}{24\pi^{3/2}}A_{1/2}
\nonumber \\     
                  &     &  +\frac{m^2}{32 \pi ^2}\left( \frac 1s +
\ln\frac{4\mu ^2 }{m^2} - \ 1\right) A_1 \ + \frac{m}{16 \pi^{3/2}}A_{3/2} \nonumber \\ 
                  &     &  -\frac{1}{32 \pi ^2}\left( \frac 1s + \ln \frac{4\mu ^2 }{m^2} - 2\right) A_2\ ,  
\end{eqnarray} 
which includes all the pole terms and all the terms proportional to non-negative powers of the mass.

In order to perform the analytical continuation $s\rightarrow 0$  we split the renormalized vacuum energy (4.9) into  two parts
\begin{equation}
{\cal E}_{ren}\ =\ {\cal E}_f + {\cal E}_{as},
\end{equation}
with 
\begin{equation}
{\cal E} _f\ =\  - \frac 14 C_s \sum_{l=-\infty}^\infty\int^\infty_m dk [k^2 -m^2]^{1-s} \frac{\partial}{\partial k} [\ln f_l(ik)- \ln f^{as}_l(ik)]\end{equation}
and
\begin{equation}
{\cal E} _{as}\ =\  -\frac 14 C_s\sum_{l=-\infty}^\infty \int^\infty_m dk [k^2 -m^2]^{1 -s} 
\frac{\partial}{\partial k}  \ln f^{as}_l(ik)-{\cal E}_{div} .
\end{equation}
Here $f^{as}_l(ik)$ is a portion of the uniform asymptotic expansion of the Jost function which must include as many orders in $l$ as it is necessary to have
\begin{equation}
\ln f_l(ik) -  \ln f^{as}_l(ik)\ = {\cal O} \left( l^{-4}\right)
\end{equation}
in the limit $l\rightarrow \infty$,  $k\rightarrow \infty$ with fixed $l/k$, which is sufficient to let the sum and the integral in (4.14) converge in the limit $s=0$. The splitting proposed in (4.13) leaves the quantity ${\cal E} _{ren}$ unchanged, while it permits the substitution $s=0$ in the finite part ${\cal E} _f$.

\section{The Jost function and its asymptotics}

We insert solution (4.3) into (4.4) and we find
\begin{equation}
\left\{ 
\begin{array}{ll}
J_{l}(kR)  =  \frac 12 \left[f_l(k) H^{(2)}_l(kR)+f^{\star}_l(k) H^{(1)}_l(kR)\right] \nonumber \\  
\left(\frac 12 \partial_r \left[f_l(k) H^{(2)}_l(kR)+f^{\star}_l(k) H^{(1)}_l(kR)\right]   \right)|_{r=R} =  \frac{\alpha}{R} J_{l}(kR) + \left( \partial_r J_{l}(kr)\right)|_{r=R}\ .
\end{array}
\right. 
\end{equation}
The system can be solved for $f_l(k)$, with the help of the Wronskian determinant of the Hankel functions \cite{Abramowitz}, the result is
\begin{eqnarray}
f_l(k) & = &  -\frac 12 i\pi kR \left( J_{l+1}(kR) H^{(1)}_l(kR)- J_{l}(kR) H^{(1)}_{l+1}(kR)\right.\nonumber \\
       &   &    \left.  -\frac{\alpha}{kR}J_{l}(kR) H^{(1)}_l(kR)\right)\ .
\end{eqnarray}
The corresponding Jost function on the imaginary axis can be written in terms of the modified Bessel I and K functions, again with the help of \cite{Abramowitz}
\begin{equation}
f_l(ik)\ =\ 1+\alpha\ I_l(kR)K_l(kR).
\end{equation}
From the Jost function (4.19) one arrives at the uniform asymptotic expansions $f^{as+}_l(ik)$ for positive $l$ and  $f^{as-}_l(ik)$ for negative $l$ and  at the asymptotic expansion $f^{as}_0(ik)$ for $l=0$, with the help of  the asymptotics of the Bessel I and K functions for large indices and large arguments available on \cite{Abramowitz}. Since the asymptotic Jost function consists  of three different contributions, the sum over $l$ appearing in (4.14) and (4.15) must also be distinguished in three  contributions: a sum over negative $l$, a sum over positive $l$ and a contribution coming from $l=0$. The first two contributions can be summed up analytically in the following way 
\begin{equation}
\sum_{l=-\infty}^{-1}...\ln f^{as-}_l(ik)...\ +\ \sum_{l=1}^{\infty}...\ln f^{as+}_l(ik)...\ =\  \sum_{l=1}^{\infty}...(\ln f^{as+}_l(ik)+ \ln f^{as-}_{-l}(ik))...\ ,
\end{equation}
where the dots represent, for simplicity, the rest of the functions in (4.14) and (4.15). Then, eq.(4.14) and (4.15) are rewritten in the form
\begin{eqnarray}
{\cal E} _{as} & = &  -\frac 14 C_s\sum_{l=1}^\infty \int^\infty_m dk [k^2 -m^2]^{1 -s} \frac{\partial}{\partial k} \ln f^{as\pm}_l(ik)  \nonumber \\
               &   &  -\frac 14 C_s \int^\infty_m dk [k^2 -m^2]^{1 -s} \frac{\partial}{\partial k} \ln f^{as}_0(ik) -{\cal E}_{div} 
\end{eqnarray}
and
\begin{eqnarray}
{\cal E} _f & = &  -\frac {1}{8\pi} \sum_{l=1}^\infty\int^\infty_m dk [k^2 -m^2] \frac{\partial}{\partial k} \left(2\ln f_l(ik)-\ \ln f^{as\pm}_l(ik)\right) \nonumber\\
            &   &  -\frac{1}{8\pi} \int^\infty_m dk [k^2 -m^2] \frac{\partial}{\partial k} \left(\ln f_0 (ik)-\ln f^{as}_0(ik)\right)\ ,
\end{eqnarray}
where  $\ln f^{as\pm}(ik)=\ln f^{as+}_l(ik)+\ln f^{as-}_{-l}(ik)$ and
we have used the property $f_l(ik)=f_{-l}(ik)$ of eq.(4.19).

 Taking the logarithm of the uniform asymptotics of the  modified Bessel functions and re-expanding in negative powers of the variable $l$ (see \cite{Marco2} for details on this procedure, see also appendix A) we find
\begin{equation}
\ln f_0^{as}\ =\ \frac{\alpha}{2kR}\ -\ \frac{\alpha^2}{8k^2R^2}\ ,
\end{equation}
\begin{equation}
\ln f^{as\pm}(ik)\ =\ \sum_{n=1}^3\sum_jX_{n,j}\frac{t^j}{l^n}\ ,
\end{equation}
where $t=(1+(kR)/l)^2)^\frac 12 $  and the non-vanishing  coefficients are
\begin{equation}
\begin{array}{l}
X_{1,1}=\alpha\ ,\ X_{2,2}=-\alpha^2/4\ , \\
X_{3,3}=\alpha/8+\alpha^3/12\ ,\ X_{3,5}=-3\alpha/4\ ,\\ X_{3,7}=5\alpha/8\ .
\end{array}
\end{equation}
In this definition we have included 3 orders in $l$ which are sufficient to satisfy condition (4.16). Substituting (4.23) and (4.24) in (4.21) and (4.22) we find
\begin{eqnarray}
{\cal E} _{as} & = &  -\frac 14 C_s\sum_{l=1}^\infty \int^\infty_m dk [k^2 -m^2]^{1 -s} \frac{\partial}{\partial k} \left(\sum_{n=1}^3\sum_jX_{n,j}\frac{t^j}{l^n}\right)  \nonumber \\
               &   &  -\frac 14 C_s \int^\infty_m dk [k^2 -m^2]^{1 -s} \frac{\partial}{\partial k}  \left(\frac{\alpha}{2kR}\ -\ \frac{\alpha^2}{8k^2R^2}\right) -{\cal E}_{div} 
\end{eqnarray}
and
\begin{eqnarray}
{\cal E} _f & = &  -\frac {1}{8\pi} \sum_{l=1}^\infty\int^\infty_m dk [k^2 -m^2] \frac{\partial}{\partial k} \left(2\ln f_l(ik)-\sum_{n=1}^3\sum_jX_{n,j}\frac{t^j}{l^n} \right) \nonumber\\
            &   &  -\frac{1}{8\pi} \int^\infty_m dk [k^2 -m^2] \frac{\partial}{\partial k} \left(\ln f_0 (ik)-\frac{\alpha}{2kR}\ -\ \frac{\alpha^2}{8k^2R^2} \right)\ .
\end{eqnarray}
We call the first addend in (4.27) ${\cal E} _{fl}  $ and the second ${\cal E} _{f0} $, that is ${\cal E} _f={\cal E} _{fl}+{\cal E} _{f0}$. 

\section{Asymptotic part of the energy}

We go forward with an analytical simplification of (4.26). We call the second addend in (4.26) ${\cal E}_{as0}$, it can be immediately calculated, giving
\begin{equation}
{\cal E} _{as0}\ =\ -\frac{\alpha m}{8 \pi R}\ -\ \frac{\alpha^2}{64 \pi R^2}\left(\frac 1s +\ln\left(\frac{4\mu^2}{m^2}\right) -2\right)\ .
\end{equation}
A simplification of the first addend in (4.26) can be achieved with the Abel-Plana formula 
\begin{equation}
\sum_{l=1}^\infty F(l)\ =\ \int_0^\infty dl\ F(l)\ \ -\ \frac 12 F(0)\  +\ \int_0^\infty\frac{dl}{1-e^{2\pi\nu}}\frac{F(il)-F(-il)}{i}\ .
\end{equation}
 in the present case  the function $F(l)$ is 
\begin{equation}
F(l)=\int_{m}^\infty dk (k^2+m^2)^{1-s}\partial_k \left(\sum_{n=1}^3\sum_jX_{n,j}\frac{t^j}{l^n}\right) \ .
\end{equation}
Then, the first addend in (4.26) turns out to be the sum of three contributions. The first contribution can be calculated by means of formula (B.1) displayed in the appendix. We find  
\begin{eqnarray}
{\cal E}_{as1} & = & \frac{\alpha m^2 }{16\pi}\left(\frac 1s +\ln\left(\frac{4\mu^2}{m^2}\right) -1\right)\ +\ \frac{\alpha^3 }{96\pi R^2}\left(\frac 1s +\ln\left(\frac{4\mu^2}{m^2}\right) -2\right)\nonumber\\ 
               &   &  +\ \frac{\alpha^2 m}{32 R}\ .
\end{eqnarray}
The integration of the second term in the Abel-Plana formula can be performed with formula (B.2), which yields immediately
\begin{eqnarray}
{\cal E}_{as2} & = & \frac{\alpha m }{8\pi R}\ +\ \frac{\alpha^2 }{64\pi R^2}\left(\frac 1s +\ln\left(\frac{4\mu^2}{m^2}\right) -2\right)\ -\ \frac{\alpha}{64\pi m R^3}\nonumber\\ 
               &   &  -\ \frac{\alpha^3 }{96\pi m R^3}\ .
\end{eqnarray}
The third contribution, after application of formula (B.2), reads
\begin{equation}
{\cal E}_{as3}\ =\ -\frac 12 C_s \sum_{n,j}^{3,7}X_{n,j}(-m_e^{2-2s}\Gamma(2-s))Z_{n,j}(mR)\ .
\end{equation}
where 
\begin{eqnarray}
Z_{n,j}(mR) & = & \frac{\Gamma(s+\frac j2 -1)}{\Gamma(\frac j2)x^j}\left[\int_0^{mR}\frac{dl}{1-e^{2\pi l}}\frac{l^{j-n}\cdot 2\sin \left[\frac{\pi}{2}(j-n)\right]}{\left(1-\frac{l^2}{mR^2}\right)^{s+\frac j2 -1}}\right.\nonumber\\
            &   & +\ \left.\int_{mR}^\infty\frac{dl}{1-e^{2\pi l}}\frac{l^{j-n}\cdot 2\sin \left[\pi(1-s-\frac n2)\right]}{\left(\frac{l^2}{mR^2}-1\right)^{s+\frac j2 -1}}\right].
\end{eqnarray}
Here, the integration over $l$ cannot be performed analytically. However the expression can be remarkably simplified by integrating several times by parts. We find finally 
\begin{eqnarray} 
{\cal E}_{as3} & = &  -\ \frac{\alpha}{2\pi R^2}\  h_1(mR)\ -\ \frac{\alpha^2}{32  R^2}\ h_2(mR)\nonumber\\
               &   &  +\ \left( \frac{\alpha}{16 \pi R^2}+\frac{\alpha^3}{24 \pi R^2}\right)\ h_3(mR)\ -\ \frac{\alpha}{8 \pi R^2}\  h_4(mR)\nonumber\\  
               &   &  +\ \frac{\alpha}{48 \pi R^2}\  h_5(mR)\ ,      
\end{eqnarray}
where the functions $h_i(x)$ are convergent integrals analogous to those found in the last chapter
\begin{eqnarray}
h_1(x) & = & \int_{x}^{\infty} \frac{dl}{1-e^{2\pi l}} \sqrt{l^2 -x^2} \nonumber \\
h_2(x) & = & \int_{x}^{\infty} dl \left(\frac{1}{1-e^{2\pi l}}\frac 1l \right)' (l^2 -x^2) \nonumber \\
h_3(x) & = & \int_{x}^{\infty} dl \left(\frac{1}{1-e^{2\pi l}}\frac 1l \right)' \sqrt{l^2 -x^2} \nonumber \\
h_4(x) & = & \int_{x}^{\infty} dl \left(\left(\frac{l}{1-e^{2\pi l}}\right)'\frac 1l \right)' \sqrt{l^2 -x^2} \nonumber \\
h_5(x) & = & \int_{x}^{\infty} dl \left( \left( \left(\frac{l^3}{1-e^{2\pi l}}\right)'\frac 1l \right)' \frac 1l\right)' \sqrt{l^2 -x^2}\ .
\end{eqnarray}
 In appendix B the reader can find more details about the derivation of (4.31), (4.32), (4.35). The contributions ${\cal E}_{as1}$ and ${\cal E}_{as2}$ contain all the pole terms (all the divergences of the vacuum energy) plus the terms proportional to non-negative powers of the mass (which do not satisfy the normalization condition). All these terms are subtracted and  are used to calculate the heat-kernel coefficients by means of definition (4.12) for ${\cal E}_{div}$. Below we give the heat kernel coefficients which we calculated up to the coefficient $A_4$ (adding four more orders in $\ln f_l^{as\pm}(ik)$), in the hope that they will be of use for future investigations on the same background 
\begin{equation}
\begin{array}{lclcccl}
A_0 & = & 0                 & , &   A_{1/2} & = & 0 \nonumber \\
A_1 & = &  -2\pi\alpha     & , &   A_{3/2} & = &  \frac{\alpha^2 \pi^{3/2}}{2R} \nonumber \\
A_2 & = &  \frac{\pi \alpha^3}{3 R^2}                & , &   A_{5/2} & = & -\frac{(3\alpha^2 +4\alpha^4)\pi^{3/2}}{192R^3} \nonumber \\
A_3 & = &  \frac{(4\alpha^3+7\alpha^5)\pi}{210 R^4}               & , &  A_{7/2}  & = & -\frac{(81\alpha^2 +120\alpha^4+128\alpha^6)\pi^{3/2}}{24576 R^5}\nonumber \\
A_4 & = &  \frac{(64\alpha^3+52\alpha^5+39\alpha^7)\pi}{16380 R^6}  & . &  &  & 
\end{array}
\end{equation}
One can note how all the integer heat-kernel coefficients depend on odd powers of the coupling constant, while the half-integer coefficients depend on even powers of $\alpha$. The same feature is present in the heat kernel coefficients of a $\delta$-potential spherical shell (section 3.4 of the preceding chapter). We note also that (4.32) is in agreement with the coefficients that one would obtain from lemma 2 of paper\cite{BordagVass-heat kernels}, where the heat kernel expansion for semi-transparent boundaries in $d$-dimensions is examined.

We perform the subtraction of the divergent portion and we obtain the final result
\begin{eqnarray}
{\cal E}_{as} & = & -\frac{\alpha}{2\pi R^2}\  h_1(mR)\ -\ \frac{\alpha^2}{32  R^2}\ h_2(mR)\nonumber \\ 
       &  & +\left( \frac{\alpha}{16 \pi R^2}+\frac{\alpha^3}{24 \pi R^2}\right)\ h_3(mR)\nonumber \\ 
       &  & -\frac{\alpha}{8 \pi R^2}\  h_4(mR)\ +\frac{\alpha}{48 \pi R^2}\  h_5(mR)\ -\ \frac{\alpha}{64 \pi m R^3}\nonumber\\
       &  &     -\frac{\alpha^3}{96 \pi m R^3}\ .      
\end{eqnarray}

\section{Finite part of the energy and numerical results}

The quantity (4.27) cannot be analytically simplified. We integrate ${\cal E} _{fl} $ and ${\cal E} _{f0} $ by parts and make the substitution $k\rightarrow k/R$ to get an explicit dependence on $R$
\begin{eqnarray}
{\cal E} _{fl} & = &  \frac {1}{4\pi R^2} \sum_{l=1}^\infty\int^\infty_{mR} dk\ k\  \left(2\ln f_l(ik)|_{k\rightarrow k/R }-\left(\sum_{n=1}^3\sum_jX_{n,j}\frac{t^j}{l^n}  \right)|_{k\rightarrow k/R }\right)\nonumber\\
{\cal E} _{f0} & = &  \frac{1}{4\pi R^2}  \int^\infty_{mR} dk\ k\ \left( \ln f_0 (ik)|_{k\rightarrow k/R }-\frac{\alpha}{2k}\ +\ \frac{\alpha^2}{8k^2} \right)\ .
\end{eqnarray}
For small values of $R$, ${\cal E} _{fl}$ behaves like $R^{-2}$, while  ${\cal E} _{f0} $ has a logarithmic behaviour
\begin{equation}
{\cal E} _{f0} \ \sim\ -\frac{\alpha^2\ln (mR)}{32\pi R^2}\ .
\end{equation}
We note that in the limit $\alpha\rightarrow 0$, the logarithm of the Jost function $f_l(ik)$ can be expanded in powers of $\alpha$, giving (from eq.(4.19))
\[
\ln (1+\alpha I_l(kR)K_l(kR)) \sim\  \alpha I_l(kR)K_l(kR)\ +\ \frac 12\alpha^2 I_l^2(kR)K_l^2(kR)\ +\ {\cal O}(\alpha^3),
\]
then, the summation over $l$ of the leading term of this expansion could be analytically performed following a method recently proposed in \cite{Klich}. This would  give us a first order approximation of ${\cal E}_{fl}$. However we will restrict us  here  to a  numerical calculation of the sum in (4.39), which can be performed for small as well as for large values of $\alpha$.
  
To find the asymptotic behaviour of ${\cal E}_{as}$ we rewrite eq.(4.38) in the following form
\begin{equation}
{\cal E}_{as}\ =\ \frac{1}{2\pi R^2}\left[\alpha w_1(mR)\ +\ \alpha^2 w_2(mR)\ +\ \alpha^3 w_3(m R)\right]\ ,
\end{equation} 
where the functions $w_1(x)$, $w_2(x)$ and $w_3(x)$ are given by
\begin{eqnarray}
w_1(x) & = & \left(-h_1(x)+\frac{1}{8}h_3(x)-\frac 14 h_4(x) +\frac{1}{24}h_5(x)-\frac{1}{32x} \right)\ ,\nonumber\\
w_2(x) & = & \left(-\frac{\pi}{16}h_2(x)\right)\ ,\nonumber\\
w_3(x) & = & \left(\frac{1}{12}h_3(x)-\frac{1}{48x}\right)\ .
\end{eqnarray}
As mentioned above, the functions $h_i(x)$ are quickly converging integrals. The behaviour of ${\cal E} _{as}$ is governed by the functions $w_{1,2,3} (x)$ and by the value of $\alpha$. For $R\rightarrow 0$ we find
\begin{eqnarray}
w_1(x) & \sim & 0.0000868\ +\ {\cal O}(x)\ ,\nonumber\\
w_2(x) & \sim & \frac{1}{16}\ln x\ +\ 0.115\ +\ {\cal O}(x) \ , \nonumber\\
w_3(x) & \sim & \frac{1}{24}\ln x\ +\ 0.00479\ +\ {\cal O}(x) \ .
\end{eqnarray}
Thus, for small values of the radius, ${\cal E} _{as}$ is proportional to $\ln R$. Summing the contribution coming from (4.40) and (4.43), we find that the leading term of the renormalized energy, for $R\rightarrow 0$, is 
\begin{equation}
{\cal E} _{ren}\ \sim \ \frac {\alpha ^3\ln (mR)}{48\pi R^2}\ . 
\end{equation}
This result is in agreement with the prediction (section II of paper \cite{master})
\begin{equation}
\lim_{R\rightarrow 0}{\cal E} _{ren} \ \sim\ \frac {A_2}{16\pi^2}\ln (mR)\ .
\end{equation}  
In the limit $R\rightarrow \infty $ the behaviour of the renormalized energy is determined by the first non-vanishing heat kernel coefficient after $A_2$. From (4.37) and eq.(4.11) we arrive at
\begin{equation}
\lim_{R\rightarrow \infty}{\cal E} _{ren} \ \sim\ -\frac {\alpha^2}{2048 mR^3}\ -\frac {\alpha^4}{1536 mR^3}\ +{\cal O}\left(\frac{1}{R^4}\right)\ .
\end{equation} 

We have numerically evaluated the quantities ${\cal E} _{as}$, ${\cal E} _{fl}$, ${\cal E} _{f0}$ and ${\cal E} _{ren}$ as functions of $R$, fixing the values of the mass to $1$. It turned out to be necessary to sum  20 terms in the variable $l$ and to integrate up to 1000 in the variable $k$  to obtain ``stable'' numerical values for the energy.
Below we give the plots of the various contributions to the vacuum energy and
the complete renormalized  energy for different values of the potential
strength. For $\alpha<0$ we found the renormalized vacuum energy to posses a
small imaginary part becoming larger when $\alpha$ approaches $-\infty$.
Particle creation accounts for this contribution. It starts when the attractive potential of the shell becomes over-critical, that is when ${\cal \epsilon}<-m<0$, where ${\cal \epsilon}$ is the energy of the bound state; in this case the effective action of the system acquires an imaginary part\footnote{For the theoretical foundations of this phenomenon see \cite{Schwinger-imaginary}. In the context of vacuum energy the appearance of an imaginary part in the renormalized energy was  observed for instance in \cite{BordagKirstenHellmund}.}, however a detailed discussion of this aspect of the theory is beyond the purpose of the present chapter. It should only be said that in the plots traced for negative values of $\alpha$ (Fig. 4.4 and 4.5) the energy is to be intended as real part of.

\begin{figure}[hp]\unitlength1cm
\begin{picture}(6,6)
\put(-0.5,0){\epsfig{file=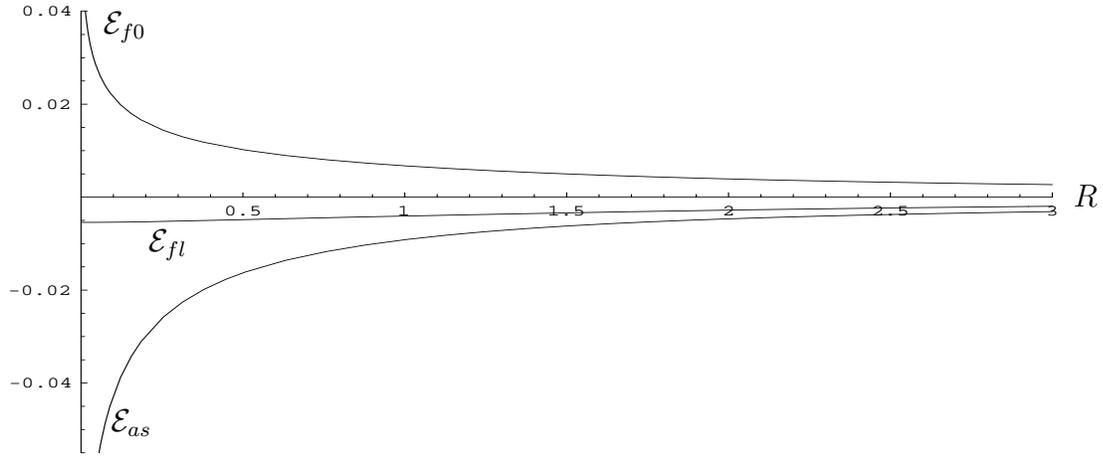,width=14cm,height=6cm}}
\put(1.4,2.7){${\cal E}_{fl} $}
\put(0.8,5.6){${\cal E}_{f0} $}
\put(0.9,0.3){${\cal E}_{as}  $}
\put(13.7,3.3){$R$}
\end{picture}
\caption{Repulsive potential. The contributions to the  renormalized vacuum energy multiplied by $R^{2}\cdot \alpha^{-2}$, for $\alpha=2.1$ .} 
\end{figure}

\begin{figure}[hp]\unitlength1cm
\begin{picture}(6,6)
\put(-0.5,0){\epsfig{file=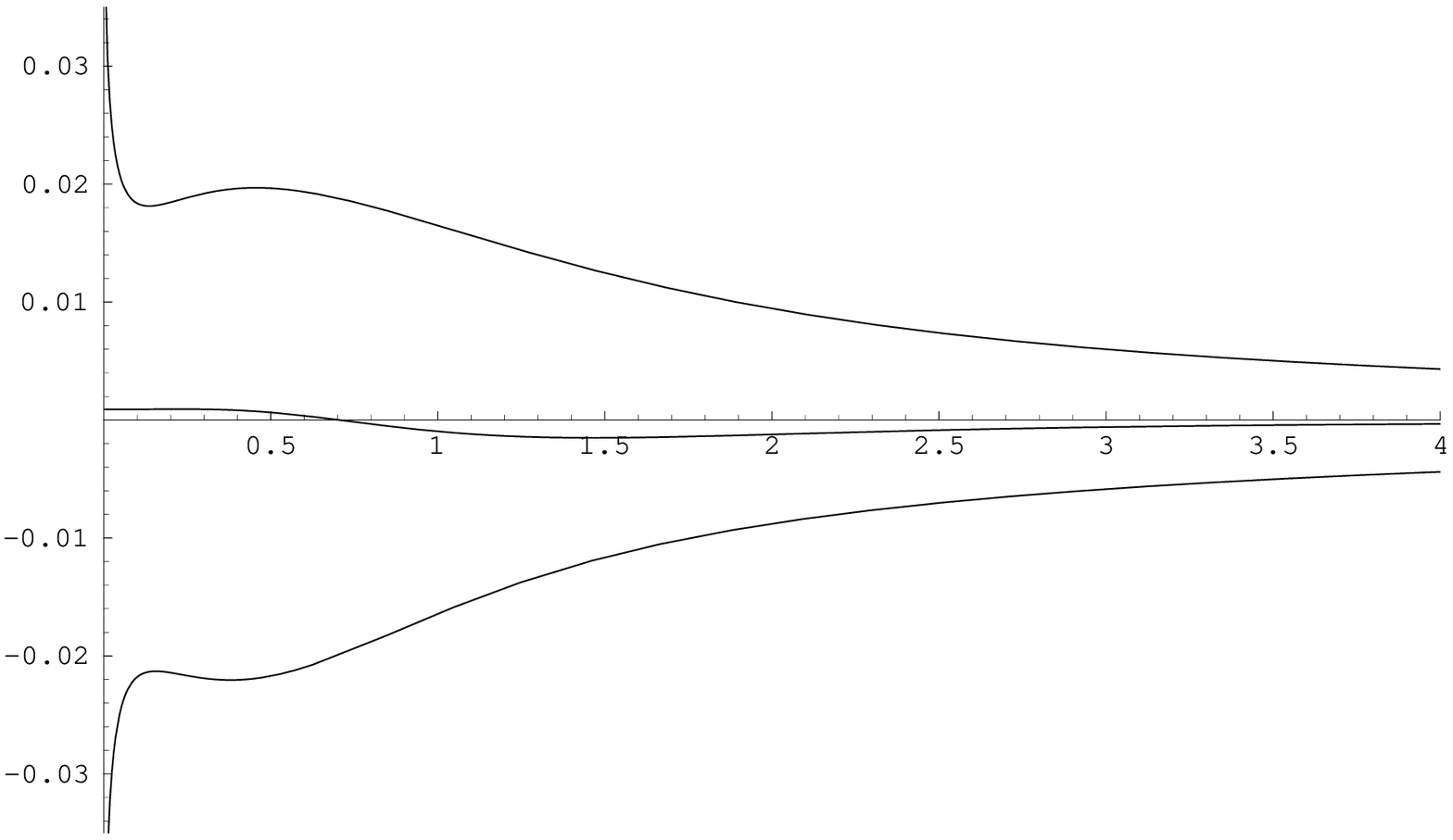,width=14cm,height=6cm}}
\put(1.6,0.7){${\cal E}_{as} $}
\put(1.6,4.9){${\cal E}_{f0} $}
\put(1.2,3.25){${\cal E}_{fl} $}
\put(13.7,2.9){$R$}
\end{picture}
\caption{Repulsive potential. The contributions to the  renormalized vacuum energy multiplied by $R^{2}\cdot \alpha^{-2}$, for $\alpha=0.3$ .} 
\end{figure}

\begin{figure}[hp]\unitlength1cm
\begin{picture}(6,6)
\put(-0.5,0){\epsfig{file=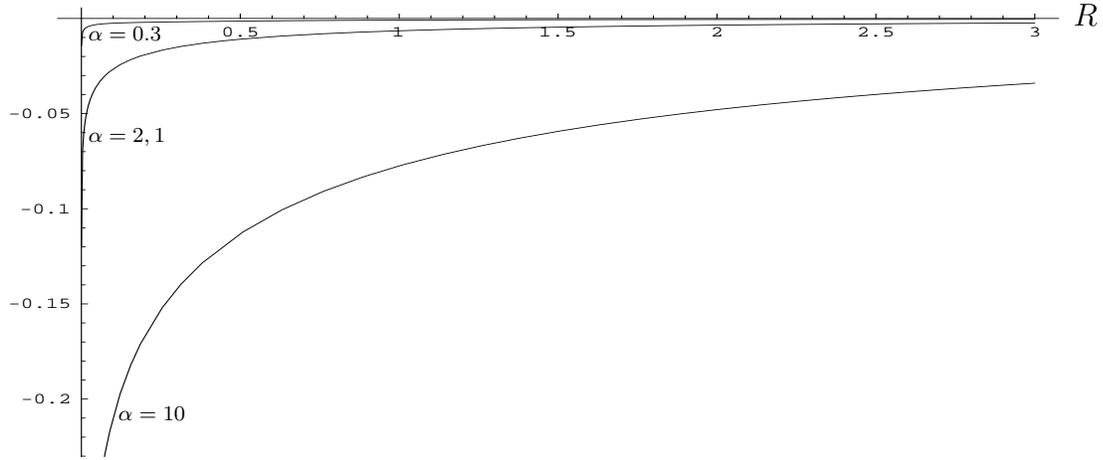,width=14cm,height=6cm}}
\put(1.0,0.5){{\scriptsize $\alpha=10$}}
\put(0.6,5.55){{\scriptsize $\alpha=0.3$}}
\put(0.6,4.2){{\scriptsize $\alpha=2,1$}}
\put(13.7,5.75){$R$}
\end{picture}
\caption{Repulsive potential. The complete renormalized vacuum energy  ${\cal E}_{ren}(R)$ multiplied by $R^{2}\cdot \alpha^{-2}$, for $\alpha=0.3$, $\alpha=2.1$ and $\alpha=10$.} 

\end{figure}
\begin{figure}[hp]\unitlength1cm
\begin{picture}(6,6)
\put(-0.5,0){\epsfig{file=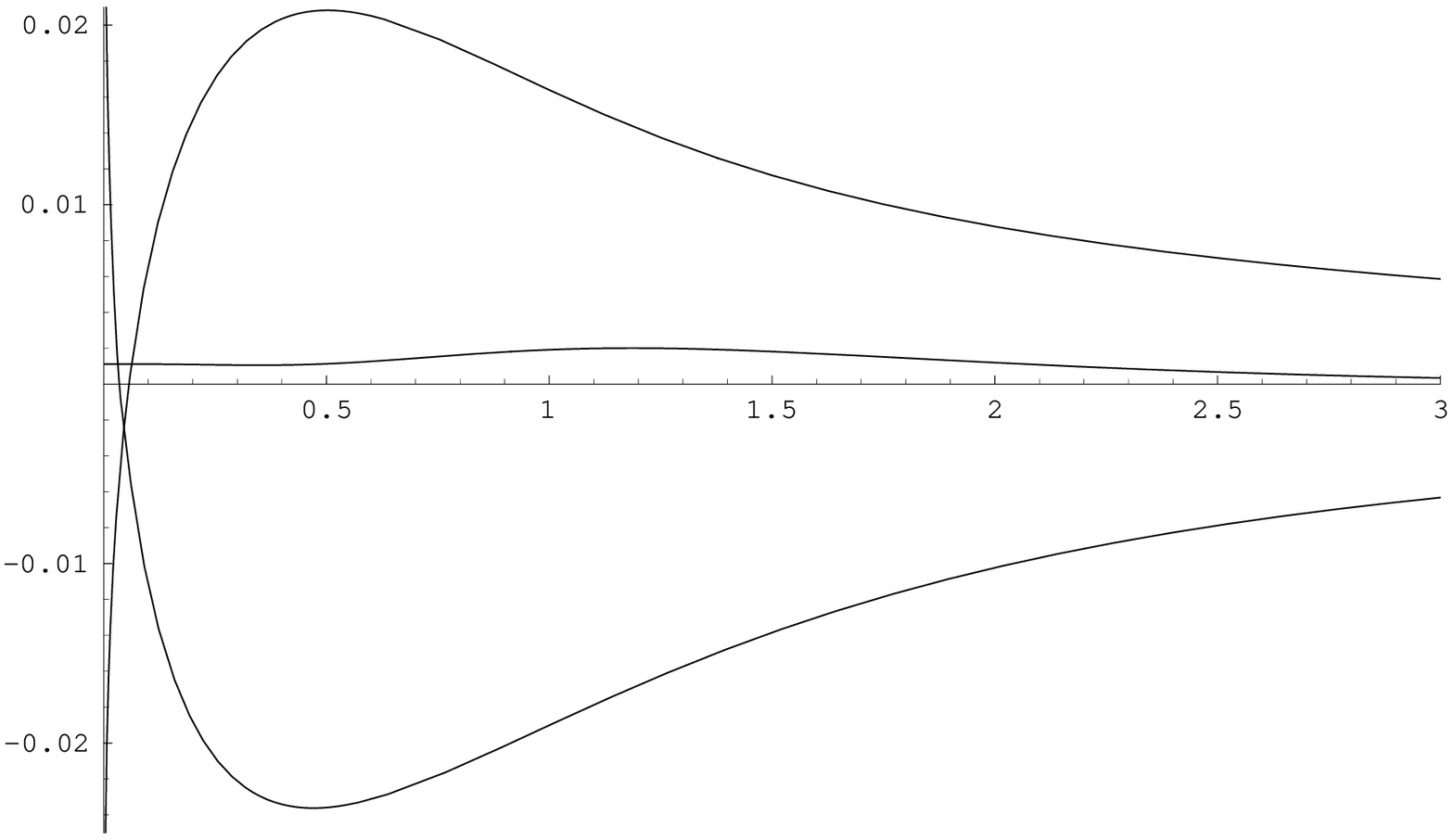,width=14cm,height=6cm}}
\put(5.6,0.7){${\cal E}_{as} $}
\put(4.4,5.7){${\cal E}_{f0} $}
\put(2.5,3.6){${\cal E}_{fl} $}
\put(13.7,3.2){$R$}
\end{picture}
\caption{Attractive potential. The contributions to the  renormalized vacuum energy multiplied by $R^{2}\cdot \alpha^{-2}$, for $\alpha=-0.3$ .}\end{figure}

\section{Discussion}

In this chapter we calculated the vacuum energy of a scalar field in the background of a
cylindrical semi-transparent shell. The formulas for the energy density per
unit length of the shell are given by equations (4.38) and (4.39). The heat kernel
coefficients (4.37) for a cylindrical potential containing a delta function, are also a relevant part of the results. A discussion of the sign of the
vacuum energy is possible by means of the data given in the numerical
section of this chapter. The energy is found to be negative in the background of repulsive potentials ($\alpha>0$) for every finite value of the radius (Fig. 4.3).  This conclusion sets a closeness between the model studied
here and that of a conducting cylinder \cite{Milton-cyl} and of a dielectric
cylinder \cite{Nesterenko}  in the electromagnetic vacuum. The latter models possess also
a  negative Casimir energy. There is also a resemblance with the $\delta$-potential spherical shell (see last chapter) investigated in paper \cite{Marco1}, where a negative energy was observed for large repulsive potentials. 

 In the background of attractive potentials ($\alpha <0$) the energy is negative on almost all the $R$-axis. As we see in Fig. 4.5, the energy becomes positive for very small values of the radius. This happens when the logarithmic term in (4.44) compensates for the term proportional to $-\alpha^2$ in (4.46) which is dominant for small values of $\alpha$ in the region $R>1$. The thin region of positive asymptotically increasing energy appearing in Fig. 4.5 is reasonably out of physical applicability. 
For both repulsive and attractive potentials the renormalized vacuum energy of the semi-transparent cylinder goes logarithmically to $\pm \infty$ in the limit $R\rightarrow 0$. This logarithmic behaviour was also found in the
semi-transparent spherical shell,  however it was not found in the $\delta$-potential flux tube \cite{Marco2} (see chapters 5 and 6), which has
many feature in common with the background investigated here. This can be explained by means of the heat-kernel coefficient $A_2$, which is a non zero coefficient here as well as in \cite{Marco1}. It is also interesting to note how in expression (4.37) for the coefficient $A_2$ the contributions proportional to $\alpha$ and $\alpha^2$ have cancelled and only a term proportional to $\alpha^3$ is present. The cancellation of the lower powers of the potential strength was also observed in \cite{Marco1} and in paper \cite{Bordag dielectric}, where a theta function profile in a dielectric spherical shell is examined. It confirms  the observation that singular profiles show weaker divergences than smoother, less singular profiles.   
\begin{figure}[hb]\unitlength1cm
\begin{picture}(6,6)
\put(-0.5,0){\epsfig{file=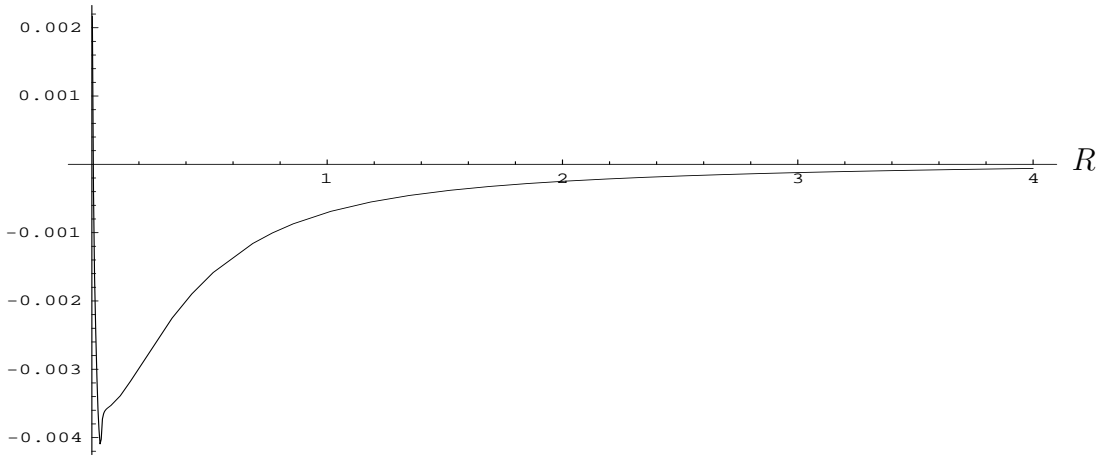,width=14cm,height=6cm}}
\put(13.7,3.8){$R$}
\end{picture}
\caption{Attractive potential. The complete renormalized vacuum energy  ${\cal E}_{ren}(R)$ multiplied by $R^{2}\cdot \alpha^{-2}$, for $\alpha=-0.3$. The peak beneath the lowest point of the curve is due to the presence of a small imaginary part in the renormalized energy causing the function to be no more analytical. The energy starts to be complex at $R \sim 0.04$.} 
\end{figure}

\chapter{Vacuum energy in the presence of a magnetic flux tube: Scalar case}

The formalism applied in the last two chapters in the context of curved material boundaries, can be extended to problems with special background fields, like classical magnetic strings.  The calculations will follow as usual, provided the Jost function is found. This method was used in paper \cite{master} to  analyze the vacuum of a spinor field in the presence of a homogeneous magnetic flux tube of finite radius. The vacuum energy was found to be negative and it did not show a minimum for any finite value of the radius. A natural question is if inhomogeneous magnetic fields can minimize the energy and render the string stable. The question was already raised in \cite{Cangemi}. The present paper extends the investigation begun in \cite{master} to an inhomogeneous magnetic string with delta function profile. The delta function, represents here a simple example of inhomogeneity, which could give an insight into the problem  of vacuum fluctuations in magnetic backgrounds. It is a  kind of ``semi-transparent  magnetic boundary''. The quantum mechanics of spinor fields in the presence of magnetic fluxes has been elaborated in early works \cite{Heisenberg},\cite{Weisskopf} while more recent  investigation in this direction has been motivated by the interest in the Aharonov-Bohm effect \cite{Serebryany}, \cite{Sitenko}.  Singular inhomogeneous magnetic fields were  examined in \cite{Fry} for the calculation of the fermion determinant and in \cite{Voropaev} for the investigation of the bound states of an electron, however the ground state energy was not calculated in those  works. In this chapter the ground state energy will be calculated for a scalar field, while the spinor field will be considered in the next chapter. In  the first part of this chapter we calculate the Jost functions and the heat-kernel coefficients. The energy will be renormalized as for the preceding chapters, imposing the vanishing of the vacuum fluctuations for fields of infinite mass. In the second part of the chapter we will work numerically on the renormalized energy finding its asymptotic behaviour for small and for large values of the radius of the string.

\section{Scalar electrodynamics}
We quantize a scalar field $\Phi$ in the presence of a classical magnetic field whose form is that of a cylindrical shell with delta function profile. The section of the string is a circle with radius $R$. The magnetic field is given by 

\begin{equation}
\vec{B}(r)\ =\ \frac{\phi}{2\pi R}\ \delta(r-R)\ \vec{e}_z   
\end{equation}  
where $\phi$ is the magnetic flux, $r=\sqrt{x^2 +y^2}$ and $z$ is the axis along which the cylindrical shell extends to infinity.
 The quantum field  obeys the Klein-Gordon equation for the scalar electrodynamics:

\begin{equation}
(D^2+m_e^2)\Phi(x)=0
\end{equation}
where $m_e$ is the mass of the field, and

\begin{equation}
\begin{array}{lcl}
D^2 & = & \partial_t^2-\vec{\nabla}^2-2ieA^0\partial_t - 2ie\vec{A}\vec{\nabla} - e^2A^0\ ^2 + e^2\vec{A}^2\ , 
\end{array}
\end{equation}
here $e$ is the electron charge, $A_\mu$ is the vector potential of the electromagnetic field and  the convention $g_{\mu\nu}=$diag$(1,-1,-1,-1)$ is used as well as the gauge $\partial^\mu A_\mu=0$.
We want to find a solution of eq.(5.2) in cylindrical coordinates $z,r$, $\varphi$, then we take the relevant operators in (5.3) in the following form
\begin{equation}
\vec{\nabla} \rightarrow\ \left(-\cos \varphi\partial_r-\frac{\sin\varphi}{r}\partial_\varphi,\ \sin\varphi\partial_r+\frac{\cos\varphi}{r}\partial_\varphi,\ \partial_z\right) 
\end{equation}
\begin{equation}
\vec{\nabla}^2 \rightarrow\ \frac1r\partial_rr\partial_r+\frac{1}{r^2}\partial_\varphi^2+\partial_z^2\ . 
\end{equation}
The potential four vector $A^\mu$ associated with the magnetic field (5.1)  contains a theta-function: 
\begin{equation}
\vec{A}=\frac{\phi}{2\pi}\frac{\Theta(r-R)}{r}\vec{e}_\varphi\ , \ \ A^0=0\ ;
\end{equation}
 where $\vec{e}_\varphi=(-\sin\phi, \cos\phi, 0)$. Therefore operator (5.3) becomes
\begin{equation}
D^2\ \rightarrow\ \partial^2_t-\left(\frac1r\partial_rr\partial_r+\frac{1}{r^2}\partial_\varphi^2+\partial_z^2\right)\ +\ \frac{2i\beta\Theta(r-R)}{r^2}\ \partial_\varphi\ +\ \frac{\beta^2\Theta^2(r-R)}{r^2}\ .
\end{equation} 
Here,  and in the rest of this chapter and of chapter 6, $\beta$ will represent the strength of the background potential
\begin{equation}
\beta\ =\ \frac{e\phi}{2\pi}\ .
\end{equation}
With the ansatz of the separation of the variables the scalar field is transformed into
\begin{equation}
\Phi(x)\ \rightarrow\ \exp(ip_0t-ip_zz+im\varphi)\ \Phi_m(p_0,p_z,r)\ ,
\end{equation}
where $p_\mu$ is the momentum four vector and $m$ is the orbital momentum quantum number.
Combining (5.2), (5.7) and (5.9) the new field equation reads
\begin{equation}
\left(p_0^2-m_e^2-p_z^2-\frac{(m-\beta\Theta(r-R))^2}{r^2}+
\frac 1r\partial_r +\partial_r^2\right)\Phi_m(p_0,p_z,r)\ \ =\ 0 \ .
\end{equation}
or
\begin{equation}
\left(k^2-\frac{(m-\beta\Theta(r-R))^2}{r^2}+
\frac 1r\partial_r +\partial_r^2\right)\Phi_m(k,r)\ =\ 0\ ,
\end{equation}
where $k=\sqrt{p^2_0-m_e^2-p_z^2}$. The solutions of this equation are Bessel and Neumann functions. The kind of function and their coefficients are to be determined by means of physical considerations. We take here the regular solution which in general scattering theory \cite{Taylor} has the known asymptotic behaviour
\begin{equation}
\Phi\ \stackrel{r\rightarrow 0}{\sim} J_m(kr) ,
\end{equation}

\begin{equation}
\Phi\ \stackrel{r\rightarrow \infty}{\sim} \frac 12 (f_m(k)H^{(2)}_{m-\beta}(kr)+f^*_m(k)H^{(1)}_{m-\beta}(kr)) ,
\end{equation}
where $J_m (kr)$ is a Bessel function of the first kind, $H^{(1)}_{m-\beta}(kr)$ and $H^{(2)}_{m-\beta}(kr)$ are Hankel functions of the first and of the second kind and the coefficients $f_m(k)$ and $f^*_m(k)$ are a Jost function and
its complex conjugate respectively. In the case of the delta function profile the regular solution reads

\begin{equation}
\Phi(r)\ =\ J_m (kr)\Theta(R-r)+ \frac 12 (f_m(k)H^{(2)}_{m-\beta}(kr)+f^*_m(k)H^{(1)}_{m-\beta}(kr))\Theta(r-R)\ ,
\end{equation}
then, we can define a field $\Phi^I$ in the region $r<R$ inside the cylinder  
\begin{equation}
\Phi^I_m(k,r)\ =\ J_m (kr)
\end{equation}
which is independent of the strength $\beta$ of the potential, and a field $\Phi^O$ in the region outside the cylinder $r>R$ 
\begin{equation}
\Phi^O_m(k,r)\ =\ \frac 12 (f_m(k)H^{(2)}_{m-\beta}(kr)+f^*_m(k)H^{(1)}_{m-\beta}(kr))
\end{equation}
which describes incoming and outgoing cylindrical waves. The conditions for the field at $r=R$ will be discussed later.

\section{Ground state energy in terms of the Jost function and normalization condition}

A regularized vacuum energy can be defined as
\begin{equation}
E^{sc}_0\ =\ \frac {\mu^{2s}}{2} \sum \epsilon_{(n,\alpha)}^{1-2s}\ ,
\end{equation}
where the $\epsilon_{(n,\alpha)}$ are the eigenvalues of the Hamiltonian operator associated with (5.2), $\alpha=\pm 1 $ being the index for the particle-anti-particle degree of freedom, while $n$ includes all oder quantum numbers. $s$ is the regularization parameter to be put to zero after the renormalization and $\mu$ is the mass parameter necessary to maintain the correct dimensions of the energy.
The string is invariant under translations along the $z$ axis, therefore the energy density per unit length of the string is
\begin{equation}
{\cal E}^{sc}\ =\ \frac 12 \mu^{2s}\int^{\infty}_{-\infty}\frac{dp_z}{2\pi}
\sum_{(n,\alpha)} (p_z^2+ \lambda_{(n)}^2)^{1/2-s}\ ,
\end{equation}
where the  $\lambda_{(n)}$ are the eigenvalues of the operator defined in (12)  with $k=\sqrt{p^2_0-m_e^2}$ . We perform the integration over $p_z$ in (5.18), getting
\begin{equation}
{\cal E}^{sc}\ =\ \frac 14 \mu^{2s}\frac{\Gamma(s-1)}{\sqrt{\pi}\Gamma(s-1/2)}\sum_{(n,\alpha)}(\lambda_{(n)}^2)^{1-s}\ .
\end{equation}
The next step is to transform the summation in (5.19) into a contour integral. We enclose temporarily the system in a large cylindrical quantization box  imposing Dirichlet boundary conditions for the field at this boundary  like we did in section 3.2. The final result will be independent of this choice, since the quantization box must be expanded to infinity. Now we can express the eigenvalues $\lambda_{(n)}$  by means of the zeros of solution (5.16), which is valid in a region far from the origin. The derivative  of the logarithm of the solution will then have poles at $k=\lambda_{(n)}$. We can express (5.19) through an integral whose contour encloses these poles which lie on the real axis $k$. The deformation of the  contour on the imaginary axis and the dropping of the Minkowski space contribution allows to reach the final form.
\begin{equation}
{\cal E} ^{sc}\ =\ -\frac 12 C_s\ \sum^\infty_{m=-\infty}\int^\infty_{m_e}dk\ (k^2+m_e^2)^{1-s} \partial_k \ln f_m(ik)\ ,
\end{equation}
where $f_m(ik)$ is the Jost function with imaginary argument and $C_s =(1+s(-1+2\ln(2\mu)))/(2\pi)$ is, as in chapter 4,  a simple function of the regularization parameter.
The renormalization of (5.20) is carried out by direct subtraction of its divergent part 
\begin{equation}
{\cal E}_{ren}^{sc}\ =\ {\cal E}^{sc}\ -\ {\cal E}_{div}^{sc}\ .
\end{equation}   
The isolation of the divergent part from the total energy will be performed via heat-kernel expansion as we  will see in a moment. The subtracted part should be added in the classical part of the energy resulting in a renormalization of the classical parameters of the string, however we do not treat the classical energy of the system here but only the vacuum contribution. For the analytical continuation $s\rightarrow 0$  we split ${\cal E}_{ren}^{sc}$ it into a ``finite'' and an ``asymptotic'' part. 
\begin{equation}
{\cal E}_{ren}^{sc}={\cal E}_f^{sc} +{\cal E}_{as}^{sc},
\end{equation}
with
\begin{equation}
{\cal E} _f^{sc}= - \frac 12 C_s \sum_{m=-\infty}^\infty\int^\infty_{m_e} dk [k^2 -m_e^2]^{1-s} \frac{\partial}{\partial k} [\ln f_m(ik)- \ln f^{as}_m(ik)] 
\end{equation}
and
\begin{equation}
{\cal E} _{as}^{sc}= -\frac 12 C_s\sum_{m=-\infty}^\infty \int^\infty_{m_e} dk [k^2 -m_e^2]^{1 -s} 
\frac{\partial}{\partial k}  \ln f^{as}_m(ik)-{\cal E}_{div}^{sc} ,
\end{equation}
where $f^{as}_m$ is a portion of the uniform asymptotic expansion of the Jost function. The number of orders to be included in  $f^{as}_m$ must be sufficient to let the function
\begin{equation}
\ln f_m(ik) -  \ln f^{as}_m(ik)
\end{equation}
fall as $m^{-4}$ (or  $k^{-4}$) for $k$ and $m$ equally large, in this case the integral and the summation in (5.23) converge for $s\rightarrow 0$. To this purpose the usual three orders in the asymptotics are enough.  The splitting proposed in (5.22) immediately
permits the analytical continuation $s=0$  in ${\cal E}^{sc}_f$, furthermore  it allows a very quick subtraction of the pole terms in the asymptotic  part (5.24). ${\cal E}_ {as}^{sc}$ is a finite quantity. For the definition of ${\cal E}_ {div}^{sc}$ we write the sum (5.19) as 
\begin{equation}
{\cal E} ^{sc}\ =\ \frac 12 \frac{\mu^{2s}}{\Gamma (s-1/2)}\ \int^\infty_0 dt\ t^{s-3/2} K(t)
\end{equation}
where $K(t)$ is the heat kernel related to the Hamilton operator , which can be expanded for small $t$
\begin{equation}
K(t)\ =\ \sum_{(n)} e^{-t \lambda_{(n)}^2}\ \sim\ \frac {e^{-tm_e^2}}{(4\pi t)^{3/2}} \sum_j^\infty A_j t^j\ , \ \ \ j=0,\frac 12 , 1...
\end{equation} 
The $A_j$ are the heat-kernel coefficients related with the Hamiltonian operator. By means of (5.26) and (5.27) we can expand the ground state energy in powers of the mass and get
\begin{equation}
{\cal E}^{sc}\ =\ \sum_j
\frac{\mu^{2s}}{32\pi^2}\frac{\Gamma(s+j-2)}{\Gamma(s+1)}m_e^{4-2(s+j)}A_j	
\end{equation}
 in which the divergent contribution is isolated
\begin{eqnarray} 
{\cal E}_{div}^{sc} &  =  & -\frac{m_e^4}{64 \pi ^2}\left( \frac 1s + \ln
\frac{4\mu ^2 }{m_e^2} - \frac 12\right) A_0\ -\frac{m_e^3}{24\pi^{3/2}}A_{1/2}
\nonumber \\     
                  &     &  +\frac{m_e^2}{32 \pi ^2}\left( \frac 1s +
\ln\frac{4\mu ^2 }{m_e^2} - \ 1\right) A_1 \ + \frac{m_e}{16 \pi^{3/2}}A_{3/2} \nonumber \\ 
                  &     &  -\frac{1}{32 \pi ^2}\left( \frac 1s + \ln \frac{4\mu ^2 }{m_e^2} - 2\right) A_2\ .  
\end{eqnarray} 
The poles are all contained in the three terms corresponding to the heat kernel coefficients $A_0, A_1, A_2$, however we included in ${\cal E}_{div}^{sc}$ two more terms in order to satisfy the normalization condition
\begin{equation} 
\lim_{m_e \rightarrow \infty }{\cal E}_{ren}^{sc} = 0\ . 
\end{equation} 
Again, the normalization condition  must  be changed if one desires to investigate the limit $m\rightarrow 0$. In paper \cite{Bordag dielectric} the interested reader can find further details and comments about this limit.

\section{The Jost function and its asymptotics}
We calculate now the Jost function $f_m(k)$ related with the delta function potential of the magnetic string. We have defined a field in the outer region $r>R$ and a field in the inner region $r<R$, we now need some matching conditions on the boundary $r=R$. As the delta function is a continuous function, we require that the field is continuous on the boundary. From this condition and from the field equation (5.11) it follows directly that the first derivative of the field must be continuous on the boundary:
\begin{equation}
\left\{
\begin{array}{lcl}
\Phi^O(r)|_{r=R} & = &\Phi^I(r)|_{r=R}\\
\partial_r\Phi^O(r)|_{r=R} & = &\partial_r\Phi^I(r)|_{r=R}
\end{array}
\right.
\end{equation}
and with the use of solution (5.15) and (5.16)
\begin{equation}
\left\{
\begin{array}{lcl}
J_m(kR) & = & \frac 12 (f_m(k)H^{(2)}_{m-\beta}(kR)+f^*_m(k)H^{(1)}_{m-\beta}(kR)) \\
\partial_r J_m(kr)|_{r=R} & = &\frac 12\partial_r\ (f_m(k)H^{(2)}_{m-\beta}(kr)+f^*_m(k)H^{(1)}_{m-\beta}(kr))|_{r=R}\ \ .
\end{array}
\right.
\end{equation}
Solving for $f_m(k)$ one finds
\begin{equation}
f_m(k)\ =\ \frac{2(\partial_r J_m(kr)|_{r=R}H^{(1)}_{m-\beta}(kR)-\partial_r H^{(1)}_{m-\beta}(kr) |_{r=R} J_m(kR))}{H^{(2)}_{m-\beta}(kR)\partial_r H^{(1)}_{m-\beta}(kr) |_{r=R}-H^{(1)}_{m-\beta}(kR)\partial_r H^{(2)}_{m-\beta}(kr) |_{r=R}}\ ,
\end{equation}
which with the use of the Wronskian determinant for the Hankel functions \cite{Abramowitz} becomes
\begin{equation}
f_m(k)\ =\ -\frac{2\pi kR}{4i}\left[J_m(kR)H^{(1)}_{m-\beta+1}-J_{m+1}(kR)H^{(1)}_{m-\beta}(kR) + \frac{\beta}{kR}J_m(kR)H^{(1)}_{m-\beta}(kR)\right]\ .
\end{equation}
The corresponding Jost function on the imaginary axis is written in terms of modified Bessel I and Bessel K functions 
\begin{equation}
f_m(ik)\ =\ i^\beta k R \left[I_m K_{m-\beta+1}+I_{m+1}K_{m-\beta}\right] + i^\beta \beta I_m K_{m-\beta}\ .
\end{equation}
where the arguments $(kR)$ of the Bessel functions are omitted for simplicity. Eq. (5.35) can be written in a more compact form
\begin{equation}
f_m(ik)\ =\ i^\beta k R\left[I'_m K_{m-\beta}+I_{m}K'_{m-\beta}\right]\ .
\end{equation}
where the prime denotes the derivative  with respect to the argument. This expression holds for positive and for negative values of  $m$. On the contrary the uniform asymptotic expansion, which we need for formulas (5.23) and (5.24), is  a different function for positive or for negative $m$. To find it we write the Bessel I and K functions of eq.(5.35) in the form 
\begin{equation}
I_{m+a}((m+a) z'),\ \ K_{m+a}((m+a) z'), \ \ \ \ \ z'=kR/(m+a),
\end{equation}
where $a$ can be $0$ or $1$ for the Bessel I function and $-\beta$ or $-\beta+1$ for the Bessel K function. Their expansions for large positive orders are well known \cite{ Abramowitz}, for instance $K_{m+a}((m+a)z')$ is expanded as
\begin{equation}
K_{m+a}((m+a)z')\ \sim\ \sqrt{\frac{\pi}{2(m+a)}}\frac{e^{-(m+a)\eta '}}{(1+z'^2)^{1/4}}\left\{ 1+\sum_{j=1}^{\infty}(-1)^j \frac{ u_j([1+z'^2]^{-1/2})}{(m+a)^j}\right\}\ ,
\end{equation}
where $\eta=\sqrt{1+z'^2}+\ln(\frac{z'}{1+\sqrt{1+z'^2}})$ and the $u_j(x)$ are the Debye polynomials in the variable $x$. However we are interested in an expansion in powers of the variable $m$ alone: an expansion of the form
\begin{equation}
K_{m+a}((m+a)z')\ \sim\ \sum_n \frac{X_n}{m^n}\ , 
\end{equation}
where the $X_n $ are some coefficients depending on $k,R$ and $\beta$. Therefore we make the substitution $z'\rightarrow z\left(\frac{m}{m+a}\right)$, with $z=KR/m$
in the argument of the Bessel function and we rewrite its  expansion as 
\begin{equation}
\begin{array}{lcl}
K_{m+a}((m+a)z')\ & \sim\ & \sqrt{\frac{\pi}{2(m+a)}}\frac{e^{-(m+a)\eta}}{(1+\left(z\frac{m}{m+a} \right)^2)^{1/4}}\\
                            &       & \left\{ 1+\sum_{j=1}^{\infty}(-1)^j \frac{ u_j([1+\left(z\frac{m}{m+a} \right) ^2]^{-1/2})}{m^j \left(1+\frac{a}{m}\right)^j}\right\}
\end{array}
\end{equation}
with the obvious change for $\eta '$. Then, re-expanding in powers $1/m^n$ we get

\begin{equation}
K_{m+a}(kR)\ \sim\ \sqrt{\frac{\pi}{2m}}\exp \{\sum_{n=-1}^3 m^{-n} S_K(n,a,t)\}\ ,
\end{equation} 
where  $t=(1+z^2)^{-1/2}$ and  the functions $S_K(n,a, t)$ are given explicitly in the Appendix. The corresponding expansion of the Bessel I function in negative powers of $m$ will be 
\begin{equation}
I_{m+a}(kR) \ \sim\ \frac{1}{\sqrt{2\pi m}}\exp \{\sum_{n=-1}^3 m^{-n} S_I(n,a, t)\}\ ,
\end{equation}
where the functions  $S_I(n,a,t)$ are given in the appendix. Inserting these expansions in (5.35) one finds an asymptotic Jost function valid for positive $m$,  we name it $f^{as+}_m(ik)$. To find the asymptotics for negative  $m$ we must take the Jost function in its form 
\begin{equation}
f_m(ik)\ =\ i^\beta k R\left[I_{-m} K_{-m+\beta-1}+I_{-m-1} K_{-m+\beta}\right] + i^\beta \beta I_{-m} K_{-m+\beta}\ ,
\end{equation}
which is identical to (5.35) because of  the property of the modified Bessel functions
\begin{equation}
I_l(x)=I_{-l}(x)\ , \ \ \ K_p(x)=I_{-p}(x)\ ,
\end{equation}  
where $l$ is any natural number and $p$ any real number. Then in (5.43) a large positive index always correspond to a large  negative $m$ and inserting (5.41) and (5.42) in (5.43) we find an asymptotic Jost function valid  for negative $m$, we call it $f^{as-}_m(ik)$. 

For $m=0$, the  asymptotic Jost function can be obtained from (5.35), using the expansions \cite{Abramowitz}
\begin{equation}
I_{\nu}(z)\ \sim\ \frac{e^z}{\sqrt{2\pi z}}\left\{1-\frac{\mu -1}{8z}+\frac{(\mu -1)(\mu -9)}{2!(8z)^2}-...\right\}\ ,\ \ \ \mu=4\nu^2\ ;
\end{equation}
\begin{equation}
K_{\nu}(z)\ \sim\ \frac{\sqrt{\pi}e^{-z}}{\sqrt{2z}}\left\{1+\frac{\mu -1}{8z}+\frac{(\mu -1)(\mu -9)}{2!(8z)^2}+...\right\}\ ,
\end{equation}
we call this contribution $f_0^{as}(ik)$. The finite and the asymptotic part of the energy  defined in (5.23),(5.24) are also split into three contributions: one for positive $m$, one for negative $m$ and one for $m=0$. The positive and negative contributions can be summed up in a single term, but the contribution coming from $m=0$ must be calculated separately and summed just numerically at the end, in fact we have
\begin{eqnarray}
{\cal E} _f^{sc}  & = & - \frac 12 C_s\left(\sum_{m=1}^\infty \int_{m_e}^{\infty} dk\ (k^2+m_e^2) \partial_k (\ln f_m^\pm (ik) -\ln f_m^{as\pm}(ik))\right.\nonumber\\
                  &   & \left. + \int_{m_e}^{\infty} dk\ (k^2+m_e^2) \partial_k (\ln f_0 (ik)- \ln f_0^{as} (ik) )\ ,\right)
\end{eqnarray}
and
\begin{eqnarray}
{\cal E} _{as}^{sc} & = &  - \frac 12 C_s \left( \sum_{m=1}^\infty \int_{m_e}^{\infty} dk\ (k^2+m_e^2)^{1-s} \partial_k \ln f_m^{as\pm}(ik)\right.\nonumber\\
     &   &  \left. +  \int_{m_e}^{\infty} dk\ (k^2+m_e^2)^{1-s} \partial_k \ln f_0^{as}(ik)\ \ -\ {\cal E}_{div}^{sc}\right)\ ,
\end{eqnarray}
where 
\[
\ln f_m^\pm= \ln f_m(ik)+\ln f_{-m}(ik) 
\]
and
\[
\ln f^{as\pm}_m(ik)=\ln f^{as+}_m(ik)+\ln f^{as-}_{-m}(ik).
\]
 Taking the logarithm of $f^{as+}_m(ik)$ and $f^{as-}_{-m}(ik)$ and re-expanding in powers of $m$ we find the functions needed in the integrand of (5.47)(5.48) up to the desired order. We are interested in  $\ln f^{as\pm}_m(ik)$ and $\ln f^{as}_0(ik)$ up to the third order, they read
\begin{equation}
\ln f_0^{as}\ =\ \frac{\beta^2}{2kR}\ ,
\end{equation}
\begin{equation}
\ln f_m^{as\pm}(ik)\ =\ \sum_n^3\sum_t X_{n,j}\frac{t^j}{m^n}\ ,
\end{equation}
where $t=1/(1+(kR/m)^2)^{\frac 12}$ and the nonzero coefficients are
\begin{equation}
\begin{array}{l}
X_{1,1}=\beta^2\ ,\ X_{2,4}=\beta^2/4\ , \  \\
X_{3,3}=\beta^2/24-\beta^4/12\ ,\ X_{3,5}=-\beta^2/2+\beta^4/4\ ,\\ X_{3,7}=\beta^4/16\ .
\end{array}
\end{equation}
As we found also in the preceding chapters, three orders are sufficient for the convergence of ${\cal E} _f^{sc}$. More orders could have been included in definition (5.49) and (5.50) to let the sum and the integral in (5.47) converge more rapidly\footnote{ We would like to stress that with the introduction of asymptotic expansions in our calculation we do {\itshape not} approximate the vacuum energy. The total energy as defined in (5.21) remains an exact quantity. The uniform asymptotics of the Jost function is just a mathematical tool which permits the analytical continuation $s\rightarrow 0$.}.

\section{The asymptotic part of the energy and the heat kernel coefficients}

Having found the Jost function related to the cylindrical delta potential an important part of the calculation is done. We proceed with the analytical simplification of ${\cal E} _{as}^{sc}$. The second term in (5.48) which we name ${\cal E}_{as0}^{sc}$, can be quickly calculated inserting in it the result (5.49). We find
\begin{equation}
{\cal E} _{as0}^{sc}\ =\ -\frac{\beta^2m_e}{4 \pi R}\ .
\end{equation}
The first term in (5.48), which we name here ${\cal E} _{as(m)}^{sc}$ contains the sum over $m$ which can be carried out with the Abel-Plana formula 
\begin{equation}
\sum^\infty_{m=1}F(m)=\int_0^\infty dm F(m)\ -\ \frac 12 F(0) +\ \int_0^\infty\frac{dm}{1-e^{2\pi m}}\frac{F(im)-F(-im)}{i}\ .
\end{equation}
In our case is
\begin{equation}
F(m)=\int_{m_e}^\infty dk (k^2+m_e^2)^{1-s}\partial_k \ln f^{as\pm}_{m}(ik) \ .
\end{equation}
Thus ${\cal E}_{as(m)}^{sc}$ is split into three addenda:
\begin{equation}
{\cal E}_{as1}^{sc} \ = -\ \frac 12 C_s \int_0^\infty dm\  \int_{m_e}^\infty dk (k^2+m_e^2)^{1-s}\partial_k \ln f^{as\pm}_{m}(ik)
\end{equation}
\begin{equation}
{\cal E}_{as2}^{sc} \ =\ \frac 14 C_s F(0)\ ,
\end{equation}
\begin{equation}
{\cal E}_{as3}^{sc}\ = -\frac 12 C_s \int_0^\infty\frac{dm}{1-e^{2\pi m}}\frac{F(im)-F(-im)}{i}\ .
\end{equation}
The contributions ${\cal E}_{as1}^{sc} $ and ${\cal E}_{as2}^{sc} $ can be calculated with the formulas given in appendix B. The result is
\begin{equation}
{\cal E}_{as1}^{sc}\ =\ \frac{\beta^2 m_e^2 }{8\pi}\left(\frac 1s +\ln\left(\frac{4\mu^2}{m_e^2}\right) -1\right)\ -\ \frac{\beta^2 m_e}{32 R}\ %+\ \frac{\beta^2-6\beta^4}{4096 m_e R^3}\ +\ \frac{27 \beta^2-100\beta^4+80\beta^6}{1572864 m_e^3 R^5}
\ ,
\end{equation}
\begin{equation}
{\cal E}_{as2}^{sc}\ =\ \frac{\beta^2 m_e }{4\pi R}\ -\ \frac{\beta^2 }{96\pi m_e R^3}\ +\ \frac{\beta^4 }{48\pi m_e R^3}\ .
\end{equation}
The divergences are all contained in the first term of ${\cal E}_{as1}^{sc}$. The first term of ${\cal E}_{as2}^{sc}$ will cancel with ${\cal E}_{as0}^{sc} $ and  it remains only one term containing a positive power of the mass. This term is the contribution to ${\cal E}^{div}_{sc}$ corresponding to the heat kernel coefficient $A_{3/2}$ (see eq.(5.29)) and therefore it will be subtracted as well as the pole term.

The last contribution to ${\cal E}_{as}^{sc}$ is ${\cal E}_{as3}^{sc}$, whose calculation demands a little more work. Using the formula displayed in  appendix B to calculate the integral over $k$, we  we find
\begin{equation}
{\cal E}_{as3}^{sc}\ =\ (-1)\frac 12 C_s \sum_{n,j} X_{n,j}\left[-m_e^{2-2s}\Gamma(2-s)\right] \Lambda_{n,j}(m_eR)\ ,
\end{equation}
where  the functions $\Lambda_{n,j}(m_e R)$ are given by
\begin{equation}
\begin{array}{lcl}
\Lambda_{n,j}(x) & = & \frac{\Gamma(s+ j/2 -1)}{\Gamma (j/2)x^j}\left[\int_0^x\frac{dm}{1-e^{2\pi m}}\frac{m^{j-n}\ 2\sin \left[\frac{\pi}{2}(j-n)\right]}{\left[1-\frac{m^2}{x^2}\right]^{s+j/2 -1}} \right.\\
                &   & +\left.\int_x^\infty\frac{dm}{1-e^{s\pi m}}\frac{m^{j-n}\ 2\sin \left[\pi (1-s- n/2)\right]}{\left[\frac{m^2}{x^2}-1\right]^{s+j/2 -1}}   \right]\ .
\end{array}
\end{equation}
We have calculated these functions by partial integration for $n\leq 3$ and $j\leq 7$ and we show them explicitly in the appendix B. By means of those functions and of the coefficients (5.51)  we find
\begin{eqnarray}
{\cal E}_{as3}^{sc} & = & -\frac{\beta^2}{\pi R^2}\ \int_{m_eR}^{\infty} \frac{dm}{1-e^{2\pi m}} \sqrt{m_e^2 -(m_eR)^2} \nonumber \\
             &   &  +\left( \frac{\beta^2}{24 \pi R^2}-\frac{\beta^4}{12 \pi R^2}\right)\ \int_{m_eR}^{\infty} dm \left(\frac{1}{1-e^{2\pi m}}\frac 1m \right)' \sqrt{m_e^2 -(m_eR)^2} \nonumber \\
             &   & +\left(- \frac{\beta^2}{6\pi R^2}+\frac{\beta^4}{12 \pi R^2}\right)\ \int_{m_eR}^{\infty} dm \left(\left(\frac{m}{1-e^{2\pi m}}\right)'\frac 1m \right)' \sqrt{m_e^2 -(m_eR)^2}\nonumber \\
             &   &  +\frac{\beta^2}{24 \pi R^2}\ \int_{m_eR}^{\infty} dm \left( \left( \left(\frac{m^3}{1-e^{2\pi m}}\right)'\frac 1m \right)' \frac 1m\right)' \sqrt{m_e^2 -(m_eR)^2}\nonumber \\    
             &   &  + \frac{\beta^2 m_e}{16 R}\frac{1}{1-e^{2\pi m_e R}}\ ,
\end{eqnarray}
where the prime in the integrand denotes derivative with respect to $m$. The last term in (5.62) goes to zero for $m_e\rightarrow \infty$ and therefore must be included in the renormalized energy, however, given its exponential behaviour,  it does not contribute to the heat kernel coefficients. The heat kernel coefficients, which we have calculated up to the coefficient $A_{7/2}$ (including four more orders in $\ln f^{as\pm}(ik)$), read
\begin{equation}
\begin{array}{lclcccl}
A_0 & = & 0                 & , &   A_{1/2} & = & 0 \nonumber \\
A_1 & = &  4\pi\beta^2    & , &   A_{3/2} & = &  \frac{\beta^2 \pi^{3/2}}{2R} \nonumber \\
A_2 & = & 0                & , &   A_{5/2} & = & \frac{[(3\pi-128)\beta^2+(256-18\pi)\beta^4]\pi^{1/2}}{384R^3} \nonumber \\
A_3 & = & 0                & , &  A_{7/2}  & = & \frac{(27\beta^2 -100\beta^4+80\beta^6)\pi^{3/2}}{24576 R^5}\nonumber\ . \\ 
\end{array}
\end{equation}
We perform the subtraction proposed in (5.21) and we obtain the final result
\begin{eqnarray}
{\cal E}_{as}^{sc} & = & -\frac{\beta^2}{\pi R^2}\  p_1(m_eR)\ +\ \left( \frac{\beta^2}{24 \pi R^2}-\frac{\beta^4}{12 \pi R^2}\right)\ p_2(m_eR)\nonumber \\ 
       &  & +\left( \frac{-\beta^2}{6 \pi R^2}+\frac{\beta^4}{12 \pi R^2}\right)\ p_3(m_eR)\nonumber \\ 
       &  & +\ \frac{\beta^2}{24 \pi R^2}\  p_4(m_eR)\ -\ \frac{\beta^2}{96 \pi m_e R^3}\ +\ \frac{\beta^4}{48 \pi m_e R^3}\nonumber\\
       &  & +\  \frac{\beta^2 m_e}{16 R}\frac{1}{1-e^{2\pi m_e R}}\ .      
\end{eqnarray}
The functions $p_n(x)$ are given by
\begin{eqnarray}
p_1(x) & = & \int_{x}^{\infty} \frac{dm}{1-e^{2\pi m}} \sqrt{m^2 -x^2} \nonumber \\
p_2(x) & = & \int_{x}^{\infty} dm \left(\frac{1}{1-e^{2\pi m}}\frac 1m \right)' \sqrt{m^2 -x^2} \nonumber \\
p_3(x) & = & \int_{x}^{\infty} dm \left(\left(\frac{m}{1-e^{2\pi m}}\right)'\frac 1m \right)' \sqrt{m^2 -x^2} \nonumber \\
p_4(x) & = & \int_{x}^{\infty} dm \left( \left( \left(\frac{m^3}{1-e^{2\pi m}}\right)'\frac 1m \right)' \frac 1m\right)' \sqrt{m^2 -x^2}\ ;
\end{eqnarray}
they are all convergent integrals which can be easily numerically calculated.
 The finite part of the ground state energy given by (5.47) can be integrated by parts giving
\begin{eqnarray}
{\cal E}_f^{sc} & = & \frac{1}{2\pi}\sum_{m=1}^{\infty}\int_{m_e}^{\infty} dk\ k \left[\ln f^{\pm}_m(ik) - \sum_n^3\sum_t X_{n,j}\frac{t^j}{m^n}\right]\nonumber \\
    &   & +\frac{1}{2\pi}\int_{m_e}^{\infty} dk\ k \left[\ln f_0(ik) - \frac{\beta^2}{2kR}\right]\ ,
\end{eqnarray}
where the coefficient $X_{n,j}$ are given by (5.51), $t=1/(1+(kR/m)^2)^{\frac 12}$ and
\begin{equation}
\ln f^{\pm}_m(ik)\ =\ \ln \left[I'_m K_{m-\beta}+I_{m}K'_{m-\beta}\right]\ +\ \ln \left[I'_m K_{m+\beta}+I_{m}K'_{m+\beta}\right]\ ,  
\end{equation}
\begin{equation}
\ln f_0(ik)\ =\  \ln \left[ I'_0 K_{-\beta}+I_{0}K'_{-\beta}\right]\ .
\end{equation}
Equations(5.64) and (5.66) are  to be considered the main analytical result of this chapter devoted to the scalar field. Their sum gives the total renormalized vacuum energy. The sum will be performed in the numerical part of this chapter.

\section{Numerical evaluations}

In this section we show some graphics of ${\cal E}_{as}^{sc}$,  ${\cal E}_{f}^{sc}$ and of the complete renormalized vacuum energy ${\cal E}_{ren}^{sc}$ as a function of the radius of the string. We calculate also the asymptotic behavior of  ${\cal E}_{as}^{sc}$ and ${\cal E}_f^{sc}$ for large and for small $R$. Since we want to study here only the dependence  on $R$ and on $\beta$, we set $m_e=1$ for all the calculations of this section.  

As a first step we rewrite  ${\cal E}^{sc}_{as}$ in a form in which the dependence on the relevant parameters is more explicit:
\begin{equation}
{\cal E}^{sc}_{as}\ =\ \frac{1}{\pi R^2}\left[\beta^2 g_1(m_eR)\ +\ \beta^4 g_2(m_e R)\right]\ ,
\end{equation} 
with
\begin{eqnarray}
g_1(x) & = & \left(-p_1(x)+\frac{1}{24}p_2(x)-\frac 16 p_3(x) +\frac{1}{24}p_4(x)-\frac{1}{96x} + \frac{x}{16(1-e^{2\pi x})}\right)\ ,\nonumber\\
g_2(x) & = & \left(-\frac{1}{12}p_2(x)+\frac{1}{12}p_3(x)+\frac{1}{48x}\right)\ .
\end{eqnarray}
the asymptotic behaviour of the $p_n(x)$ functions for $x\rightarrow 0$ is found to be
\begin{eqnarray}
p_1(x) & \sim & -1/24\ +{\cal O}(x)\ ,\nonumber\\
p_2(x) & \sim & \frac{1}{4x}+\frac 12 \ln x -0.0575\ + {\cal O}(x)\ ,\nonumber\\
p_3(x) & \sim & \frac 12 \ln x +0.442\ +{\cal O} (x)\ ,\nonumber\\
p_4(x) & \sim & \frac 32\ln x + 1.826\ +{\cal O} (x)\ ;
\end{eqnarray}
the corresponding behaviour of the functions $g_n(x)$ is
\begin{eqnarray}
g_1(x) & \sim & 0.0317 +\ {\cal O}(x)\ ,\nonumber\\
g_2(x) & \sim & 0.0417   +\ {\cal O}(x) \ .
\end{eqnarray}
The logarithmic contributions have cancelled. This was to expect also from the vanishing of the heat kernel coefficient $A_2$. ${\cal E}^{sc}_{as}$ is proportional to  $R^{-2}$ for $R\rightarrow 0$ and for an arbitrary $\beta$. For $R\rightarrow \infty$ all the $p_n(x)$ functions fall exponentially and so does ${\cal E}^{sc}_{as}$.
 
The finite part ${\cal E}^{sc}_f$ is also proportional to $R^{-2}$ in the limit $R\rightarrow 0$, it can be seen substituting $k\rightarrow k/R$ in the integrands of expression (5.66)
\begin{eqnarray}
{\cal E}^{sc}_f & = & \frac{1}{2\pi R^2}\sum_{m=1}^{\infty}\int_{m_eR}^{\infty} dk k \left[\ln f^{\pm}_m(ik)|_{k\rightarrow k/R} - \sum_n^3\sum_t X_{n,j}\frac{t^j}{m^n}\right]\nonumber \\
    &   & +\frac{1}{2\pi R^2}\int_{m_e}^{\infty} dk\ k \left[\ln f_0(ik)|_{k\rightarrow k/R} - \frac{\beta^2}{2k}\right]\ ,
\end{eqnarray}
where the Jost functions are now independent of $R$. We name the first addend in (5.73) ${\cal E}^{sc}_{fm}$ and second addend ${\cal E}^{sc}_{f0}$, in the plots we will display them separately. For large $R$ we found numerically ${\cal E}^{sc}_f\sim R^{-3}$, which is in agreement with the heat kernel coefficient $A_{5/2}$ shown in (5.63), in fact the first non vanishing heat kernel coefficient after $A_2$ determines the behaviour of the renormalized energy for $R\rightarrow \infty$. Below we show the plots of 
all the contributions to the renormalized ground state energy. The functions $g1(x)$ and  $g1(x)$ are shown as well. Each contribution has been multiplied by $R^2$ so that all the curves take a finite value at $R=0$. We found necessary to sum up to $20$ in the parameter $m$ an to integrate up to $1000$ in the $k$ variable in order to obtain reliable plots. All the calculations where performed with computer programming relying on a precision of $34$ digits.
\begin{figure}[h]\unitlength1cm
\begin{picture}(6,6)
\put(-0.5,0){\epsfig{file=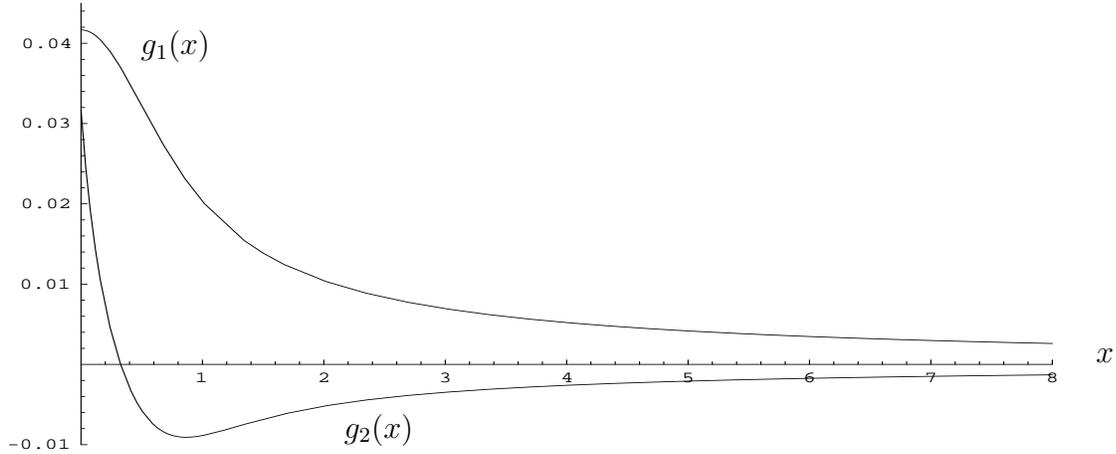,width=14cm,height=6cm}}
\put(1.3,5.3){$g_1(x)$}
\put(14,1.2){$x$}
\put(4,0.2){$g_2(x)$}
\end{picture}
\caption{ Scalar field. The functions $g_1(x)$ and $g_2(x)$ contributing to the asymptotic part of the ground state energy.} 
\end{figure}

\begin{figure}[h]\unitlength1cm
\begin{picture}(6,6)
\put(-0.5,0){\epsfig{file=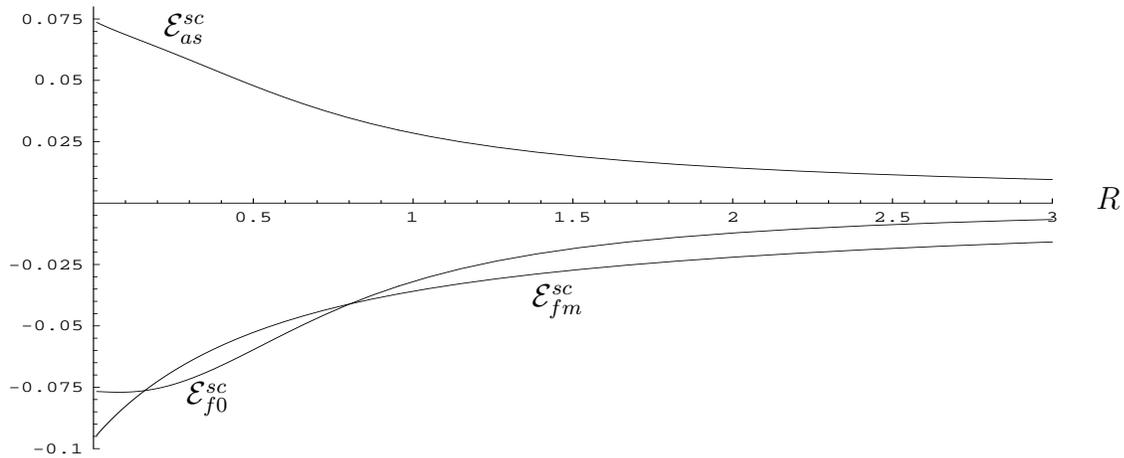,width=14cm,height=6cm}}
\put(1.6,5.6){${\cal E}^{sc}_{as}$}
\put(6.5,2.0){${\cal E}^{sc}_{fm}$}
\put(1.9,0.7){${\cal E}^{sc}_{f0}$}
\put(14,3.3){$R$}
\end{picture}
\caption{ Scalar field. ${\cal E}^{sc}_{as}$, ${\cal E}^{sc}_{fm}$ and ${\cal E}^{sc}_{f0}$ multiplied by $R^{2}\cdot \beta^{-4}$, for $\beta =2.2$.} 
\end{figure}

\begin{figure}[h]\unitlength1cm
\begin{picture}(6,6)
\put(-0.5,0){\epsfig{file=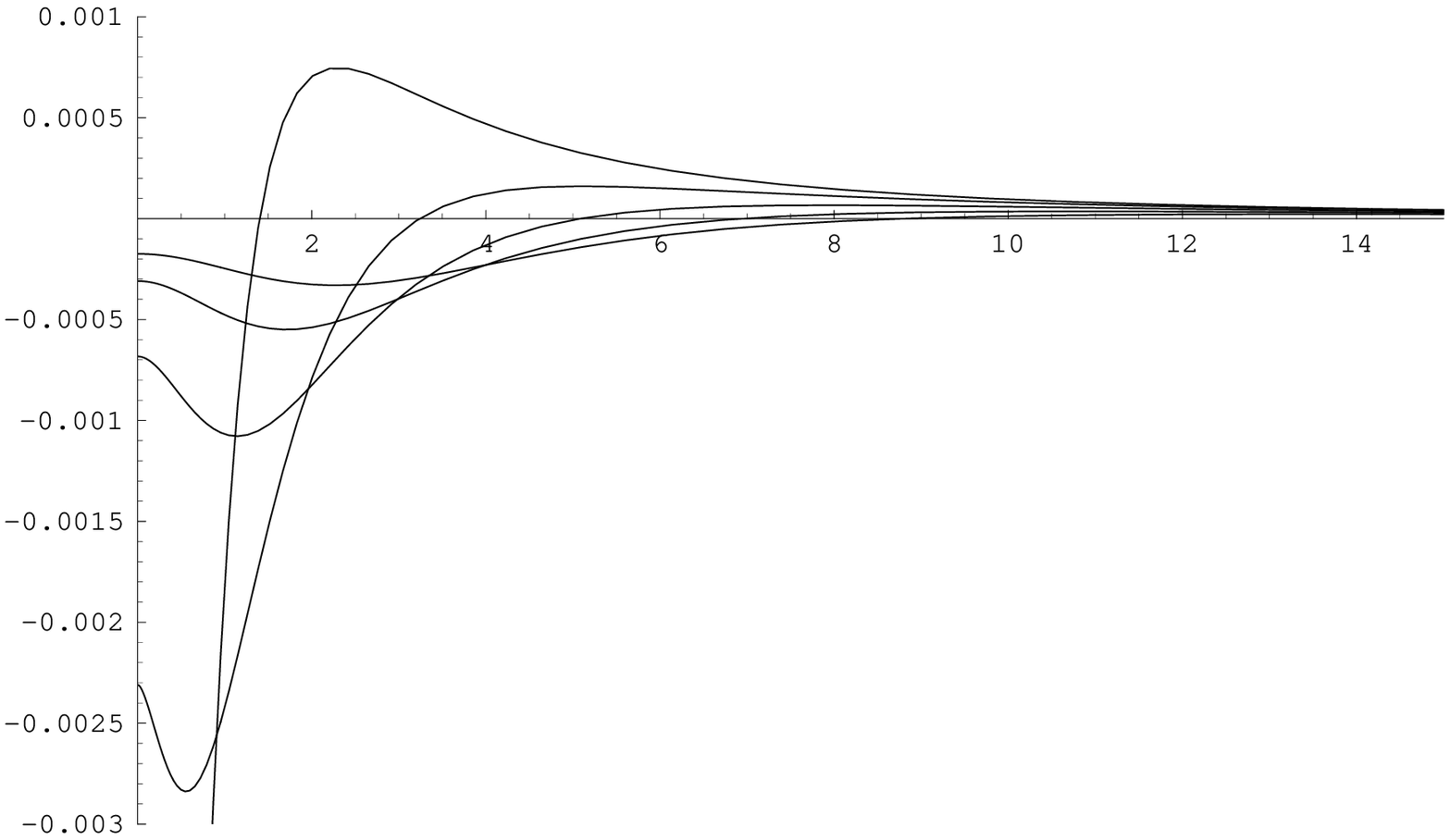,width=14cm,height=6cm}}
\put(2.3,2.3){$m=2$}
\put(3.2,5.6){$m=1$}
\put(14,4.3){$R$}
\end{picture}
\caption{ Scalar field. The contributions $m=1,2...5$ to the finite part of the energy multiplied by $R^{2}\cdot \beta^{-4}$, for $\beta =2.2$.}
\end{figure}

\section{Discussion}
In this chapter we have carried out a complete calculation of the vacuum energy of a scalar field in the background of a magnetic string with delta function profile. The renormalized vacuum energy is given in terms of convergent integrals (eq.(5.64) and (5.66)). A first remark can be made about the vanishing of the heat kernel coefficient $A_2$. This coefficient, contributing to ${\cal E}^{sc}_{div}$, is not zero for a generic background potential. The vanishing of $A_2$ is also observed in a dielectric spherical shell with a squared profile \cite{Bordag dielectric}. Here, as in the preceding chapters it could be argued that singular profiles possess less ultraviolet divergences than smooth profiles. Such a statement is also confirmed by the heat kernel coefficients calculated in \cite{BordagVass-heat kernels}. 

The dependence of the sign of the energy on the radius $R$ of the string and on the potential strength $\beta$ is non trivial. The energy is negative only for large values of the potential strength, while for $\beta$ smaller than one, the energy shows a maximum (Fig. 5.5).  For large $R$ the vacuum energy is negative in the background of strong fluxes $\beta>1$ and positive in the background of  weak fluxes $\beta<1$. This strong dependence on the parameter $\beta $ was not observed in paper \cite{master}, where an homogeneous field inside the flux tube was investigated. In fact the most relevant result of our calculation is that the energy numerically shows a dependence on $\beta^{4}$ for large $\beta$.

\begin{figure}[h]\unitlength1cm
\begin{picture}(6,6)
\put(-0.5,0){\epsfig{file=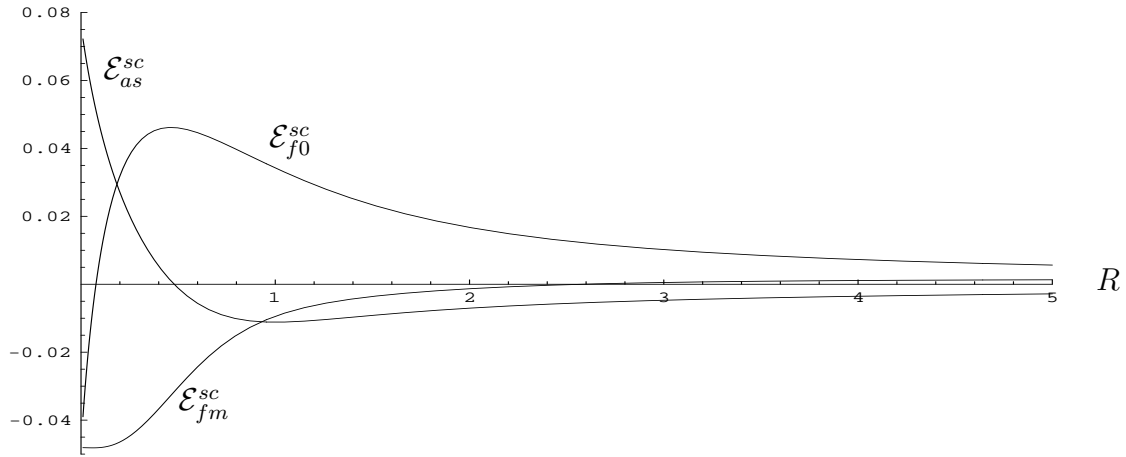,width=14cm,height=6cm}}
\put(3.0,4.1){${\cal E}^{sc}_{f0}$}
\put(1.8,0.6){${\cal E}^{sc}_{fm}$}
\put(0.8,5.0){${\cal E}^{sc}_{as}$}
\put(14,2.2){$R$}
\end{picture}
\caption{ Scalar field. The  three  contributions to the renormalized vacuum energy multiplied by $R^{2}\cdot \beta^{-4}$, for $\beta =0.4$. The curve ${\cal E}^{sc}_{f0}$ shows the contribution to the finite part of the energy coming from the orbital momentum $m=0$, while ${\cal E}^{sc}_{fm}$ displays the contribution coming from the sum of all other $m$'s.} 
\end{figure} 

\begin{figure}[tp]\unitlength1cm
\begin{picture}(6,6)
\put(-0.5,0){\epsfig{file=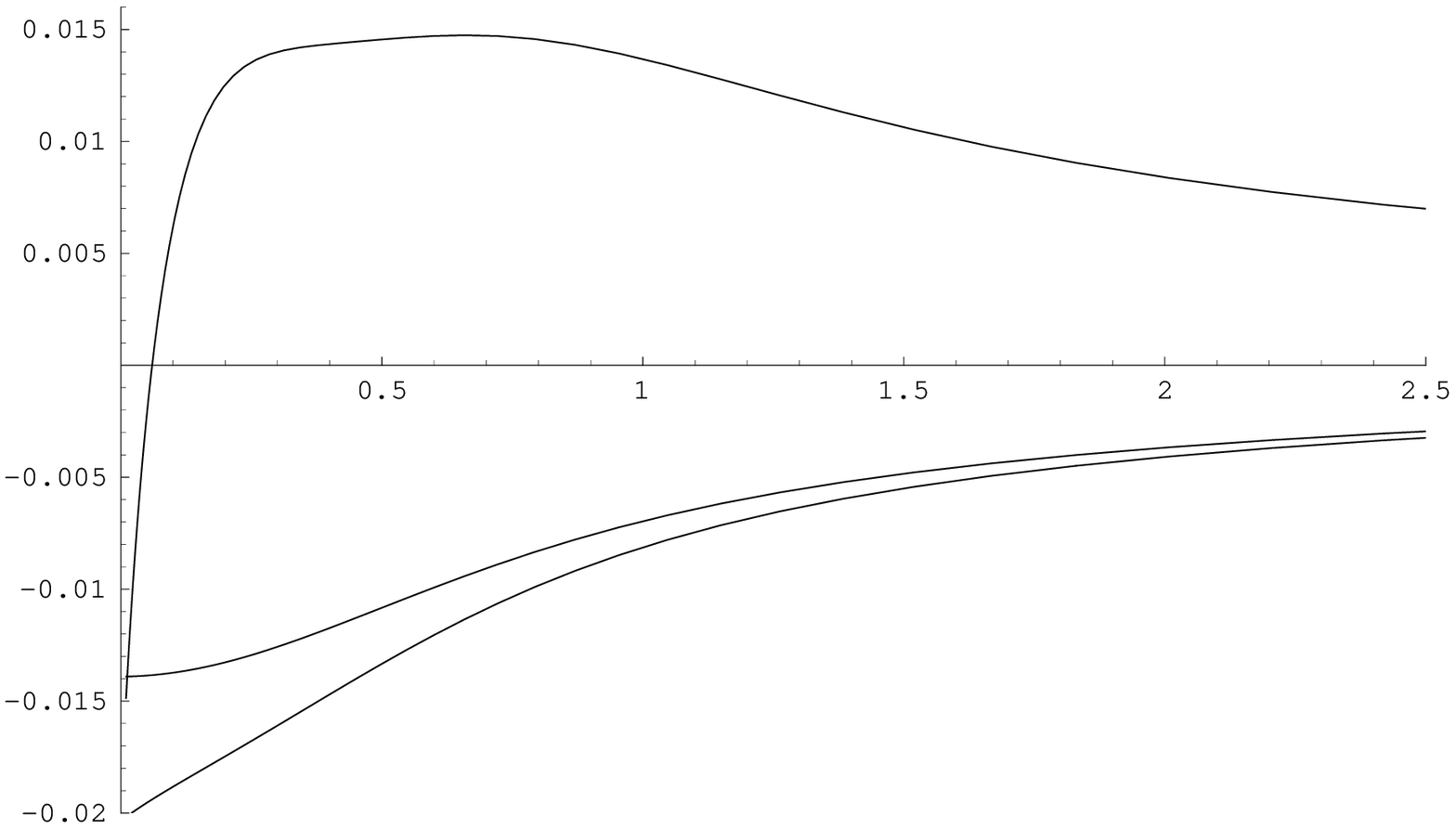,width=14cm,height=6cm}}
\put(2.5,0.5){$\beta=20$}
\put(2.3,5.9){$\beta=0.4$}
\put(3.3,2.2){$\beta=2,2$}
\put(14,3.3){$R$}
\end{picture}
\caption{ Scalar field. The complete renormalized vacuum energy  ${\cal E}_{ren}^{sc}(R)$ multiplied by $R^{2}\cdot \beta^{-4}$, for different values of strength of the potential.} 
\end{figure}

\chapter{Vacuum energy in the presence of a magnetic flux tube: Spinor case}

From the toy model analyzed  in the last chapter, we pass here to  the more interesting situation of a spinor field in the background of a magnetic string. 
The mathematical setup is analogous to that of the preceding chapter. The quantum field theoretical problem must be reformulated for the Fermi-Dirac statistic and for a quantum field with an additional degree of freedom. The Jost function will turn out to be simpler as that derived in the scalar situation. Furthermore the  contribution to the vacuum energy coming from the orbital momentum $m=0$ will not need to be treated separately.

\section{Solution of the field equation}

The analysis of a spinor field in the background of a cylindrical magnetic field with an arbitrary profile has been performed in \cite{master}. The field equation for a spinor field
\begin{equation}
\Psi(r)\ = \left( \begin{array}{c} g_1(r)\\ g_2(r) \end{array}\right)
\end{equation}
in the background of a translationally invariant potential, with delta function profile is
\begin{equation}
\left( \begin{array}{ll}
p_0-m_e & \partial_r-\frac{m-\beta\Theta(R-r)}{r}\\
-\partial_r-\frac{m+1-\beta\Theta(R-r)}{r} & p_0+m_e
\end{array}\right)
\left( \begin{array}{c} g_1(r)\\ g_2(r) \end{array} \right)\ =\ 0\ .
\end{equation}
The reader is referred to paper \cite{master} for a derivation of this equation. 
Let us find the solution to (6.2) for one component of the spinor\footnote{We are not interested here in the complete set of solutions to eq.(6.2), the solution for $g_2(r)$ is sufficient to find  the Jost function of the scattering problem. The decoupled equation for the component $g_1(r)$ is however indispensable and it will be used later.}. The decoupled equation for the component $g_2$ is
\begin{equation}
\left( k^2-\frac{\left[ m-\beta\Theta(R-r)\right]^2}{r^2}+\frac{\beta}{r}\delta(R-r) +\frac 1r\partial_r +\partial_r^2\right)g_2(r)\ =\ 0\ ,
\end{equation}
where $k=\sqrt{p_0^2-m_e^2}$. The solution in the region $r<R$ is
\begin{equation}
g^I_2(r)\ =\ J_m(kr) .
\end{equation}
and in the region $r>R$
\begin{equation}
g^O_2(r)\ =\ \frac 12 \left[f^{spin}_m(k)H^{(2)}_{m-\beta}(kr)+f^{spin^*}_m(k)H^{(1)}_{m-\beta}(kr)\right]\ ,
\end{equation}
here $f^{spin}_m(k)$ and $f^{spin^*}_m(k)$ are the Jost function and its conjugate related to the scattering problem for the spinor field. 
Solutions (6.4) and (6.5) have the same form as those found in the scalar case for the inner and outer space. However at $r=R$ the field has not the same form, as we discuss below, owing to the presence of the term $\frac{\beta}{r}\delta(R-r)$ in the field equation (6.3).

The ground state energy of the spinor field in the background of the magnetic string is
\begin{equation}
E_0\ =\ -\frac{\mu^2}{2}\sum_{n,\alpha,\sigma} \epsilon_{(n,\alpha,\sigma)}^{1-2s}\ ,
\end{equation}
where the minus sign accounts for the change of the statistics, and the $\epsilon_{(n,\alpha,\sigma)}$ are the eigenvalues of the Hamiltonian
\begin{equation}
H\ =\ -i\gamma^0\gamma^l\left(\partial_{x^l}-ieA_l(x)\right)+\gamma^0m_e\ .
\end{equation}
The degree of freedom $\sigma$ accounts for the two independent spin states.
As in the scalar case we calculate the energy for a section of the string. The ground state energy density per unit length of the string in terms of the Jost function is given by
\begin{equation}
{\cal E}^{spin}\ =\ C_s\ \sum^\infty_{m=-\infty}\int^\infty_{m_e}dk\ (k^2-m_e^2)^{1-s} \partial_k \ln f^{spin}_{m}(ik)\ .
\end{equation}
The renormalization scheme is the same we introduced for the scalar case. The expansion of the ground state energy in powers of the mass and the definition of ${\cal E}^{spin}_{div}$ are the same as  in (5.28), (5.29), apart from a factor $-1$ coming from the change of the statistics. The heat kernel coefficients will be of course not the same, we call them $B_n$. The normalization condition (5.30) remains unchanged. The ground state energy is split into the two parts
\begin{equation}
{\cal E}^{spin}_f= \frac{1}{2\pi} \sum_{m=-\infty}^\infty\int^\infty_{m_e} dk [k^2 -m_e^2] \frac{\partial}{\partial k} [\ln f^{spin}_m(ik)- \ln f^{as-spin}_m(ik)] 
\end{equation}
and
\begin{equation}
{\cal E}^{spin}_{as}= C_s\sum_{m=-\infty}^\infty \int^\infty_{m_e} dk [k^2 -m_e^2]^{1 -s} 
\frac{\partial}{\partial k}  \ln f^{as-spin}_m(ik)-{\cal E}_{div}^{spin}\ ,
\end{equation}
where $f^{as-spin}_m(ik)$ is the uniform asymptotic expansion of the Jost function taken up to the third order in $m$. In (6.9) the analytical continuation $s\rightarrow 0$ has already been performed, while in ${\cal E}^{spin}_{as}$ it will be performed after the subtraction of the divergent portion. (6.10) is a finite quantity for $s=0$.

\section{Matching conditions and  Jost function}
The matching conditions for the field on the surface of the string are not the same as in the scalar case. More exactly the condition for the first derivative of the field at $r=R$ is different from the one in the scalar case. The field is free inside the magnetic cylinder i.e. independent from $\beta$, then from eq.(6.2) and (6.4) we have
\begin{equation} 
k g_1^I(r)+\partial_r g_2^I(r)-\frac mr g_2^I(r)\ =\ 0
\end{equation}
and outside  the cylinder
\begin{equation} 
k g_1^O(r)+\partial_r g_2^O(r)-\frac {m-\beta}{r}g_2^O(r)\ =\ 0\ ,
\end{equation}
the continuity of the field on the boundary is required as in the scalar case:
\begin{equation}
g_2^I(r)|_{r=R}\ =\ g_2^O(r)|_{r=R}\ , \ \ g_1^I(r)|_{r=R}\ =\ g_1^O(r)|_{r=R}\ , 
\end{equation}
therefore, combining (6.11),(6.12) and (6.13) we find
\begin{equation}
(\partial_r g_2^I(r))|_{r=R}-(\partial_r g_2^O(r))|_{r=R}\ =\ \frac{\beta}{r}g_2^I(r)\ .
\end{equation}
Finally, the matching conditions at $r=R$ read
\begin{equation}
\left\{
\begin{array}{l}
g_2^I(r)|_{r=R}\  =\ g_2^O(r)|_{r=R}\\
(\partial_r g_2^I(r))|_{r=R} -(\partial_r g_2^O(r))|_{r=R}\ =\  \frac {\beta}{r} g_2^I(r)|_{r=R}\ .
\end{array}
\right.
\end{equation}
Inserting in (6.15) solutions (6.4) and (6.5) we get the system 
\begin{equation}
\left\{
\begin{array}{lcl}
J_m'- \frac 12 \left[f_m(k)H'^{(2)}_{m-\beta}+f^*_m(k)H'^{(1)}_{m-\beta}\right]\ =\ \frac{\beta}{kR} J_m\\
J_m\ =\ \frac 12 \left[f_m(k)H^{(2)}_{m-\beta}+f^*_m(k)H^{(1)}_{m-\beta}\right]
\end{array}
\right.
\end{equation}
for positive $m$,  and the system
\begin{equation}
\left\{
\begin{array}{lcl}
J_{-m}'- \frac 12 \left[f_m(k)H'^{(2)}_{\beta-m}+f^*_m(k)H'^{(1)}_{\beta-m}\right]\ =\ \frac{\beta}{kR} J_{-m}\\
J_{-m}\ =\ \frac 12 \left[f_m(k)H^{(2)}_{\beta-m}+f^*_m(k)H^{(1)}_{\beta-m}\right]
\end{array}
\right.
\end{equation}
for negative $m$. In (6.16) and (6.17) the argument $(kR)$ of the Bessel and of the Hankel functions has been omitted for simplicity and the prime symbol indicates derivative with respect to this  argument. Much in the same way as we did in the scalar case we find the Jost functions on the imaginary axis
\begin{equation}
f_m^{spin+}(ik)\ =\ i^\beta kR \left[I_mK_{m-\beta+1}+I_{m+1}K_{m-\beta}\right], \ \ m>0  
\end{equation}
\begin{equation}
f_m^{spin-}(ik)\ =\ i^{-\beta} kR \left[I_mK_{m-\beta+1}+I_{m+1}K_{m-\beta}\right], \ \ m<0\ .  
\end{equation}
The two Jost functions are identical apart form the sign on the exponent of the imaginary factor. However  formula (6.8) for the ground state energy contains the logarithm of the Jost function and the derivative with respect to $k$, then the imaginary factor $i^{\pm\beta}$ which is independent of $k$ will not contribute to the energy. 

For the calculation of the asymptotic Jost function we found more convenient to use instead of $m$ the expansion parameter $\nu$ given by
\begin{equation}
\nu\ =\ \left\{ \begin{array}{rcl}
m+ 1/2  & {rm for} & m=0,1,2...\\
-m- 1/2 & {rm for} &  m=-1,-2,... 
\end{array}
\right.
\end{equation} 
with $\nu=\frac 12, \frac 32,...$ in both cases. Then the Jost functions (6.18) and (6.19) become
\begin{equation}
f_{\nu}^+(ik)\ =\ i^\beta kR \left[I_{\nu+\frac 12} K_{\nu-\frac 12-\beta}+I_{\nu-\frac 12}K_{\nu+\frac 12 -\beta}\right],\ \ m\geq 0  \end{equation}
which can be expanded for large positive $m$
\begin{equation}
f_{\nu}^-(ik)\ =\ i^{-\beta} kR \left[I_{\nu+\frac 12} K_{\nu-\frac 12 +\beta}+I_{\nu-\frac 12}K_{\nu+\frac 12 + \beta}\right]\ ,\ \ m<0  
\end{equation}
which can be expanded for large negative $m$.
We recall the formulas for the asymptotic expansions of the Bessel I and K functions for $\nu$ and $k$ equally large, given by
\begin{equation}
I_{\nu+a}\ \sim\ \frac{1}{\sqrt{2\pi\nu}}\exp \left\{ \sum_{n=-1}x^n S_I(n,a,t) \right\}\ ,
\end{equation}
\begin{equation}
K_{\nu+a\pm \beta}\ \sim\ \frac{\sqrt{\pi}}{\sqrt{2\nu}}\exp \left\{ \sum_{n=-1}x^n S_K(n,a,t) \right\}\ ,
\end{equation}
where $x\equiv 1/\nu$, the functions  $S_I(n,a,t)$, $S_K(n,a,t)$  are the same as in the scalar case  and $a$ takes the values $\pm \frac 12$ for the Bessel I function and $\pm\frac 12\pm\beta$ for the Bessel K function. From these formulas the logarithm of the asymptotic Jost function can be easily calculated up to the third order and we define
\begin{equation}
\ln f^{as-spin}_{\nu}(ik)\ =\ \sum_{j,n}^3 Y_{j,n}\frac{t^j}{\nu^n}\ ,
\end{equation}
where  $t=1/(1+(kR/\nu)^2)^{\frac 12}$ and the nonzero coefficients are
\begin{equation}
\begin{array}{l}
Y_{1,1}=\beta^2\ ,\ Y_{2,2}=-\beta^2/4\ ,\ Y_{2,4}=\beta^2/4\ , \  \\
Y_{3,3}=\beta^2/6-\beta^4/12\ ,\ Y_{3,5}=-7\beta^2/8+\beta^4/4\ ,\\ Y_{3,7}=5\beta^2/8\ .
\end{array}
\end{equation}

\section{The asymptotic and the finite part of the energy}

The asymptotic part of the energy can be written, using result (6.26), as
\begin{equation}
{\cal E}^{spin}_{as}=  C_s\sum_{\nu=\frac 12}^\infty \int^\infty_{m_e} dk [k^2 -m_e^2]^{1 -s} 
\frac{\partial}{\partial k}\sum_{j,n}^3 Y_{j,n}\frac{t^j}{\nu^n}  -{\cal E}_{div}^{spin} ,
\end{equation}
we calculate the sum over $\nu$ with the help of the Abel-Plana formula for half integer variables which can be found in section 3.3 of the chapter devoted to the spherical shell. The case $m=0$ (i.e. $\nu=1/2$) does not need to be treated separately. We have only the two contributions
\begin{equation}
{\cal E}^{spin}_{as1}\ =\ \frac{\beta^2m_e^2}{4\pi}\left(\frac 1s +\ln\left(\frac{4\mu^2}{m_e^2}\right) -1\right)\-\ -\frac{\beta^2m_e}{16R} 
\end{equation}
and
\begin{equation} 
{\cal E}^{spin}_{as2}\ =\ C_s \sum_{n,j}^{3,7}Y_{n,j}(-m_e^{2-2s}\Gamma(2-s))\Sigma_{n,j}(m_eR)\ .
\end{equation}
The functions $\Sigma_{n,j}(x)$ are given in the appendix. They correspond to the functions $\Sigma_{n,j}(x)$  found in paper \cite{master} for a generic smooth background potential.
The only pole term is contained in ${\cal E}^{spin}_{as1}$ and the term proportional to $m_e$ will be subtracted as well as the pole term, thus (6.28) cancels  completely with the subtraction of ${\cal E}_{div}^{spin}$. The heat kernel coefficients which we have calculated up to the coefficient $B_4$, read
\begin{equation}
\begin{array}{lclcccl}
B_0 & = & 0                 & , &   B_{1/2} & = & 0 \nonumber \\
B_1 & = &  8\pi\beta^2    & , &   B_{3/2} & = &  -\beta^2 \pi^{3/2}/R \nonumber \\
B_2 & = & 0                & , &   B_{5/2} & = & \frac{(3\beta^2+2\beta^4)\pi^{3/2}}{64R^3} \nonumber \\
B_3 & = & 0                & , &  B_{7/2}  & = & -\frac{(135\beta^2 -68\beta^4+16\beta^6)\pi^{3/2}}{12288 R^5}\nonumber \\
B_4 & = &  \frac{5\beta^8}{1281 R^6}  & . &  &  & 
\end{array}
\end{equation}
After the subtraction of ${\cal E}_{div}^{spin}$ the asymptotic part of the energy reads
\begin{eqnarray}
{\cal E}^{spin}_{as} & = & \frac{2\beta^2}{\pi R^2}\  q_1(m_R)\ +\ \left( -\frac{\beta^2}{3 \pi R^2}+\frac{\beta^4}{6 \pi R^2}\right)\ q_2(m_eR)\nonumber \\ 
       &  & +\left( \frac{7\beta^2}{12 \pi R^2}+\frac{\beta^4}{6 \pi R^2}\right)\ q_3(m_eR)\ -\ \frac{\beta^2}{12 \pi R^2}\  q_4(m_eR)\ ,      
\end{eqnarray}
the functions $q_n(x)$ are
\begin{eqnarray}
q_1(x) & = & \int_{x}^{\infty} \frac{d\nu}{1+e^{2\pi \nu}} \sqrt{\nu^2 -x^2} \nonumber \\
q_2(x) & = & \int_{x}^{\infty} d\nu \left(\frac{1}{1+e^{2\pi \nu}}\frac {1}{\nu} \right)' \sqrt{\nu^2 -x^2} \nonumber \\
q_3(x) & = & \int_{x}^{\infty} d\nu \left(\left(\frac{\nu}{1+e^{2\pi \nu}}\right)'\frac {1}{\nu} \right)' \sqrt{\nu^2 -x^2} \nonumber \\
q_4(x) & = & \int_{x}^{\infty} d\nu \left( \left( \left(\frac{\nu^3}{1+e^{2\pi \nu}}\right)'\frac {1}{\nu} \right)' \frac {1}{\nu}\right)' \sqrt{\nu^2 -x^2}\ .
\end{eqnarray}
It is interesting to note how in both scalar and spinor cases we found the final expression for the asymptotic part of the energy to depend only on even powers of $\beta$. This was to expect for physical reasons, in fact, inverting the direction of the magnetic flux $\phi$  the ground state energy should not change.
 
The finite part of the energy  ${\cal E}^{spin}_f$ can be hardly analytically simplified. We can only integrate expression (6.9) by parts to obtain a final form which is suitable for the numerical calculation:
\begin{eqnarray}
{\cal E}^{spin}_f & = & -\frac{1}{\pi}\sum_{\nu=\frac 12}^{\infty}\int_{m_e}^{\infty} dk\ k \left[\ln f^{\pm}_{\nu}(ik) - \sum_{n,j}^{3,7} Y_{n,j}\frac{t^j}{\nu^n}\right]\ ,
\end{eqnarray}
where
\begin{eqnarray}
\ln f^{\pm}_{\nu}(ik) & = &  \ln \left[ I_{\nu+\frac 12} K_{\nu-\frac 12-\beta}+I_{\nu-\frac 12}K_{\nu+\frac 12 -\beta}\right]\nonumber \\                            &   & +\ \ln \left[I_{\nu+\frac 12} K_{\nu-\frac 12 +\beta}+I_{\nu-\frac 12}K_{\nu+\frac 12 + \beta} \right]\ .
\end{eqnarray}

\section{Numerical results}

We study numerically the behaviour of equations (6.33) and (6.31) and of their sum, which expresses the total renormalized energy, as functions of the radius of the string. The asymptotic behaviour of the two contribution is first found.
The asymptotic part of the energy is rewritten in the form
\begin{equation}
{\cal E}^{spin}_{as}\ =\ \frac{1}{12\pi R^2}\left[\beta^2 e_1(m_eR)\ +\ \beta^4 e_2(m_e R)\right]\ ,
\end{equation}
where
\begin{eqnarray}
e_1(x) & = & \left(24q_1(x)-4q_2(x)+7 q_3(x)-q_4(x)\right)\ ,\nonumber\\
e_2(x) & = & 2\left(q_2(x)-q_3(x)\right)\ .
\end{eqnarray}
The asymptotic behaviour of the $q_n(x)$ functions for $x\rightarrow\ 0$ is
\begin{eqnarray}
q_1(x) & \sim & 1/48\ + {\cal O}(x)\ ,\nonumber\\
q_2(x) & \sim & \frac 12 \ln x +0.635\ +{\cal O}(x)\ ,\nonumber\\
q_3(x) & \sim & \frac 12 \ln x +1.135\ +{\cal O}(x)\ ,\nonumber\\
q_4(x) & \sim & \frac 32\ln x + 3.906\ +{\cal O}(x)
\end{eqnarray}
and the corresponding behaviour of the $e_n(x)$ functions is
\begin{eqnarray}
e_1(x) & \sim & 2 +\ {\cal O}(x)\ ,\nonumber\\
e_2(x) & \sim & -1 +\ {\cal O}(x) \ ;
\end{eqnarray}
therefore ${\cal E}^{spin}_{as}$ behaves like $R^{-2}$ for small values of $R$.
For $R\rightarrow\infty$ ${\cal E}^{spin}_{as}$ falls like $e^{-R}$.

The finite part of the energy can be rewritten as
\begin{eqnarray}
{\cal E}^{spin}_f & = & -\frac{1}{\pi R^2}\sum_{\nu=\frac 12}^{\infty}\int_{m_eR}^{\infty} dk\ k \left[\ln f^{\pm}_{\nu}(ik)|_{k\rightarrow k/R} - \sum_{n,j}^{3,7} Y_{n,j}\frac{t^j}{\nu^n}\right]\ ,
\end{eqnarray}
which is proportional to $R^{-2}$ for $R\rightarrow 0$ and to $R^{-3}$ for $R\rightarrow \infty$ in agreement with the heat kernel coefficient $B_{5/2}$. We give the plots of the functions $e_1(x)$, $e_2(x)$ of ${\cal E}^{spin}_{as}$, ${\cal E}^{spin}_f$ and of ${\cal E}^{spin}_{ren}$, for small and for large values of the potential strength. The same remarks which we made in the scalar case about the numerical limits of summation and integration and about the precision employed in the computer graphics are valid here.  

\begin{figure}[ht]\unitlength1cm
\begin{picture}(6,6)
\put(-0.5,0){\epsfig{file=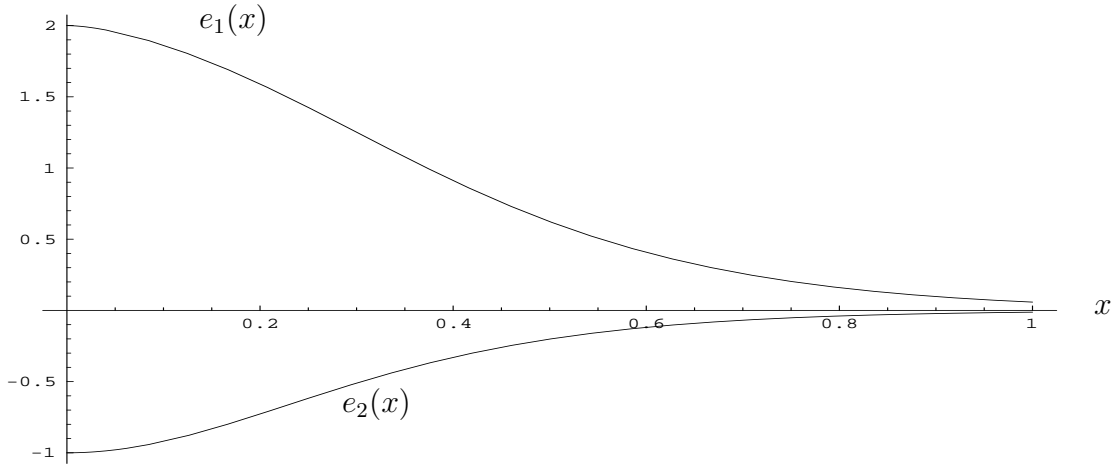,width=14cm,height=6cm}}
\put(2.1,5.8){$e_1(x)$}
\put(14,2){$x$}
\put(4,0.7){$e_2(x)$}
\end{picture}
\caption{ Spinor field. The functions $e_1(x)$ and $e_2(x)$ contributing to the asymptotic part of the ground state energy.} 
\end{figure}

\begin{figure}[ht]\unitlength1cm
\begin{picture}(6,6)
\put(-0.5,0){\epsfig{file=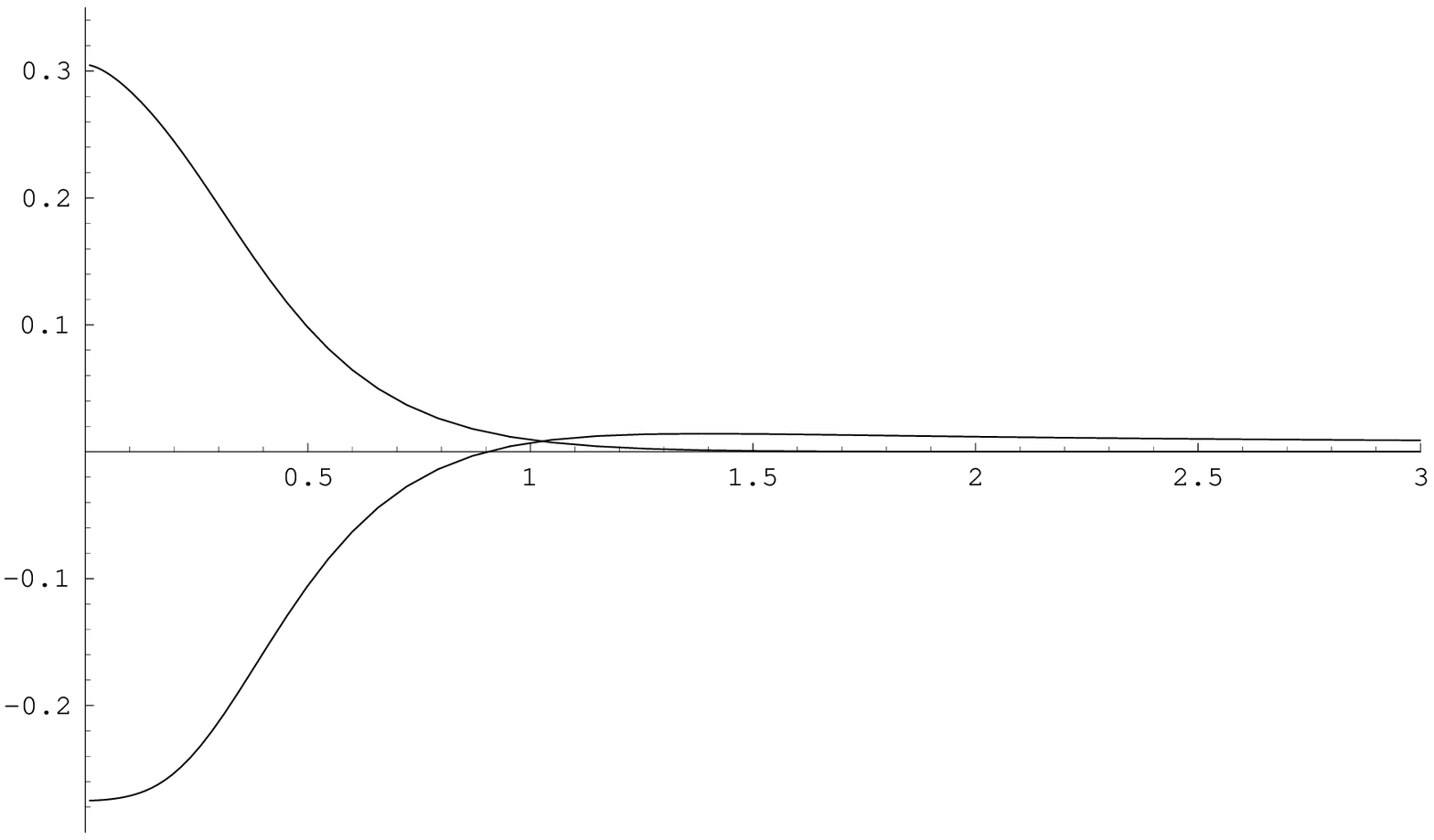,width=14cm,height=6cm}}
\put(1.5,5.0){${\cal E}^{spin}_{as}$}
\put(2,0.75){${\cal E}^{spin}_f$}
\put(14,2.7){$R$}
\end{picture}
\caption{ Spinor field. The curves of  the asymptotic and of the finite part of the energy multiplied by $R^{2}\cdot \beta^{-4}$, for $\beta =0.4$.} 
\end{figure}

\begin{figure}[ht]\unitlength1cm
\begin{picture}(6,6)
\put(-0.5,0){\epsfig{file=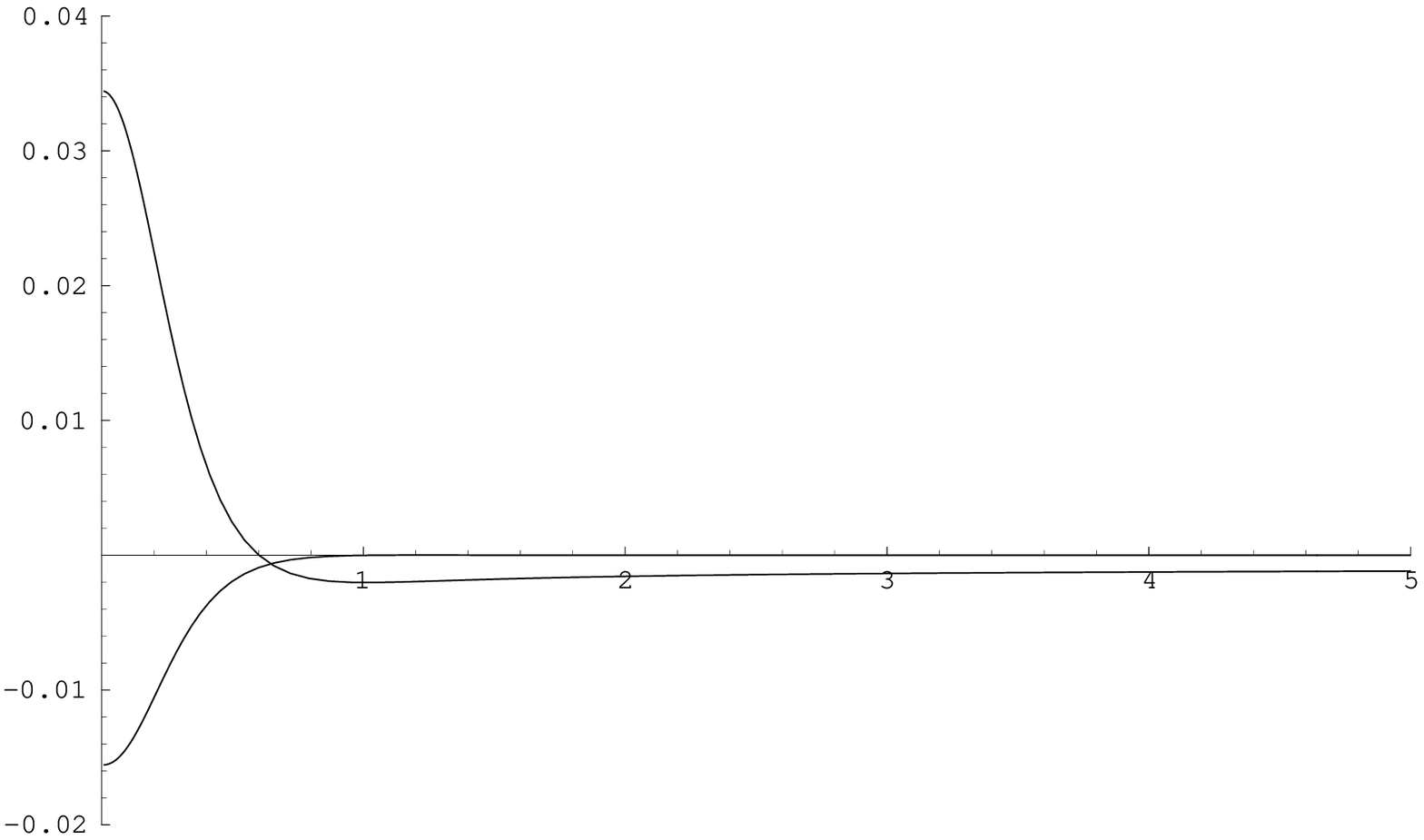,width=14cm,height=6cm}}
\put(1.1,5.0){${\cal E}^{spin}_{as}$}
\put(1.1,0.75){${\cal E}^{spin}_f$}
\put(14,2){$R$}
\end{picture}
\caption{Spinor field. The curves of  the asymptotic and of the finite part of the energy multiplied by $R^{2}\cdot \beta^{-4}$, for $\beta =2.2$.} 
\end{figure}

\begin{figure}[ht]\unitlength1cm
\begin{picture}(6,6)
\put(-0.5,0){\epsfig{file=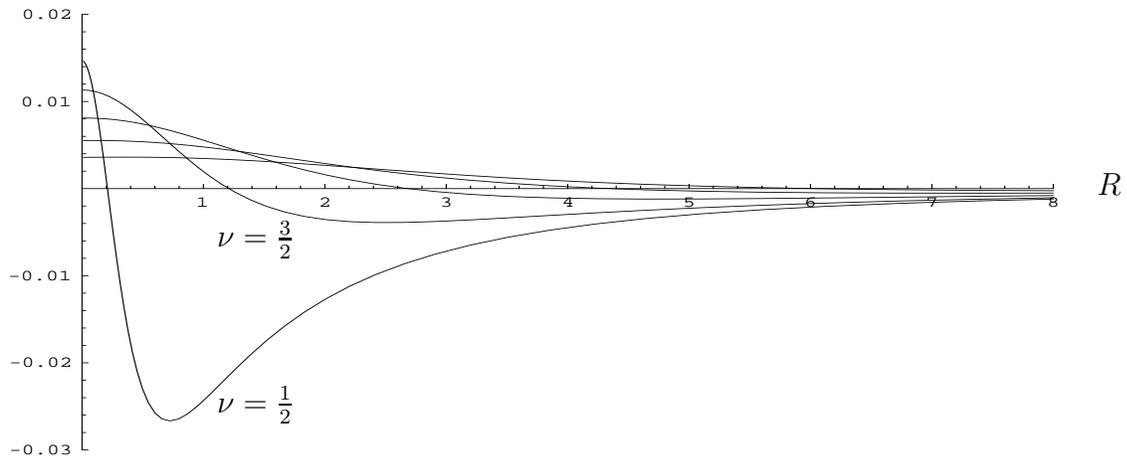,width=14cm,height=6cm}}
\put(2.3,2.8){$\nu=\frac 32$}
\put(2.3,0.60){$\nu=\frac 12$}
\put(14,3.5){$R$}
\end{picture}
\caption{Spinor field. The contributions $\nu=\frac 12,...\frac 92$ to the finite part of the energy multiplied by $R^{2}\cdot \beta^{-4}$, for $\beta =10$.} 
\end{figure}
\begin{figure}[ht]\unitlength1cm
\begin{picture}(6,6)
\put(-0.5,0){\epsfig{file=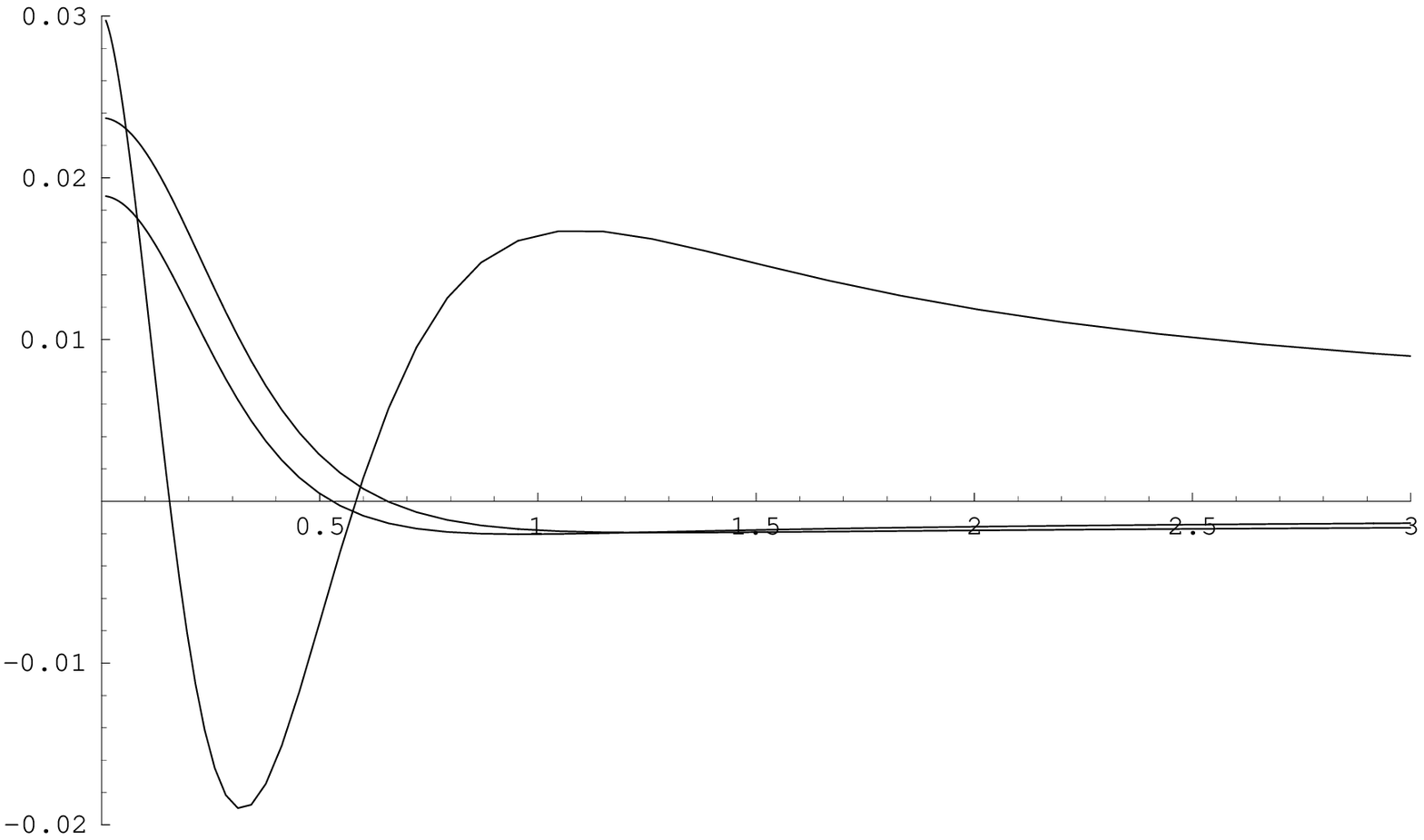,width=14cm,height=6cm}}
\put(1.5,4.2){$\beta=10$}
\put(6.3,4.35){$\beta=0.4$}
\put(3.3,1.8){$\beta=2,2$}
\put(14,2.4){$R$}
\end{picture}
\caption{Spinor field. The complete renormalized vacuum energy  ${\cal E}^{spin}(R)$ multiplied by $R^{2}\cdot \beta^{-4}$, for different values of strength of the 
potential.} 
\end{figure}

\section{Discussion}

In this chapter the vacuum energy of a spinor field in the presence of a semi-transparent magnetic was calculated, analytically in the first part of the chapter, and numerically in the remaining part. 
As it is clear from the plots, the renormalized vacuum energy changes with the radius, being not definitely positive or negative.

With respect to the scalar case, we have an opposite behaviour  in the region $R<1$ (Fig. 6.5), where the energy shows a minimum for small values of $\beta$, being  positive for $\beta > 1 $. However, when the radius become large, the vacuum energy shows the same behaviour in the scalar and in the spinor case: it is negative for large $\beta$ and positive for small $\beta$. 
In the limit $\beta \rightarrow \infty$ we found the energy to be proportional to $\beta^4$. This opens the interesting possibility that the vacuum energy could overwhelm the classical energy of the system when the flux is sufficiently strong, in fact we have
\begin{equation}
E_{TOT}\ \sim \ \underbrace{\frac{\beta^2 {\cal A}}{\alpha R^2}}_{E_{class}}\ +\ \underbrace{\frac{\beta^2 {\cal B} +\beta^4 {\cal C}}{R^2}}_{E_{vacuum}}\ ,
\end{equation}
where ${\cal A}$, ${\cal B}$ and ${\cal C}$ are numbers and $\alpha $ is the fine structure constant. This possibility is, however, not directly applicable to the model studied here. In fact,  owing to the delta function the classical energy has not a finite value.

\chapter{Conclusion} 

We have studied the vacuum energy in the presence of curved semi-transparent
boundaries represented by delta functions. The model and the results give rise
to a number of discussions in the context of the existing literature. First, the
sign of the Casimir energy in some of the plots provided in this work (Fig. 3.5,
chapter 3, Fig. 6.5 chapter 6) is not in agreement with that found in calculations with similar models \cite{Bordag Elizalde,master}. Secondly, it seems arduous to extrapolate a general rule for the sign of the Casimir energy in the presence of a non-ideal boundary like a non-ideal sphere or a non-ideal cylinder. More exactly, in almost all of the situations investigated the sign of the energy is not fix, but it varies with the dimensions of the cavity. This  cannot  depend simply on the peculiarity of the potential profile chosen (which is singular and therefore unphysical) in fact, in papers \cite{cricca,Jou31}, the vacuum energy of a scalar field in the presence of a spherical shell with Dirichlet boundary conditions and of a spinor field  in a bag, was calculated and the sign  was also found to depend on the radius. This result may express the interesting feature that the Casimir energy is in general a non monotonous function. It is also interesting the possibility that the energy could have a minimum in some situations, like it was found in chapter 3, for spherical shells with attractive potential or in chapter 6 for the magnetic string. This would imply that the Casimir force exerted on a hollow non-ideal object could vanish for some particular size of the object, and that the object could find some equilibrium in vacuum. However one should be careful in generalizing the results found here. We remeber that we have dealed for 3/4 of the work with the toy model of a scalar field, which does not have a counterpart in reality. Furthermore, in  all the analyzed systems the classical energy has not been taken in to account. The classical contribution, which must undergo a renormalization, could compensate for the above mentioned effects. Only in the background of the magnetic string we found that the vacuum contribution could dominate the total energy of the system. However, to investigate extensively this aspect one should apply the method to a inhomogeneous magnetic string with non-singular potential, possibly taking also into account the anomalous magnetic moment. As a matter of fact, the results found in chapter 5 are far from those found in  \cite{master}, where a homogeneous flux tube with square well profile is considered with much the same procedure employed here. In that work the vacuum energy was found to be negative in all the $R$-axes. This does not take place in our computations. However, in the concluding section of paper \cite{master} it was supposed that the presence of a singular background could considerably change the results.  

A common feature of all the investigated backgrounds is that the vacuum energy shows very weak UV-divergences. A crucial term in ${\cal E}_{div}$, namely that given by the heat-kernel coefficient $A_2$ was found to depend only on the third power of the coupling, while all the lower powers have cancelled. We had consequently a small number of pole terms in the calculations. The heat-kernel coefficients, which we calculated in the problem of the cylinder and of the magnetic string up to the coefficient $A_{5/2}$, are confirmed by those calculated in other works \cite{Bordag dielectric,BordagVass-heat kernels}. 

In the background of the magnetic string we  found the coefficient $A_2$ to be zero. Here the renormalized vacuum energy possesses no logarithmic behaviour. A logarithmic behaviour for small values of $R$ was found in the background of a spherical shell (eq.(3.46)) and of a cylindrical shell (eq.(4.44)).

\appendix
\chapter{Asymptotic expansion of the modified Bessel functions}

For the problems of the cylindrical shell  and  of the magnetic string, the expansions of the Bessel functions in negative powers of the parameters $l$, $m$ and $\nu$ (chapters 4, 5 and 6,  respectively) are obtained from the following formulas
\begin{equation}
I_{l+a}(kR) \ \sim\ \frac{1}{\sqrt{2\pi l}}\exp \{\sum_{n=-1}^3 l^{-n} S_I(n,a,t)\}\ ,
\end{equation}
\begin{equation}
K_{l+a}(kR)\ \sim\ \sqrt{\frac{\pi}{2l}}\exp \{\sum_{n=-1}^3 l^{-n} S_K(n,a,t)\}\ ,
\end{equation}
where $a$ takes the values $1$ and $0$ in the case of the cylinder  and the values specified in sections 5.3 and 6.2 in the case of the string. The functions $S_I(n,a,t)$ and $S_K(n,a,t)$ are Deby  polinomials. Up to the third order they are given by
\begin{eqnarray}
S_I(-1,a,t) & = & t^{-1} + \frac 12 \ln\left(\frac{1 - t}{1 + t}\right)\ ,\nonumber \\
 S_I(0,a,t) & = & \frac 12\ln t -\frac{a}{2} \ln\left(\frac{1 + t}{1 - t}\right)\ ,\nonumber \\
S_I(1,a,t)& = & -\frac {t}{24} (-3 + 12a^2 + 12a t + 5 t^2)\ ,\nonumber \\
S_I(2,a,t) & = & \frac{t^2}{48} [8a^3 t + 12a^2 (-1 + 2 t^2) + 
       a(-26 t + 30 t^3) + 3 (1 - 6 t^2 + 5 t^4)]\ ,\nonumber \\
S_I(3,a,t) & = & \frac {1}{128}(((25 - 104a^2 + 16a^4) t^3)/3 + 
      16 a (-7 + 4a^2) t^4\nonumber\\
               &   & - (531/5 - 224a^2 + 16a^4)t^5 - (32a (-33 + 8a^2) t^6)/3\nonumber \\
               &   & - (-221 + 200a^2) t^7 - 240a t^8 - (1105 t^9)/9)\ ;
\end{eqnarray}
\begin{eqnarray}
S_K(-1,a,t) & = & -t^{-1} -\frac 12 \ln\left(\frac{1 - t}{1 + t}\right)\ ,\nonumber \\
 S_K(0,a,t) & = & \frac 12\ln t +\frac{a}{2} \ln\left(\frac{1 + t}{1 - t}\right)\ ,\nonumber \\
S_K(1,a,t)& = & -\frac {t}{24} (-3 + 12a^2 + 12a t + 5 t^2)\ ,\nonumber \\
S_K(2,a,t) & = & \frac{t^2}{48} [-8a^3 t + 12a^2 (-1 + 2 t^2) + 
       a(-26 t + 30 t^3) + 3 (1 - 6 t^2 + 5 t^4)]\ ,\nonumber \\
S_K(3,a,t) & = & -\frac {1}{128}((25 - 104a^2 + 16a^4) t^3)/3 + 
      16 a (-7 + 4a^2) t^4\nonumber\\
               &   & - (-531/5 + 224a^2 - 16a^4)t^5- (32a (-33 + 8a^2) t^6)/3\nonumber\\
               &   & - (221 - 200a^2) t^7 - 240a t^8 + (1105 t^9)/9\ ,
\end{eqnarray}
where $t=(1+(kR/l)^2)^{-\frac 12}$ for the problem of the cylindrical shell,  $t=(1+(kR/m)^2)^{-\frac 12}$ for problem of a scalar field in the background of a string and  $t=(1+(kR/\nu)^2)^{-\frac 12}$ for a spinor field in the background  of a string.

\chapter{Calculation of the integrals}

In chapters 4, 5 and 6 the contributions to the asymptotic part of the energy coming from the Abel-Plana formula are calculated  by means of the formulas given below, which were first derived in \cite{master}. The integration variable $l$ is that for the cylinder. In the problem of the magnetic string it corresponds to $m$ (scalar case) or to $\nu$ (spinor case). The following formula
\begin{equation}
\int_0^\infty dl\ \int_{m}^\infty dk\ (k^2-m^2)^{1-s}\partial_k\frac{t^j}{l^n}=-\frac{m^{2-2s}}{2}\frac{\Gamma(2-s)\Gamma\left(\frac{1+j-n}{2}\right)\Gamma(s+\frac{n-3}{2})}{(Rm)^{n-1}\Gamma(j/2)}\ .
\end{equation}
has been used to obtain equations (4.31),(5.58),(5.59) and (6.28). Formula
\begin{equation}
\int_{m}^\infty dk\ (k^2-m^2)^{1-s}\partial_k\frac{t^j}{l^n}=-m^{2-2s}\frac{\Gamma(2-s)\Gamma\left(s+\frac j2-1\right)l^{j-n}}{\Gamma \left(\frac j2\right) (Rm)^{j}\left( 1+\left(\frac{l}{mR}\right)^2\right)^{s+\frac j2-1}}\ ,
\end{equation}
has been applied to arrive at the results (4.32), (4.35), (5.62), (6.29). The functions $Z_{n,j},\ \Lambda_{n,j}(x)$  and $\Sigma_{n,j}(x)$ calculated for  $n\leq 3,j\leq 7$ read 
\begin{eqnarray}
Z_{1,1}(x) & = & -\frac{4}{x^2}\int_x^\infty \frac{dl}{1-e^{2\pi l}} \sqrt{l^2 -x^2} \nonumber \\
Z_{2,2}(x) & = & +\frac{\pi}{x^2}\int_{x}^{\infty} dl \left(\frac{1}{1-e^{2\pi l}}\frac 1l \right)' (l^2 -x^2) \nonumber \\  
Z_{3,3}(x) & = & +\frac{4}{x^2}\int_{x}^{\infty} dl \left(\frac{1}{1-e^{2\pi l}}\frac 1l \right)' \sqrt{l^2 -x^2} \nonumber \\
Z_{3,5}(x) & = & +\frac{4}{x^2}\int_{x}^{\infty} dl\ \frac 13 \left(\left(\frac{l}{1-e^{2\pi l}}\right)'\frac 1l \right)' \sqrt{l^2 -x^2} \nonumber \\
Z_{3,7}(x) & = & -\frac{4}{x^2}\int_{x}^{\infty} dl \ \frac{1}{15}\left( \left( \left(\frac{l^3}{1-e^{2\pi l}}\right)'\frac 1l \right)' \frac 1l\right)' \sqrt{l^2 -x^2}\ ;
\end{eqnarray}
\begin{eqnarray}
\Lambda_{1,1}(x) & = & -\frac{4}{x^2}\int_x^\infty \frac{dm}{1-e^{2\pi m}} \sqrt{m^2 -x^2} \nonumber \\
\Lambda_{2,4}(x) & = & -\frac{\pi}{x}\frac{1}{1-e^{2\pi x}}\nonumber\\
\Lambda_{3,3}(x) & = & -\frac{4}{x^2}\int_{x}^{\infty} dm \left(\frac{1}{1-e^{2\pi m}}\frac 1m \right)' \sqrt{m^2 -x^2} \nonumber \\
\Lambda_{3,5}(x) & = & -\frac{4}{x^2}\int_{x}^{\infty} dm\ \frac 13 \left(\left(\frac{m}{1-e^{2\pi m}}\right)'\frac 1m \right)' \sqrt{m^2 -x^2} \nonumber \\
\Lambda_{3,7}(x) & = & -\frac{4}{x^2}\int_{x}^{\infty} dm \ \frac{1}{15}\left( \left( \left(\frac{m^3}{1-e^{2\pi m}}\right)'\frac 1m \right)' \frac 1m\right)' \sqrt{m^2 -x^2}\ ;
\end{eqnarray}
\begin{eqnarray}
\Sigma_{1,1}(x) & = & -\frac{4}{x^2}\int_x^\infty \frac{d\nu}{1+e^{2\pi \nu}} \sqrt{\nu^2 -x^2} \nonumber \\
\Sigma_{3,3}(x) & = & -\frac{4}{x^2}\int_{x}^{\infty} d\nu \left(\frac{1}{1+e^{2\pi \nu}}\frac 1\nu \right)' \sqrt{\nu^2 -x^2} \nonumber \\
\Sigma_{3,5}(x) & = & -\frac{4}{x^2}\int_{x}^{\infty} d\nu\ \frac 13 \left(\left(\frac{\nu}{1+e^{2\pi m}}\right)'\frac 1\nu \right)' \sqrt{\nu^2 -x^2} \nonumber \\
\Sigma_{3,7}(x) & = &  -\frac{4}{x^2}\int_{x}^{\infty} d\nu \ \frac{1}{15}\left( \left( \left(\frac{\nu^3}{1+e^{2\pi \nu}}\right)'\frac 1\nu \right)' \frac 1\nu \right)' \sqrt{\nu^2 -x^2}\ ;
\end{eqnarray}

\vspace{3cm}
{\Large \bf Danksagung}
\vskip0.2in  
Mein Dank geht vor allem an Herrn Dr. Bordag f\"ur seine wertvolle  Betreuung. Nur dank seiner Anregungen und Hilfen konnte ich meine Untersuchungen ausf\"uhren, die Ergebnisse ver\"offentlichen und diese Dissertation anfertigen. Weiterhin danke ich allen Mitarbeitern des Institutes f\"ur Theoretische Physik der Universit\"at Leipzig und Klaus Kirsten f\"ur seinen wertvollen Rat, sowie allen Doktoranden im Graduiertenkolleg Quantenfeldtheorie.

\end{document}